\documentclass[aps,prd,twocolumn,superscriptaddress, nofootinbib, longbibliography, notitlepage]{revtex4-1}
\usepackage{epsf,epsfig,graphicx}
\usepackage{amsmath}
\usepackage{amssymb}
\usepackage{tikz}
\usepackage{subfigure}
\usepackage{multirow}
\usepackage{float}

\makeatletter
\DeclareFontFamily{OMX}{MnSymbolE}{}
\DeclareSymbolFont{MnLargeSymbols}{OMX}{MnSymbolE}{m}{n}
\SetSymbolFont{MnLargeSymbols}{bold}{OMX}{MnSymbolE}{b}{n}
\DeclareFontShape{OMX}{MnSymbolE}{m}{n}{
    <-6>  MnSymbolE5
   <6-7>  MnSymbolE6
   <7-8>  MnSymbolE7
   <8-9>  MnSymbolE8
   <9-10> MnSymbolE9
  <10-12> MnSymbolE10
  <12->   MnSymbolE12
}{}
\DeclareFontShape{OMX}{MnSymbolE}{b}{n}{
    <-6>  MnSymbolE-Bold5
   <6-7>  MnSymbolE-Bold6
   <7-8>  MnSymbolE-Bold7
   <8-9>  MnSymbolE-Bold8
   <9-10> MnSymbolE-Bold9
  <10-12> MnSymbolE-Bold10
  <12->   MnSymbolE-Bold12
}{}

\let\llangle\@undefined
\let\rrangle\@undefined
\DeclareMathDelimiter{\llangle}{\mathopen}%
                     {MnLargeSymbols}{'164}{MnLargeSymbols}{'164}
\DeclareMathDelimiter{\rrangle}{\mathclose}%
                     {MnLargeSymbols}{'171}{MnLargeSymbols}{'171}
\makeatother

\RequirePackage[colorlinks,citecolor=red,urlcolor=red,linkcolor=red]{hyperref}

\def\edth{\;\raise1.0pt\hbox{$'$}\hskip-6pt\partial\;}
\def\baredth{\;\overline{\raise1.0pt\hbox{$'$}\hskip-6pt
\partial}\;}
\def\gsim{~\rlap{$>$}{\lower 1.0ex\hbox{$\sim$}}}

\newcommand{\be}{\begin{equation}}
\newcommand{\ee}{\end{equation}}
\newcommand{\bw}{\begin{widetext}}
\newcommand{\ew}{\end{widetext}}

\newcommand{\intinf}{\int_{-\infty}^{\infty}}
\newcommand{\intzeroinf}{\int_{0}^{\infty}}
\newcommand{\suml}{\sum_{\ell=0}^{\infty}}
\newcommand{\summ}{\sum_{m=-\ell}^{\ell}}

\usepackage{xcolor}

\usepackage[normalem]{ulem}

\AtBeginDocument{}

\begin{document}

\title{Simulating a Gaussian stochastic gravitational wave background signal \\ in pulsar timing arrays}

\author{Reginald Christian Bernardo}
\email{reginald.christian.bernardo@aei.mpg.de}
\affiliation{Max Planck Institute for Gravitational Physics (Albert Einstein Institute), Hannover D-30167, Germany}
\affiliation{Asia Pacific Center for Theoretical Physics, Pohang 37673, Korea}

\author{Kin-Wang Ng}
\email{nkw@phys.sinica.edu.tw}
\affiliation{Institute of Physics, Academia Sinica, Taipei 11529, Taiwan}
\affiliation{Institute of Astronomy and Astrophysics, Academia Sinica, Taipei 11529, Taiwan}


\begin{abstract}
We revisit the theoretical modeling and simulation of a Gaussian stochastic gravitational wave background (SGWB) signal in a pulsar timing array (PTA). We show that the correlation between Fourier components of pulsar timing residuals can be expressed using transfer functions; that are indicative of characteristic temporal correlations in a SGWB signal observed in a finite time window. These transfer functions, when convolved with the SGWB power spectrum and spatial correlation (Hellings \& Downs curve), describe the variances and correlations of the pulsar timing residuals' Fourier coefficients. The convolutions are the exact frequency- and Fourier-domain representations of the time-domain covariance function. We derive explicit forms for the transfer functions for unpolarized and circularly polarized SGWB signals. We validate our results by comparing Gaussian theoretical expectation values with standard simulations based on point sources and our own covariance-matrix-based approach. The unified frequency- and Fourier-domain formalism provides a robust foundation for future PTA precision analyses and highlights the importance of temporal correlations in interpreting GW signals.
\end{abstract}

\maketitle

\section{Introduction}
\label{sec:introduction}

The compelling evidence of a nanohertz stochastic gravitational wave background (SGWB) provided by pulsar timing arrays (PTAs) \cite{NANOGrav:2023gor, Reardon:2023gzh, EPTA:2023fyk, Xu:2023wog, Miles:2024seg} is a milestone in gravitational wave (GW) astronomy \cite{LIGOScientific:2016aoc, KAGRA:2021vkt, LIGOScientific:2021sio}. Unlike the signals captured by ground-based GW detectors, which originate from individual compact binary mergers, the leading GW signal observed by PTAs is expected to arise from the incoherent superposition of GWs emitted by a vast population of sources, each too weak to be individually resolved \cite{Joshi:2013at, McLaughlin:2014wna, Manchester:2015mda, Lommen:2015gbz, Romano:2016dpx, Becker:2017yyc, Verbiest:2021kmt, Taylor:2021yjx, Verbiest:2024nid, Yunes:2024lzm, Domcke:2024soc}. While searches for a SGWB in the hectohertz regime remain a primary goal of ground-based detectors, and similar efforts are planned in the millihertz band by future space-based observatories \cite{Caprini:2018mtu, Christensen:2018iqi, Romano:2019yrj, Moore:2021ibq, NANOGrav:2020spf, Pol:2022sjn, Staelens:2023xjn, Chen:2024xzw}, the primary source of the SGWB may vary across frequency bands. In the nanohertz regime accessible to PTAs, the dominant contribution is widely considered to come from an astrophysical population of supermassive black hole binaries (SMBHBs) \cite{Sazhin:1978myk, Detweiler:1979wn, Phinney:2001di, Wyithe:2002ep, Sesana:2004sp, Sesana:2008mz, Vigeland:2016nmm, Burke-Spolaor:2018bvk, Liu:2021ytq, Sato-Polito:2023gym, Sato-Polito:2023spo, Bi:2023tib, Sato-Polito:2024lew, Sah:2024oyg, Raidal:2024odr, Sah:2024etc}, producing the characteristic Hellings \& Downs (HD) correlation \cite{Hellings:1983fr, Jenet:2014bea, Romano:2023zhb}. However, uncertainties in the observed signal's spectrum leave room for a variety of alternative interpretations, including scenarios involving beyond-standard-model cosmological processes \cite{Chen:2019xse, Ellis:2020ena, Vagnozzi:2020gtf, Samanta:2020cdk, Benetti:2021uea, NANOGrav:2021flc, Buchmuller:2021mbb, Deng:2021gkx, Xue:2021gyq, Sharma:2021rot, EPTA:2023xxk, Ben-Dayan:2023lwd, Vagnozzi:2023lwo, Figueroa:2023zhu, Ellis:2023dgf, Saeedzadeh:2023biq, Liu:2023pau, Liu:2023hpw, Chen:2023bms, Liu:2023ymk, Jin:2023wri, Huang:2023chx, Ye:2023tpz, Wang:2023sij, Wang:2023ost, Zhu:2023lbf, Jiang:2023gfe, Bian:2023dnv, Datta:2023vbs, Datta:2023xpr, Chowdhury:2023xvy, Jiang:2024dxj, Winkler:2024olr, Kumar:2024bvp, Agazie:2024kdi, Calza:2024qxn, Papanikolaou:2024fzf, Papanikolaou:2024cwr, Basilakos:2024diz, Yin:2024ccm, Datta:2024bqp,
Athron:2024fcj, Lopez:2025gfu, Bernardo:2025lie, Choi:2025rfr, Datta:2025owx, Kumar:2025jfi}.

The ability of PTAs to constrain the SGWB through timing data improves with the inclusion of additional millisecond pulsars and longer observation times \cite{Siemens:2013zla, Moore:2014eua, Vigeland:2016nmm, Pol:2022sjn}. A natural path forward is to continue expanding current PTA experiments, increasing sky coverage and sensitivity to refine our understanding of the signal's origin. At the same time, it is crucial to revisit the theoretical foundations of the analysis \cite{vanHaasteren:2014faa, vanHaasteren:2014qva, Qin:2018yhy, Ng:2021waj, Liu:2022skj, Allen:2022dzg, Caliskan:2023cqm, Kehagias:2024plp, Liang:2024mex, Hu:2024wub, Xue:2024qtx, Allen:2024bnk, Allen:2024uqs, Madison:2024fup, Allen:2024cgm, Pitrou:2024scp}, utilizing simulations and improved data processing techniques to extract physical insights while mitigating statistical and systematic uncertainties \cite{Hazboun:2019vhv, DiMarco:2024irz, NANOGrav:2024xhc, Goncharov:2024htb, Goncharov:2024fsi, Bernardo:2024tde, Crisostomi:2025vue}. This paper follows the latter approach by revisiting the theoretical framework of a Gaussian SGWB signal in the frequency- and Fourier-domain,\footnote{
{We are reminding the subtle distinction between the terms frequency- and Fourier-domains, since a random signal can always be interpreted in the frequency-domain with an arbitrary choice of orthogonal basis vectors other than with sines and cosines. See \cite{Allen:2025waa} for an example of using a Legendre polynomial basis to look into PTA signals.}
} providing a perspective that complements the time-domain methods commonly adopted in PTA analyses.

A Gaussian reference model provides a meaningful baseline for analyzing stochastic signals. In cosmology, the cosmic microwave background (CMB) serves as a prime example: standard inflation predicts nearly Gaussian primordial fluctuations, and deviations from this baseline may point to new physics such as non-standard inflation or late-time effects like lensing \cite{Planck:2018jri, Planck:2019evm}. The Gaussian model thus defines a null hypothesis, against which any statistically significant deviations can be tested for physical implications. A similar logic applies to PTAs, where a Gaussian SGWB is typically assumed as a starting point in both theoretical modeling and data analysis. However, unlike the CMB, where the signal dominates over noise, PTAs face the challenge of identifying weak signals amid strong pulsar-intrinsic red noise \cite{Bernardo:2024uiq} and other systematics. Despite this, establishing and testing the Gaussian null hypothesis remains a promising approach---it may uncover subtle statistical imprints that help distinguish between astrophysical and cosmological origins of the SGWB. This motivates a careful re-examination of the signal's statistical structure, particularly how Gaussianity manifests in correlations and spectral features.

A (Universe-age old) question we address in this work is: given that a SGWB source emits over a broad range of frequencies with a characteristic timescale of order of a million to a billion years, how can such a signal be characterized observationally when the PTA baseline is only a few decades? The answer has been provided in Ref. \cite{Allen:2024uqs}, and later in Refs. \cite{Bernardo:2024tde, Crisostomi:2025vue}. A signal with such a characteristic time scale would have made a very large number of cycles and have continuous support at all frequencies in the observation band. Any Fourier basis with a discrete set of frequencies determined by the observation time scale is inevitably misaligned with the signal's intrinsic harmonics, leading to correlations among frequency components.

Building on Ref.~\cite{Bernardo:2024tde}, we show that the correlation of the Fourier coefficients of pulsar timing residuals can be expressed using `transfer functions'\footnote{
Filters are suggestive of a function chosen to suit ones purposes, e.g., polaroid. Instead the ones introduced in Ref.~\cite{Bernardo:2024tde} are a property of any time correlated process. Hence, we are now calling them transfer functions. We thank Bruce Allen for the suggestion.}.
In general, transfer functions describe how intrinsic properties of a signal map onto observable quantities. In PTAs, they encode temporal correlations due to GW propagation in time and frequency correlations induced by observing the signal in a finite time window\footnote{
Misconceptions in the community \cite{Maggiore:2018sht, Cruz:2024svc, Cruz:2024esk} lead to a dismissal of the significance of the temporal correlation \cite{vanHaasteren:2014faa, vanHaasteren:2014qva}.
}. As we demonstrate, this structure manifests in the frequency- and Fourier-domain as a convolution between the power spectrum and the transfer functions \cite{Allen:2024uqs, Bernardo:2024tde}. In this work, we derive the transfer functions for both isotropic unpolarized [Eqs. (\ref{eq:alpha_filter}-\ref{eq:beta_filter})] and circularly polarized [Eq. (\ref{eq:cp_filter})] SGWB signals in a PTA. We validate the theoretical expectation values for a Gaussian SGWB by their agreement with standard numerical simulations of a GW signal produced by a finite number of point sources \cite{Hobbs:2009yn} and with our own covariance-matrix-based simulations. These results clarify how Gaussian SGWB signals whether unpolarized or circularly polarized should be modeled within the PTA observation window. This, in turn, enables more efficient simulations and lays a foundational step toward testing the Gaussianity of the SGWB signal and possibly utilizing simulation-based inference methods in PTA science.

The remainder of this paper is organized as follows. Section~\ref{sec:standard_analysis_of_the_signal} provides a brief summary of the standard interpretation of the SGWB signal in terms of the Fourier coefficients of pulsar timing residuals. In Section~\ref{sec:the_signal}, we present our first main result: by revisiting the derivation of the pulsar timing residual correlation [Eq.~\eqref{eq:timing_residual_correlation_general}], we extract the correlations between Fourier coefficients, leading to transfer functions. Section~\ref{sec:comparison_with_simulation} introduces a numerical framework for simulating a Gaussian SGWB and validates our analytical predictions against ensemble-averaged quantities from both theoretical expectations and standard point-source based PTA simulations. In Section~\ref{sec:circular_polarization_in_pta}, we extend our analysis to accommodate a circularly polarized SGWB, and further validate these results through simulation in Section~\ref{sec:comparison_with_simulation_cp}, highlighting subtleties related to the computational frame~\cite{Mingarelli:2013dsa, Kato:2015bye, Belgacem:2020nda, Sato-Polito:2021efu, Chu:2021krj}. Then, in Section \ref{sec:discussion}, we provide a brief discussion of what we think are the most conceptually interesting aspects of the work. Finally, Section~\ref{sec:conclusions} summarizes our main findings and outlines directions for future work.

Appendix \ref{sec:a_different_basis} establishes the equivalence between the sine-cosine and exponential interpretations of a time series through the Fourier components of the timing residual correlation and the corresponding transfer functions.
Appendix~\ref{sec:spinweight} collects useful spherical harmonic identities. Appendix~\ref{sec:simulation} details our procedure for simulating a Gaussian SGWB signal in PTAs based on covariance matrices and Gaussian random variables. Appendix~\ref{sec:number_pulsars} presents results for varying the number of pulsars in a PTA. {Appendix \ref{sec:summary_covariance} provides a quick recipe for computing the elements of the covariance matrix in the frequency- and Fourier-domain.}

Throughout this work, we adopt the mostly-plus metric signature $(-,+,+,+)$ and use geometrized units with $G = c = 1$. Pulsars are indexed by lowercase Latin letters $a, b$, and frequency bins by $j, k$. For brevity, we denote $ \langle X_{aj} Y_{bk} \rangle \equiv \langle X_j(\hat{e}_a)\, Y_k(\hat{e}_b) \rangle $
where $X_j(\hat{e}_a)$ and $Y_k(\hat{e}_b)$ are the Fourier coefficients of the timing residuals in directions $\hat{e}_a$ and $\hat{e}_b$, respectively [cf.~Eq.~\eqref{eq:residual_fourier_series}], and $\langle \cdots \rangle$ denotes ensemble averaging.

\section{{Timing residual analysis in the frequency- and Fourier-domain}}
\label{sec:standard_analysis_of_the_signal}

The pulsar timing residual can be written as a Fourier series\footnote{
An equivalent way to write the Fourier series of the timing residual is $r(t,\hat{e})=\sum_{k=-\infty}^\infty c_k(\hat{e}) e^{i \omega_k t}$ \cite{Allen:2024uqs}.
See Appendix \ref{sec:a_different_basis}.
} \cite{Taylor:2021yjx, Bernardo:2024bdc},
\begin{equation}
\label{eq:residual_fourier_series}
r(t,\hat{e}) = \frac{\alpha_0(\hat{e})}{2} + \sum_{k=1}^{\infty} \alpha_k(\hat{e}) \sin(\omega_k t) + \sum_{k=1}^{\infty} \beta_k(\hat{e}) \cos(\omega_k t) \,,
\end{equation}
where $\omega_k = 2\pi f_k$, $f_k = k/T$, $k$ is a positive integer, $T$ is the total observation time, and $\hat{e}$ is the unit vector pointing in the pulsar’s direction. Then, the Fourier components of the timing residual are given by\footnote{
In PTA analysis \cite{vanHaasteren:2008yh, Lentati:2012xb}, the Fourier coefficients are inferred by Bayesian parameter estimation, accounting for marginalization over the Fourier mode phases and spectral leakage mitigation through fitting to pulsar spin frequency and its first derivative, which are expected to make variances equal. The Bayesian inferred coefficients are generally different compared to Eq. \eqref{eq:residual_fourier_components} which are the mathematically exact inverse transform of the linear operation \eqref{eq:residual_fourier_series} when interpreting a stochastic process in a finite time window.
}
\begin{align}
\alpha_0(\hat{e}) &= \frac{2}{T} \int_0^T dt \ r(t,\hat{e}) \,, \nonumber \\
\alpha_j(\hat{e}) &= \frac{2}{T} \int_0^T dt \ r(t,\hat{e}) \sin(\omega_j t) \,, \nonumber \\
\beta_j(\hat{e}) & = \frac{2}{T} \int_0^T dt \ r(t,\hat{e}) \cos(\omega_j t) \,. \label{eq:residual_fourier_components}
\end{align}

PTAs measure the spectrum of the signal at discrete frequencies $f_k = k/T = k f_1$, where $T$ is the total observation time. For $T \sim n_{\rm yr}$ years, the lowest accessible frequency is $f_1 \sim 31.8/n_{\rm yr}$ nHz. In contrast, the characteristic time scale of the stochastic GW signal is much longer, $T_{\rm gw} \sim 10^9$ years. The transfer functions (\ref{eq:alpha_filter}-\ref{eq:beta_filter}) and \eqref{eq:cp_filter} determine how a broadband signal is mapped onto the $\alpha$- and $\beta$-observational Fourier bins of the timing residual correlation. Crucially, these transfer functions are {\it not} Kronecker deltas; instead, their finite width means that each observed frequency bin receives contributions from a broad range of frequencies. This leads to inter-frequency correlations, which must be accounted for in PTA analyses. For a discussion of the observational implications, see Refs.~\cite{Allen:2024uqs, Bernardo:2024tde, Crisostomi:2025vue} and Section~\ref{sec:discussion}.

\section{The unpolarized signal: Spectrum $\times$ HD curve $\times$ Transfer functions}
\label{sec:the_signal}

In this section, we re-derive the SGWB signal and correlation in the pulsar timing residuals.

\subsection{Pulsar Timing Residual from GWs}
\label{subsec:gw_residual}

We begin by considering a stochastic superposition of GWs, characterized by various frequencies $f$ and propagation directions $\hat{k}$: 
\begin{equation}
\label{eq:gw_general}
\begin{split}
    h_{ij}\left(\eta, \vec{x}\right) = & \sum_{A={+,\times}} \int_{-\infty}^\infty df \int_{S^2} d\hat{k} \\
    & \,\,\,\, \times h_A\left(f, \hat{k}\right) \varepsilon_{ij}^A e^{-2\pi i f \left( \eta - \hat{k} \cdot \vec{x} \right)} \,,
\end{split}
\end{equation}
where $\eta$ denotes conformal time, $\vec{x}$ represents the position vector, and $\varepsilon_{ij}^+$ and $\varepsilon_{ij}^\times$ are the transverse-traceless polarization basis tensors. Here $h_{ij}$ is real; its Fourier components satisfy $h_A(-f,\hat{k})=h_A^*(f,\hat{k})$. We define a SGWB as a collection of GWs such that $h_{ij}$ are Gaussian random fields with a statistical behavior completely characterized by
the two-point correlation function, $\langle h_{ij}(\eta,\vec{x}_1) h_{{kl}}(\eta,\vec{x}_2) \rangle$, where angle brackets denote ensemble averages.

For a GW propagating along a direction {$\hat{k}$}, the polarization tensors can be written as $\varepsilon^{+} = \hat{m} \otimes \hat{m} - \hat{n} \otimes \hat{n}$ and $\varepsilon^{\times} = \hat{m} \otimes \hat{n} + \hat{n} \otimes \hat{m}$, where {$( \hat{m}, \hat{n}, \hat{k} )$} form an orthonormal triad \cite{Boitier:2021rmb}. The influence of a GW on pulsar timing residuals, $r(t, \hat{e})$, can be expressed as an integral over redshift space fluctuations, $z(t, \hat{e})$:
\begin{equation}
\label{eq:timing_residual}
    r\left(t, \hat{e}\right) = \int_0^t dt' \ z\left(t', \hat{e}\right) \,,
\end{equation}
where $t$ is the time of observation. Note that this considers the GW effects in the locations of the Earth and the pulsar; traditionally referred to as Earth and pulsar terms, respectively. It is worth {pointing} out that the HD curve is derived from the Earth term. For a GW, such as $h_{ij}\left(\eta, \vec{x}\right)$ given by Eq. \eqref{eq:gw_general}, with photons emitted at conformal time $\eta_e$ and received at $\eta_r$, the redshift space fluctuation can be described by a Sachs-Wolfe integral \cite{Sachs:1967er}:
\begin{equation}
\label{eq:z_swolf}
    z\left(t', \hat{e}\right) = - \dfrac{1}{2} \int_{t' + \eta_e}^{t' + \eta_r} d\eta \, {\hat{e}^i \hat{e}^j} \, \partial_\eta h_{ij} \left( \eta, \vec{x} \right) \,.
\end{equation}
Here, $\hat{e}^i \otimes \hat{e}^j$ is commonly referred to as a `detector tensor'.

By substituting Eq.~\eqref{eq:gw_general} into Eqs.~(\ref{eq:timing_residual}-\ref{eq:z_swolf}), we obtain
\begin{widetext}
\begin{equation}
\begin{split}
    r\left(t, \hat{e}\right) = & \int_0^t dt' \int_{t' + \eta_e}^{t' + \eta_r} d \eta  \sum_{A} \int_{-\infty}^\infty df \int_{S^2} d\hat{k} \ \pi i f \, {\hat{e}^i \hat{e}^j} \, h_A \left(f, \hat{k}\right) \varepsilon_{ij}^A\left(\hat{k}\right) e^{-2\pi i f \left( \eta - \hat{k} \cdot \vec{x} \right)} \,.
\end{split}
\end{equation}
Then, by expanding the plane wave in spherical harmonics and performing the integrals over $t'$ and $\eta$, we find the expression of the pulsar timing residual as a harmonic series,
\begin{equation}
\label{eq:gwb_residual_harmonic_series}
    r\left(t, \hat{e}\right) = \sum_{lm} a_{lm} (t) Y_{lm} \left( \hat{e} \right) \,,
\end{equation}
where the coefficients $a_{lm}(t)$'s are given by
\begin{equation}
\label{eq:gwb_residual_harmonic_coefficients}
\begin{split}
    a_{lm}(t) = & \sum_{A=+,\times} \int_{-\infty}^\infty d\omega \int_{S^2} d\hat{k} \ \left( 1 - e^{-i \omega t} \right) \left( \dfrac{e^{- i \omega \eta_r}}{\omega} \right) \, {\hat{e}^i \hat{e}^j} \, h_A\left(\dfrac{\omega}{2\pi}, \hat{k}\right) \varepsilon_{ij}^A\left(\hat{k}\right) \int_0^{\omega D} dx \ e^{ix} i^l j_l\left(x\right) Y^*_{lm}\left(\hat{k}\right) \,,
\end{split}
\end{equation}
$\omega=2\pi f$ is the angular frequency, and $j_l(x)$'s are the spherical Bessel functions of the first kind.
\end{widetext}

We use this result to compute the pulsar timing residual correlation.

\subsection{Timing Residual Correlation due to a SGWB}
\label{subsec:timing_and_orf}

We consider a standard Gaussian, isotropic, and stationary SGWB; in symbols, we write down
\begin{equation}
\label{eq:gwb_power_spectrum}
    \langle h_A\left(f, \hat{k}\right) h_{A'}^* \left( f', \hat{k}'\right) \rangle = I(f) \delta_{AA'} \delta \left( f - f' \right) \delta \left( \hat{k} {,} \hat{k}' \right) \,,
\end{equation}
where $I(f)$ is the SGWB power spectrum. The Dirac delta functions are dictated by temporal and spatial translational invariance.

Using Eqs.~(\ref{eq:gwb_residual_harmonic_series}-\ref{eq:gwb_residual_harmonic_coefficients}), we can express the timing residual cross correlation between a pair of pulsars to be
\begin{equation}
\begin{split}
    & \langle r\left(t_a, \hat{e}_a\right) r\left(t_b, \hat{e}_b\right) \rangle \\
    & \,\,\,\, = \sum_{l_1 m_1} \sum_{l_2 m_2} \langle a_{l_1 m_1}(t_a) a^*_{l_2 m_2}(t_b) \rangle Y_{l_1 m_1}\left( \hat{e}_a \right) Y^*_{l_2 m_2}\left( \hat{e}_b \right) \,,
\end{split}
\end{equation}
where
\begin{equation}
\label{eq:two_point_lm}
\begin{split}
    & \langle a_{l_1 m_1}(t_a) a^*_{l_2 m_2}(t_b) \rangle \\
    & \,\,\,\, = \int_{-\infty}^\infty \dfrac{df}{\left(2\pi f\right)^2} C(f, t_a, t_b) \\
    & \,\,\,\, \times \sum_{A_1 A_2} \int_{S^2} d\hat{k} \ I\left( f \right) J^{A_1}_{l_1 m_1} \left( f D_a, \hat{k} \right) J^{A_2 *}_{l_2 m_2} \left( f D_b, \hat{k} \right) \,,
\end{split}
\end{equation}
\begin{equation}
\label{eq:C_filter_def}
C(f,t_a, t_b)=\left( 1 - e^{-2\pi i f t_a} \right) \left( 1 - e^{2\pi i f t_b} \right) \,,
\end{equation}
and
\begin{equation}
\small
\label{eq:Jlm_def}
\begin{split}
    J_{lm}^A \left( fD, \hat{k} \right) = & \, 2\pi \int_0^{2\pi f D} dx \ e^{i x} \sum_{LM} i^L Y^*_{LM} \left( \hat{k} \right) j_L(x) \\
    & \times \int_{S^2} d\hat{e} \, {\hat{e}^i \hat{e}^j} \, \varepsilon_{ij}^A\left(\hat{k}\right) Y_{LM}\left( \hat{e} \right) Y_{lm}^*\left(\hat{e}\right) \,.
\end{split}
\end{equation}
Then, after straightforward manipulation, we obtain the timing residual correlation as
\begin{equation}
\label{eq:timing_residual_correlation_general}
\begin{split}
    \langle r\left(t_a, \hat{e}_a\right) r\left(t_b, \hat{e}_b\right) \rangle = & \int_0^\infty \dfrac{2 \, df}{\left(2\pi f\right)^2} {\rm Re} \left[ C(f, t_a, t_b) \right] I(f) \\
    & \times \sum_{A=+,\times}\gamma^A(fD_a, fD_b, \hat{e}_a\cdot\hat{e}_b) \,,
\end{split}
\end{equation}
where the $\gamma^A(fD_a, fD_b, \hat{e}_a\cdot\hat{e}_b)$'s are given by
\begin{equation}
\label{eq:orf_general}
\begin{split}
    & \gamma^A \left(f D_a, fD_b, \hat{e}_a\cdot\hat{e}_b \right) \\
    & \,\,\,\, = \sum_{l_1 m_1} \sum_{l_2 m_2} Y_{l_1 m_1}\left( \hat{e}_a \right) Y^*_{l_2 m_2} \left( \hat{e}_b \right) \\
    & \,\,\,\, \times \int_{S^2} d\hat{k} \ Y_{00}\left(\hat{k}\right) J^A_{l_1 m_1} \left( f D_a, \hat{k} \right) J^{A*}_{l_2 m_2} \left( f D_b, \hat{k} \right) \,.
\end{split}
\end{equation}
The timing residual cross correlation between a pair of pulsars depends only on the angular separation. The angular dependence, $\hat{e}_a \cdot \hat{e}_b$, depending only upon on the relative positions of pulsars with one another, is reflective of isotropy; $\gamma \left(f D_a, fD_b, \hat{e}_a, \hat{e}_b \right) \equiv \gamma \left(f D_a, fD_b, \hat{e}_a\cdot\hat{e}_b \right)$. The above result \eqref{eq:timing_residual_correlation_general} shows that the SGWB signal in the timing residual correlation has three parts: a time-frequency kernel $C(f,t,t')$, a power spectrum, $I(f)$, and a spatial correlation (HD curve), $\gamma \left(f D_a, fD_b, \hat{e}_a\cdot\hat{e}_b \right)$. Setting apart the terms associated with the lower limit of the integral \eqref{eq:timing_residual} (which fixes the initial phase of the timing residuals), the result \eqref{eq:timing_residual_correlation_general} could be read as a statement of the Wiener-Khinchin theorem for a stationary stochastic process. We emphasize that the temporal correlation $C(f,t,t')$ is associated with the GW propagation in time. Implications of this factor for interpreting a SGWB observation in PTAs were discussed in Ref.~\cite{Bernardo:2024tde}.

{Adding to our last thought, we may remind that a stationary colored process with a covariance function $C(|t-t'|)$ at two different times $t\neq t'$ can be described by the Wiener-Khinchin theorem: $C(\tau) = \int df \, S(f) \cos(2\pi f \tau)$. This can be derived by treating the signal in the Fourier basis. In a physical process, such as GWs that behave as $h \sim \exp\left({-i \left( \omega t - \vec{k}\cdot \vec{x} \right)}\right)$ apart from the $+$ and $\times$ polarization tensors, the Wiener-Khinchin theorem is realized through a large sum of deterministic contributions that render the description of the overall signal a statistical sense. Part of what gives the stochastic signal its temporally correlated behavior are the phase factors $\exp\left({-i \omega t}\right)$ which describe a physical wave's temporal propagation; e.g., the oscillating motion of an infinitesimal piece of a vibrating drum.}

For general relativity and most of this work, the spatial correlation can be regarded as simply the HD curve, 
\begin{equation}
\label{eq:spatial_corr_hd_limit}
\begin{split}
    & \sum_{A=+,\times}\gamma^A(fD_a, fD_b, \hat{e}_a\cdot\hat{e}_b) \\
    & \,\,\,\, \sim \gamma^{\rm HD} (\hat{e}_a\cdot\hat{e}_b) = \overline{\cal N} \times \left( {\mathbf{\Gamma}}(\hat{e}_a \cdot \hat{e}_b) + \dfrac{\delta_{ab}}{2} \right) \,,
\end{split}
\end{equation}
where $\overline{\cal N}$ is a constant and ${\mathbf{\Gamma}}(x)$ is
\begin{equation}
\label{eq:hd_curve}
    {\mathbf{\Gamma}}(x) = \dfrac{1}{2} - \dfrac{1}{4} \left( \dfrac{1-x}{2} \right) + \dfrac{3}{2} \left( \dfrac{1-x}{2} \right) \ln \left( \dfrac{1-x}{2} \right) \,.
\end{equation}
This approximation holds well in the long arm limit \cite{Romano:2023zhb, Domenech:2024pow}, valid for PTAs where the wavelength of GWs are $\lambda_{\rm gw} \sim {\cal O}(1)$ pc and the distances to the pulsars are $D \sim {\cal O}(1)$ kpc. The term $\delta_{ab}$ in Eq.~\eqref{eq:spatial_corr_hd_limit} is added ad hoc to account for the power at small angular scales. A derivation of the spatial correlation starting with Eq.~\eqref{eq:orf_general} can found in Refs.~\cite{Liu:2022skj, Bernardo:2024bdc}, among other recent references \cite{Gair:2014rwa, Jenet:2014bea, Qin:2018yhy, Mingarelli:2018kgp}. The quantities $\gamma(x_a, x_b, \cos\theta)$ and ${\mathbf\Gamma}(\cos\theta)$ pertaining to spatial correlation in PTA are referred to as ORFs.

\subsection{Transfer Functions}
\label{subsec:filters}

When reading the timing residual data as Eq.~\eqref{eq:residual_fourier_series}, the SGWB signal \eqref{eq:timing_residual_correlation_general} can be interpreted as a convolution between the power spectrum, $I(f)$, the spatial correlation, $\gamma(f D_a, fD_b, \hat{e}_a \cdot \hat{e}_b)$, and transfer functions, tailored to each frequency bin and Fourier component of the correlation \cite{Allen:2024uqs, Bernardo:2024tde}.

We derive the transfer functions in the following. Using Eq.~\eqref{eq:timing_residual_correlation_general}, for the $\langle \alpha_{aj} \alpha_{bk}\rangle$ and $\langle \beta_{aj} \beta_{bk} \rangle$ components, we obtain
\begin{align}
& \langle \alpha_{aj} \alpha_{bk}\rangle \nonumber \\
&\equiv \left(2\over T\right)^2 \iint_0^T dt \, dt' 
\langle r(t,\hat{e}_a)r(t',\hat{e}_b) \rangle \sin(\omega_j t) \sin(\omega_k t') \nonumber \\
&=  \int_{f_{\rm min}}^{f_{\rm max}} \frac{2\,df}{(2\pi f)^2}\, {\cal C}_{j,k}^\alpha (f) I(f) \gamma(fD_a, fD_b,\hat{e}_a,\hat{e}_b)\,,
\label{aiaj}
\end{align}
\begin{align}
& \langle \beta_{aj} \beta_{bk} \rangle \nonumber \\
& \equiv \left(2\over T\right)^2 \iint_0^T dt \, dt'
\langle r(t,\hat{e}_a)r(t',\hat{e}_b) \rangle \cos(\omega_j t) \cos(\omega_k t') \nonumber \\
&=  \int_{f_{\rm min}}^{f_{\rm max}} \frac{2\,df}{(2\pi f)^2}\, {\cal C}_{j,k}^\beta (f) I(f) \gamma(fD_a, fD_b,\hat{e}_a,\hat{e}_b)\,,
\label{bibj}
\end{align}
where {we use the notation $\iint_0^T = \int_0^T \int_0^T$ for the double integral, and the} transfer functions, ${\cal C}_{j,k}^\alpha (f)$ and ${\cal C}_{j,k}^\beta (f)$, have been defined as
\begin{align}
& {\cal C}_{j,k}^\alpha (f) \nonumber \\
&=\left(2\over T\right)^2 \iint_0^T dt \, dt' \,{\rm Re}\left[C(f,t,t')\right] \sin(\omega_j t) \sin(\omega_k t')\,,  \\
& {\cal C}_{j,k}^\beta (f) \nonumber \\
& =\left(2\over T\right)^2 \iint_0^T dt \, dt' \,{\rm Re}\left[ C(f,t,t') \right] \cos(\omega_j t) \cos(\omega_k t')\,. 
\end{align}
The integral limits have been truncated at $f_{\rm min}\sim1/T$ to $f_{\rm max}={\cal O}(10)/T$ to correspond to the PTA sensitivity range. By substituting \eqref{eq:C_filter_def}, we obtain\footnote{
Another way this can be written is
\begin{align*}
{\cal C}_{j,k}^\alpha (f) &= \dfrac{4 (-1)^{j+k} f_j f_k }{(f+f_j)(f+f_k)}{\rm sinc}\left( \pi T \left( f-f_j\right) \right){\rm sinc}\left( \pi T \left( f-f_k\right) \right) \,, \\
{\cal C}_{j,k}^\beta (f) &= \dfrac{4 (-1)^{j+k} f^2 }{(f+f_j)(f+f_k)}{\rm sinc}\left( \pi T \left( f-f_j\right) \right){\rm sinc}\left( \pi T \left( f-f_k\right) \right) \,,
\end{align*}
where ${\rm sinc}(x)=\sin(x)/x$. Note that $(-1)^k \sin(\pi f T)/(\pi T (f - f_k)) = {\rm sinc}( \pi T (f - f_k))$. This form makes the equivalence between the trigonometric (sine and cosine) and the exponential Fourier representations of a stochastic process transparent. See Appendix \ref{sec:a_different_basis}. 
}
\begin{align}
\label{eq:alpha_filter}
{\cal C}_{j,k}^\alpha (f) &= \dfrac{4 f_j f_k }{\pi^2 T^2} \dfrac{\sin^2(\pi f T)}{(f^2 - f_j^2)(f^2 - f_k^2)} \,, \\
\label{eq:beta_filter}
{\cal C}_{j,k}^\beta (f) &= \dfrac{4 f^2 }{\pi^2 T^2} \dfrac{\sin^2(\pi f T)}{(f^2 - f_j^2)(f^2 - f_k^2)} \,.
\end{align}
As the name suggests, the transfer functions, ${\cal C}_{jk}^\alpha (f)$ and ${\cal C}_{jk}^\beta (f)$, carry the properties of the signal to the observable quantity, or in this case the Fourier components of the timing residual correlation \cite{Bernardo:2024tde}.

\begin{figure*}[t]
    \centering
	\subfigure[ \ $\alpha$- and $\beta$-bin transfer functions (\ref{eq:alpha_filter}-\ref{eq:beta_filter}) for $j=k$; and the DC transfer function \eqref{eq:0_filter} ]{
		\includegraphics[width = 0.45 \textwidth]{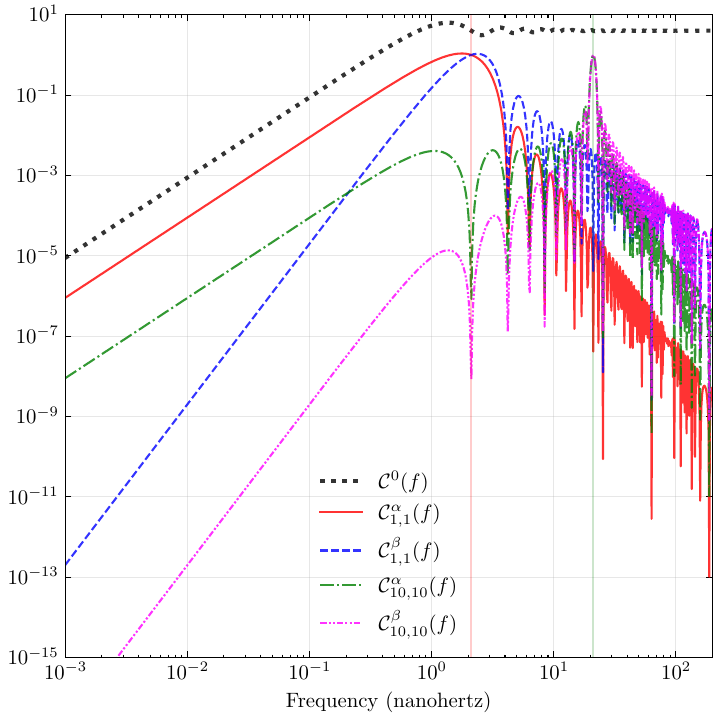}}
	\subfigure[ \ $\alpha$-DC- and $\beta$-DC-bin transfer functions (\ref{eq:alpha0_filter}-\ref{eq:beta0_filter}) ]{
		\includegraphics[width = 0.45 \textwidth]{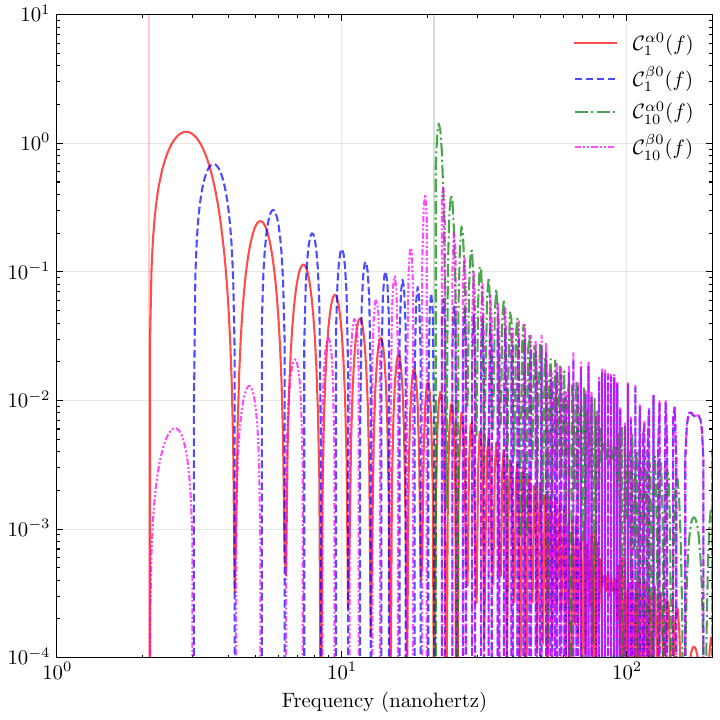}}
	\subfigure[ \ $\alpha$-bin transfer functions \eqref{eq:alpha_filter} for $j\neq k$ ]{
		\includegraphics[width = 0.45 \textwidth]{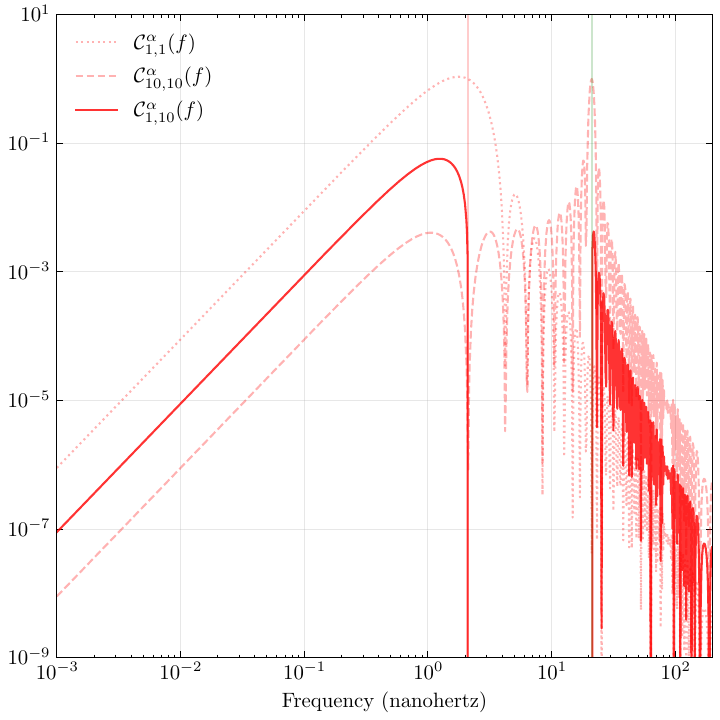}}
	\subfigure[ \ $\beta$-bin transfer functions \eqref{eq:beta_filter} for $j\neq k$ ]{
		\includegraphics[width = 0.45 \textwidth]{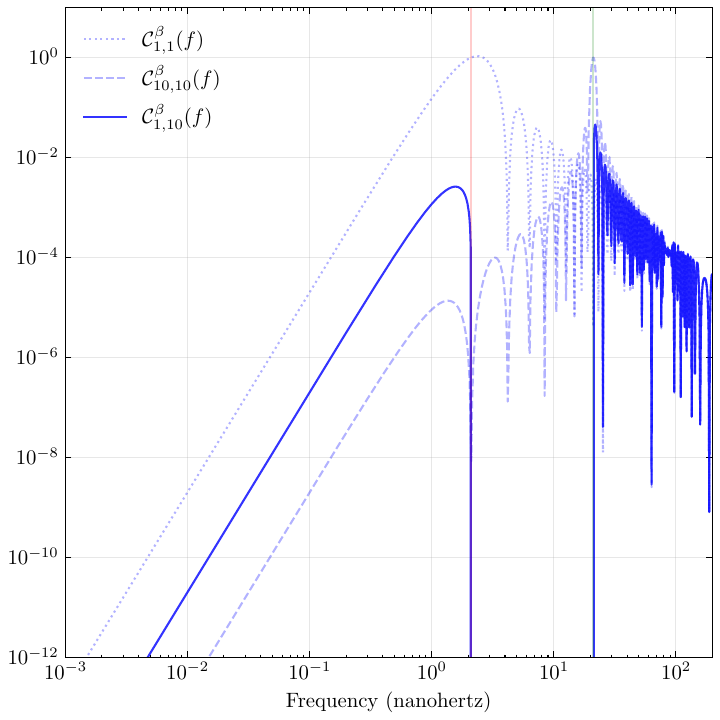}}

    \caption{Representative transfer functions (\ref{eq:alpha_filter}-\ref{eq:beta0_filter}) for an SGWB signal in PTA observation with $T=15$ years, $f_1 \sim 2.1$ nanohertz, and $f_k \sim k f_1$.}
    \label{fig:filters}
\end{figure*}

It is worth mentioning that the important piece of the temporal correlation \eqref{eq:C_filter_def} that gives Eqs.~(\ref{eq:alpha_filter}-\ref{eq:beta_filter}) is the last term, $\exp\left(-2\pi i f (t-t')\right)$, pertaining to stationarity. The other terms which pertain to initial phases obtained in the lower limit of the redshift integral \eqref{eq:timing_residual} vanishes after time integration.

One implication of the transfer functions is that different frequency bins, $j \neq k$, are correlated, although very weakly. By following the same steps, it can be shown that the transfer function for $\langle \alpha_{aj}\beta_{bk} \rangle$ exactly vanishes. Thus, the $\alpha$- and $\beta$-Fourier bins of Eq.~\eqref{eq:residual_fourier_series} are uncorrelated. We leave it as a quick exercise to compute the transfer functions for $\langle \alpha_{a0} \alpha_{b0} \rangle$, $\langle \alpha_{a0} \alpha_{bk} \rangle$, and $\langle \alpha_{a0} \beta_{bk} \rangle$; the results are given by
\begin{align}
\label{eq:0_filter} {\cal C}^{0}(f) &=\bigg[4 \sin (\pi  f T) \sin (\pi  {\rm FP}[f T]) \cos (\pi  {\rm IP}[f T]) \nonumber \\
& \,\,\,\,\,\,\,\,\,\,\,\, +4 \pi  f T [\pi  f T-\sin (2 \pi  f T)] \bigg]/\left(\pi ^2 f^2 T^2\right) \,, \\
\label{eq:alpha0_filter} {\cal C}_k^{\alpha 0}(f) &= -\frac{4 k \sin ^2(\pi  f T)}{ \pi  \left(k^2-f^2 T^2\right) } \,, \\
\label{eq:beta0_filter} {\cal C}_k^{\beta 0}(f) &= \frac{2 [\pi  f T \sin (2 \pi  f T)+\cos (2 \pi  f T)-1]}{\pi ^2 \left(k^2-f^2 T^2\right)} \,,
\end{align}
respectively, where ${\rm IP}[x]$ and ${\rm FP}[x]$ are symbolic representations of the integer and fractional parts of a real-valued input $x$, such that ${\rm IP}[x] + {\rm FP}[x] = x$. Representative transfer functions are shown in Figure~\ref{fig:filters}.

We will refer to Eqs.~\eqref{eq:alpha_filter}, \eqref{eq:beta_filter}, \eqref{eq:0_filter}, \eqref{eq:alpha0_filter}, and \eqref{eq:beta0_filter} as an $\alpha$-, $\beta$-, DC, $\alpha$-DC, and $\beta$-DC transfer functions, respectively.

Figure~\ref{fig:filters} shows that ${\cal C}^\alpha_{j,k}(f)$ and ${\cal C}^\beta_{j,k}(f)$ accommodate more low- and high-frequency components, respectively, relative to each other\footnote{
For this reason they were referred to as low pass and high pass filters in Refs.~\cite{Bernardo:2024tde, Bernardo:2024bdc}. However, the ${\cal C}(f)$'s are not arbitrarily chosen as `filters' might misleadingly suggest.
}. Mainly, it can be seen that the $\alpha$-transfer function accepts more low-frequency contributions to the bins (shown are the 1st and 10th ones), compared with the corresponding $\beta$-transfer function which accepts more high-frequency contributions. This can be teased by the relation
\begin{equation}
\label{eq:alpha_beta_relation}
    {\cal C}^\beta_{j,k}(f) = {\cal C}^\alpha_{j,k}(f) f^2/(f_j f_k) \,.
\end{equation}
The shapes of the transfer functions are reminiscent of a mainlobe and sidelobes of a far-field radiation pattern \cite{Allen:2024uqs}. Noticeably, the mainlobe carries the siginificant contribution; the sidelobes support subdominant contributions at all frequencies, except at integer multiples of $1/T$ \cite{Allen:2024uqs}. In addition, shown in Figure~\ref{fig:filters}(a) is the static or DC transfer function. We emphasize that the DC bin $\alpha_0$ (see Eq.~\eqref{eq:residual_fourier_series}) carries no frequency information, but is relevant when reconstructing nonprojected signal covariances with the right amplitudes. The transfer functions for $\langle \alpha_{a0} \alpha_{b0} \rangle$ (dotted-line in Figure~\ref{fig:filters}(a)), $\langle \alpha_{a0} \alpha_{bk} \rangle$, and $\langle \alpha_{a0} \beta_{bk} \rangle$ (Figure~\ref{fig:filters}(b)) were shown for completeness. It is also useful to look at the $\alpha$- and $\beta$-bin transfer functions for different frequency components ($j \neq k$) (Figure~\ref{fig:filters}(c-d)). This supports our previous statement that Fourier components of the correlation at different frequency bins are very weakly correlated. The transfer functions for different frequency bins $j\neq k$ are orders of magnitude lower compared to their $j=k$ counterparts, and are missing the significant power between the two frequency components, $f_j$ and $f_k \neq f_j$. Consequently, the contributions of $\langle \alpha_{aj} \alpha_{bk} \rangle$ and $\langle \beta_{aj} \beta_{bk} \rangle$ with $j \neq k$ are subdominant compared to the $j=k$ contributions. Nonetheless, the off-diagonal terms' actual contribution can be quantified by an efficient reconstruction of the time-domain covariance (see Ref. \cite{Crisostomi:2025vue}).

The transfer functions provide a unified formalism for computing the PTA correlation in each Fourier bin due to the SGWB signal. All the correlation components can now be expressed as
\begin{equation}
\label{eq:xy_correlation}
    \langle X_{aj} Y_{bk} \rangle = {\cal F}[ I(f) \gamma(fD_a, fD_b,\hat{e}_a,\hat{e}_b), \mathcal{C}_{j,k}^{(XY)}(f) ] \,,
\end{equation}
where the functional ${\cal F}$ is given by
\begin{equation}
\label{eq:F_functional}
    {\cal F}[G(f, y) , {\cal C}(f)] = \int_{f_{\rm min}}^{f_{\rm max}} \frac{2\,df}{(2\pi f)^2} G(f, y) {\cal C}(f) \,,
\end{equation}
and ${\cal C}_{jk}^{(XY)}(f)$ are transfer functions given by Eqs.~(\ref{eq:alpha_filter}-\ref{eq:beta0_filter}). 

It is worth emphasizing that the components $\langle X_{aj} Y_{bk} \rangle$ are the frequency- and Fourier-domain representations of the stationary covariance function. In this regard, the timing residual correlation can be written in Fourier space as
\begin{equation}
\label{eq:timing_residual_correlation_fourier_decomposition}
\begin{split}
    & \langle r(t, \hat{e}_a,) r(t', \hat{e}_b) \rangle
    \\
    &\,\,\,\, = \sum_{jk} \bigg[ \langle \alpha_{aj} \alpha_{bk} \rangle \sin(\omega_j t)\sin(\omega_k t') \\
    & \,\,\,\,\,\,\,\,\,\,\,\,\,\,\,\,\,\,\,\,\,\,\,\, + \langle \beta_{aj} \beta_{bk} \rangle \cos (\omega_j t) \cos(\omega_k t') \bigg] + \cdots
\end{split}
\end{equation}
where the ellipses are all terms correlated with the DC component, i.e.,
\begin{equation}
\label{eq:timing_residual_correlation_fourier_decomposition_dc}
\begin{split}
    \dfrac{ \langle \alpha_{a0} \alpha_{b0} \rangle }{ 4 } & \\
    + \,\,\,\, \dfrac{1}{2} \sum_k \bigg[ & \langle \alpha_{ak} \alpha_{b0} \rangle \sin (\omega_k t) + \langle \alpha_{bk} \alpha_{a0} \rangle \sin (\omega_k t') \\
    & + \langle \beta_{ak} \alpha_{b0} \rangle \cos (\omega_k t) + \langle \beta_{b k} \alpha_{a0} \rangle \cos(\omega_k t') \bigg] \,.
\end{split}
\end{equation}
This highlights that the main contributions due to the SGWB come from the Fourier components $\langle \alpha_{ak} \alpha_{bk} \rangle$ and $\langle \beta_{ak} \beta_{bk} \rangle$. Note that the variances of the sine- and cosine-Fourier coefficients are different, because the Fourier basis $\{ \sin(2\pi k t/T), \cos(2\pi k t/T), {\rm for }\, k \in \mathbb{N} \}$ does not diagonalize the covariance of a signal with a time scale that is much larger compared to the observation period (see Section \ref{sec:discussion} for details). Turning this around, it is straightforward to show that if the signal's harmonic frequencies align with the observational frequecies (e.g., take a source with a power spectral density ${\cal S}(f) = \sum_k s_k \delta(f - k/T)$), then the covariance is going to be diagonalized and the variances of the sine- and cosine-Fourier coefficients are going to have the same probability distribution function. The decomposition (\ref{eq:timing_residual_correlation_fourier_decomposition}-\ref{eq:timing_residual_correlation_fourier_decomposition_dc}) can also be taken as the Green's function expansion in a Fourier basis with a finite time window.\footnote{
{{
We elaborate on `diagonalization'. A stationary random process $y(t)$ can be represented as
$y(t) = \sum_{k} \tilde{y}_k \phi_k(t)$
using a complete set of orthogonal basis functions $\{ \phi_k(t) \} $. In this representation, the covariance $\langle y(t) y(t') \rangle$ can be shown to be
$\langle y(t) y(t') \rangle
= \sum_{k} \, \langle \tilde{y}_k^2 \rangle \phi_k(t) \phi_k(t') + \sum_{j \neq k} \, \langle \tilde{y}_j  \tilde{y}_k \rangle \phi_j(t) \phi_k(t')$ \,,
which is broken down into diagonal ($j=k$) and off-diagonal ($j \neq k$) contributions; the quantities $\langle \tilde{y}_j  \tilde{y}_k \rangle$ represent $\langle y(t) y(t') \rangle$ in this basis. The Wiener-Khinchin theorem reveals that there is a preferred set of basis functions $\{ \phi^\star_k (t) \} \equiv \{ \sin(2\pi k t/ T_{\rm s}), \cos(2\pi k t/ T_{\rm s}) \}$ (which happens to be a Fourier basis) that `diagonalizes' the covariance as
\begin{equation}
\nonumber
    \langle y(t) y(t') \rangle
    \xrightarrow{\phi = \phi^\star}  \sum_{k} \, \langle \tilde{y}_k^2 \rangle \phi^\star_k(t) \phi^\star_k(t') \,,
\end{equation}
or $\langle \tilde{y}_j  \tilde{y}_k \rangle = 0$ for $j \neq k$. {$T_{\rm s}$ is a time scale, such that a stationary signal's harmonic frequencies are separated by $T_{\rm s}^{-1}$.}
An important point is that in the {\it preferred} basis, the variance of the sine and cosine pairs in each frequency bin are equal, and that different frequency bins are completely uncorrelated. Because of this preferentiality, there is no reason to expect that the same outcome will be realized in any arbitrary basis, such as in a Fourier basis with a different time window.
}}
}

{
Mathematically, the relation between the variances can be teased out using \eqref{eq:alpha_beta_relation}. {The windowing of a stationary signal} can also be used to physically argue why the variances of the sine and cosine coefficients will be different. Sines and cosines differ by fixed phases. However, in a stationary signal, sine and cosine modes have equal probability distribution functions (and hence moments, including the variance), because the fixed phase difference should not be distinguishable under stationarity. The outcome of an observation however can be strikingly different when only a short train of the signal is measured. In this case, the short period of observation or windowing breaks stationarity, and as a consequence distinguishes sines and cosines by their fixed phases.
}

It might be useful to note that the lower limit of \eqref{eq:timing_residual} implies that the DC and non-DC coefficients are related mathematically; it can be shown that $\alpha_{a0}/2 = -\sum_{k>=1} \beta_{ak}$, $\langle \alpha_{a0} \alpha_{b0} \rangle/4 = \sum_{jk} \langle \beta_{aj} \beta_{bk} \rangle$, and $0 = \langle \alpha_{a0} \alpha_{b0} \rangle/2 + \sum_k \left( \langle \beta_{ak} \alpha_{b0} \rangle + \langle \beta_{bk} \alpha_{a0} \rangle \right)/2 + \sum_{jk} \langle \beta_{aj} \beta_{bk} \rangle$.
The same arbitrary lower limit produces the extra phase terms in the temporal correlation \eqref{eq:C_filter_def}, whereas only the stationary term, i.e., $\exp\left({-2\pi f (t-t')}\right)$, is physical. See also Appendix A of \cite{Depta:2024ykq} for a related discussion.

\section{Comparison with simulation}
\label{sec:comparison_with_simulation}

We motivate the results---Eqs.~(\ref{eq:xy_correlation}-\ref{eq:F_functional}) and Eqs.~(\ref{eq:alpha_filter}-\ref{eq:beta0_filter})---by comparing with simulations, based on GWs emitted by a finite number of binaries. We use the standard \texttt{TEMPO2} \cite{Hobbs:2006cd, Edwards:2006zg, Hobbs:2009yn} pulsar timing software and its python wrapper \texttt{libstempo} to simulate a clean SGWB signal (noise-free); we choose $A_{\rm gw}=10^{-15}$, $\gamma_{\rm gw}=13/3$, and $n_{\rm bhb}=10^3$ for the SGWB characteristic stain, spectral index, and number of sources/SMBHBs, respectively. In addition, we take the GW emission frequencies to be $f \in (1, 100)$ nHz. For the reference positions on the sphere, we refer to the NANOGrav 15-year data set's 67 millisecond pulsars.

\begin{figure}[h!]
\centering
\subfigure[ \ signal ]{
    \includegraphics[width=0.45\textwidth]{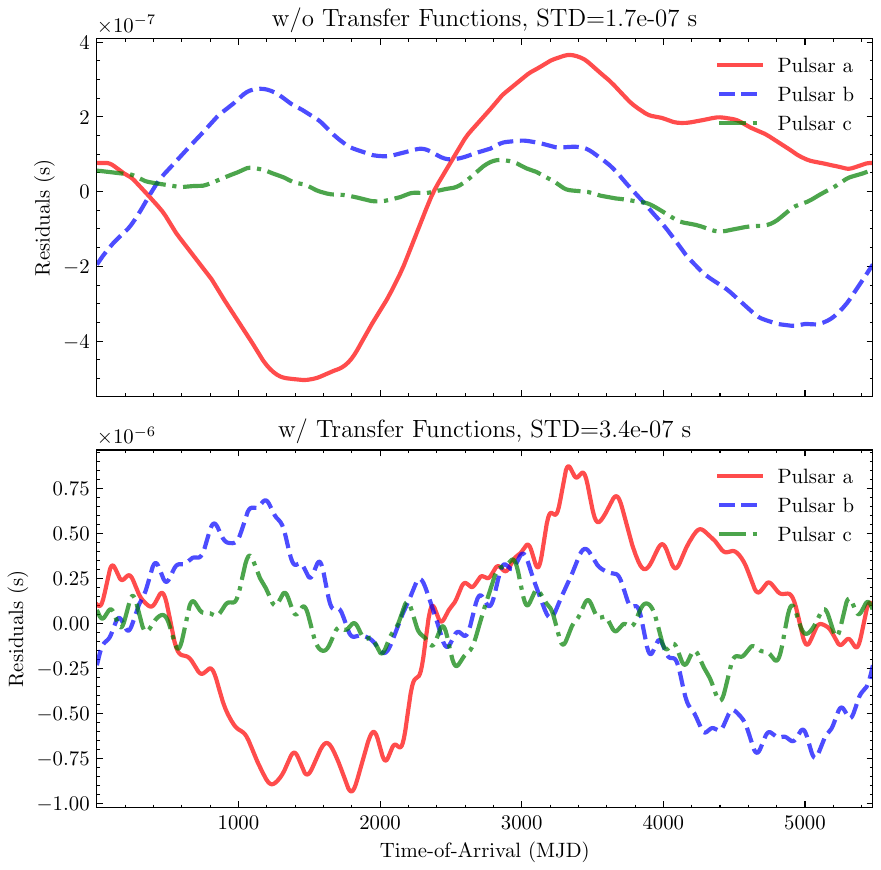}
}
\subfigure[ \ signal + white noise ]{
    \includegraphics[width=0.45\textwidth]{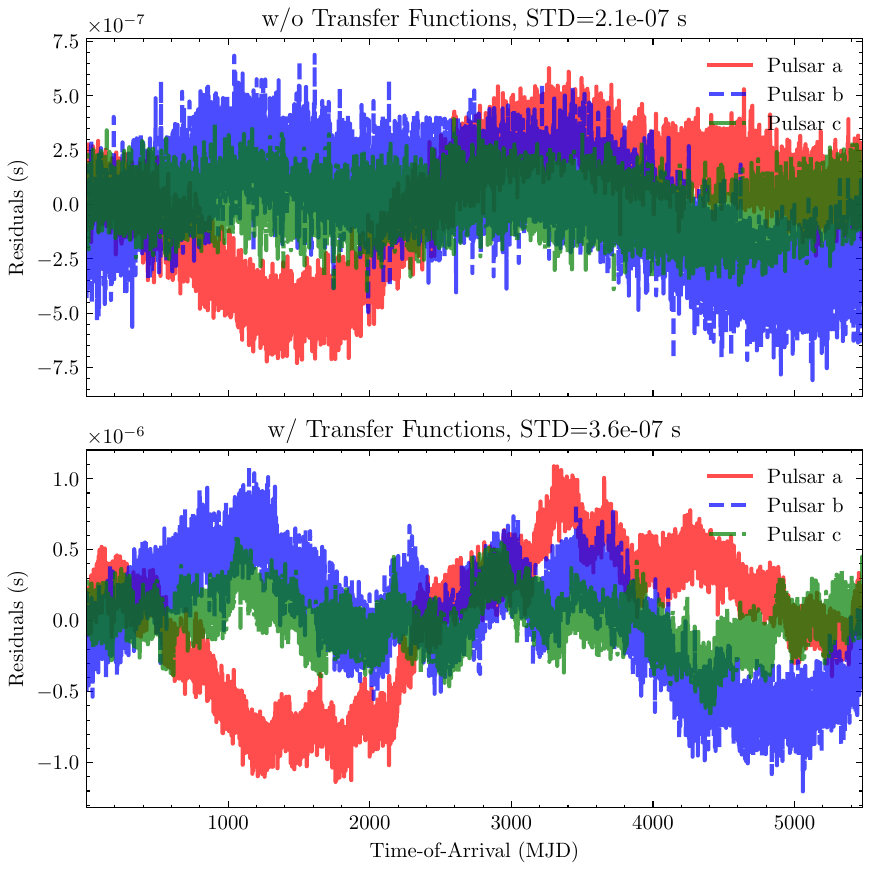}
}
\caption{Simulated pulsar timing residuals due to a Gaussian SGWB signal with a power-law spectrum \eqref{eq:power_spectral_density_power_law}; signal parameters: $A_{\rm gw}=10^{-15}$, $\gamma_{\rm gw}=13/3$, and $f_{\rm gw} \in (1, 100)$ nHz. Transfer functions (\ref{eq:alpha_filter}-\ref{eq:beta_filter}) are implemented in \texttt{PTAfast}. The STD shown are the average standard deviations of the timing residuals of the pulsar sample $\{a,b,c\}$; STD$\,=\{ \sqrt{ E[r(t)^2]-E[r(t)]^2 } \}\propto$ Amplitude of the time series, where $E$ is a time average operator and $\{ \cdots \}$ is the sample average over $a, b, c$. Note that $E[r(t)] = 0$ or that $\alpha_0=0$ in all cases.}
\label{fig:residuals_gauss}
\end{figure}

In addition, we employ our own simulation of a Gaussian SGWB signal in a PTA; which we refer to as \texttt{PTAfast} in the labels; building on a software that we developed for deriving correlation functions for PTA\footnote{This foreshadows a version of our open-source code \texttt{PTAfast} \cite{2022ascl.soft11001B} equipped with pulsar timing residual simulations.}. This was done by first generating a set of pulsar positions on the celestial sphere, $\hat{e}_a$, for $a=1, 2, \cdots n_{\rm psrs}$ pulsars; we generate $n_{\rm psrs}=100$ pulsars and scatter them across the sphere. Then, for each pulsar, labeled $a$, we generate a set of Fourier coefficients, $\alpha_{ak}$ and $\beta_{ak}$, for $k=1, 2, \cdots k_{\rm max}$ frequency bins, where $k_{\rm max}$ is determined by the highest frequency of the signal. To generate the signal contribution, for each frequency bin, we draw independent random numbers, $\xi$, from a Gaussian distribution with zero mean and unit variance. We correlate the pulsars according to Eqs.~(\ref{eq:xy_correlation}-\ref{eq:F_functional}) and Eqs.~(\ref{eq:alpha_filter}-\ref{eq:beta0_filter}) by constructing the covariance matrix ${\bf\Sigma}$, given by
\begin{equation}
\label{eq:covariance_matrix_unpolarized}
    \Sigma^k_{ab} \equiv \left( \begin{array}{cc}
    \langle \alpha_{ak} \alpha_{bk} \rangle & 0 \\[0.2cm]
    0 & \langle \beta_{ak} \beta_{bk} \rangle
    \end{array} \right)  \,,
\end{equation}
for all pulsar pairs $(a,b)$ and frequency bins $k$. Note that for a fixed frequency bin $k$, the covariance matrix ${\bf \Sigma}$ is a $2n_{\rm psrs}\times 2n_{\rm psrs}$ matrix, with the diagonal elements given by the variances of the Fourier coefficients. The off-diagonal elements in the upper left and lower right block matrices, $\langle \alpha_{ak} \alpha_{bk} \rangle$ and $\langle \beta_{ak} \beta_{bk} \rangle$, factor in pulsar pair correlations (HD curve). The rest of the off-diagonal elements vanish because the $\alpha$- and $\beta$-Fourier coefficients are uncorrelated. The Fourier components $\langle \alpha_{ak} \alpha_{bk} \rangle$ and $\langle \beta_{ak} \beta_{bk} \rangle$ are computed by convolving the transfer functions (\ref{eq:alpha_filter}-\ref{eq:beta0_filter}) with a power-law power spectral density,
\begin{equation}
\label{eq:power_spectral_density_power_law}
    S(f) \equiv \frac{A_{\rm gw}^2}{12 \pi^2} \left( \dfrac{f}{f_{\rm yr}} \right)^{-\gamma_{\rm gw}} f_{\rm yr}^{-3} \,,
\end{equation}
over the frequency range of the SGWB signal, $f \in (1, 100)$ nHz; $f_{\rm yr}= {\rm yr}^{-1}$. The coefficients $\alpha_{ak}$ and $\beta_{ak}$ are then obtained by applying the Cholesky decomposition to the covariance matrix ${\bf \Sigma} \equiv {\bf L} {\bf L}^{\rm T}$ and multiplying the lower triangular matrix ${\bf L}$ to the vector ${\bf \xi}$, i.e., ${\bf \alpha}, {\bf \beta} \sim {\bf L} {\bf \xi}$ \cite{Sato-Polito:2021efu}. The resulting Fourier coefficients are used to simulate the timing residuals via Eq.~\eqref{eq:residual_fourier_series}, i.e., $r(t, \hat{e}_a) \equiv \sum_k \left[ \alpha_{ak} \sin(\omega_k t) + \beta_{ak} \cos(\omega_k t) \right]$. {A sample of correlated residuals simulated by accounting for transfer functions are shown in Figure \ref{fig:residuals_gauss}. Timing residual simulations generated with and without transfer functions are visually distinguishable. Time correlated signals can be easily generated in the time-domain using Gaussian processes given the corresponding covariance matrix. The resulting time series for any signal with a continuous support exhibits all frequency components. In the frequency-domain, these frequency components are accounted for by transfer functions.} We refer the interested reader to Appendix \ref{sec:simulation} for more details.

With either \texttt{TEMPO2} or \texttt{PTAfast} PTA simulations, we process the timing data as follows. Each PTA realization gives a time series\footnote{
In general, each PTA realization will have signal and noise. If the noise has only a white component, then our results will hold in the first few frequency bins. On the other hand, if there are other correlated noises (a red process) in the data, a full PTA analysis can be done to distinguish the correlated components from the spatially-correlated GWB signal. 
}: $r(t, \hat{e}_a)$, for $a=1, 2, \cdots n_{\rm psrs}$ pulsars. The Fourier coefficients $\alpha_{ak}, \beta_{ak}$, for each pulsars, are extracted via a standard discrete Fourier transform routine, i.e., discretized Eq.~\eqref{eq:residual_fourier_components}. This function is available in many scientific computing libraries such as \texttt{scipy}. Once the Fourier coefficients are available, for each frequency bin (index $k$), products of $\alpha_{a k}$ and $\beta_{b k}$ are binned spatially (based on the pulsar-pair angular separation $\hat{e}_a \cdot \hat{e}_b$). This leads to a set
\begin{equation}
\label{eq:binned_data}
\begin{split}
\bigg \{ \alpha_{a k} \alpha_{b k}, \beta_{a k} \beta_{b k}, \alpha_{a k} \beta_{b k}; & \, k = 1, \cdots, n_{\rm bins}; \\
& (a,b)\equiv {\rm all \, pulsar \, pairs} \bigg\} \,.
\end{split}
\end{equation}
The sample averages are computed over each set \eqref{eq:binned_data}, fixed $k$ and fixed angular bin $\hat{e}_a \cdot \hat{e}_b \pm \Delta (\hat{e}_a \cdot \hat{e}_b)$ where $\Delta (\hat{e}_a \cdot \hat{e}_b)$ is an angular bin width, e.g., $\Delta (\hat{e}_a \cdot \hat{e}_b) \sim \cos (12^\circ)$ if we take 15 angular bins ($a\neq b$). We associate the terms $a=b$ (auto-correlation) with a zeroth angular bin at $\zeta = 0$. The processing of one PTA realization ends when the sample statistics are computed in all frequency and angular bins. For our tests, we consider 14 frequency bins, centered at $f_k = k/T$ for $k=1, \cdots, 15$, and 16 angular bins, centered at $\zeta_q = q \times (180^\circ / 15)$ for $q=0, \cdots, 15$. Then, the entire process is repeated numerous times with the same input; each realization turns in sample statistics in a PTA realization, representative of an underlying ensemble probability distribution function (PDF) across realizations.

\begin{figure*}[t]
    \centering
	\subfigure[]{
		\includegraphics[width = 0.45 \textwidth]{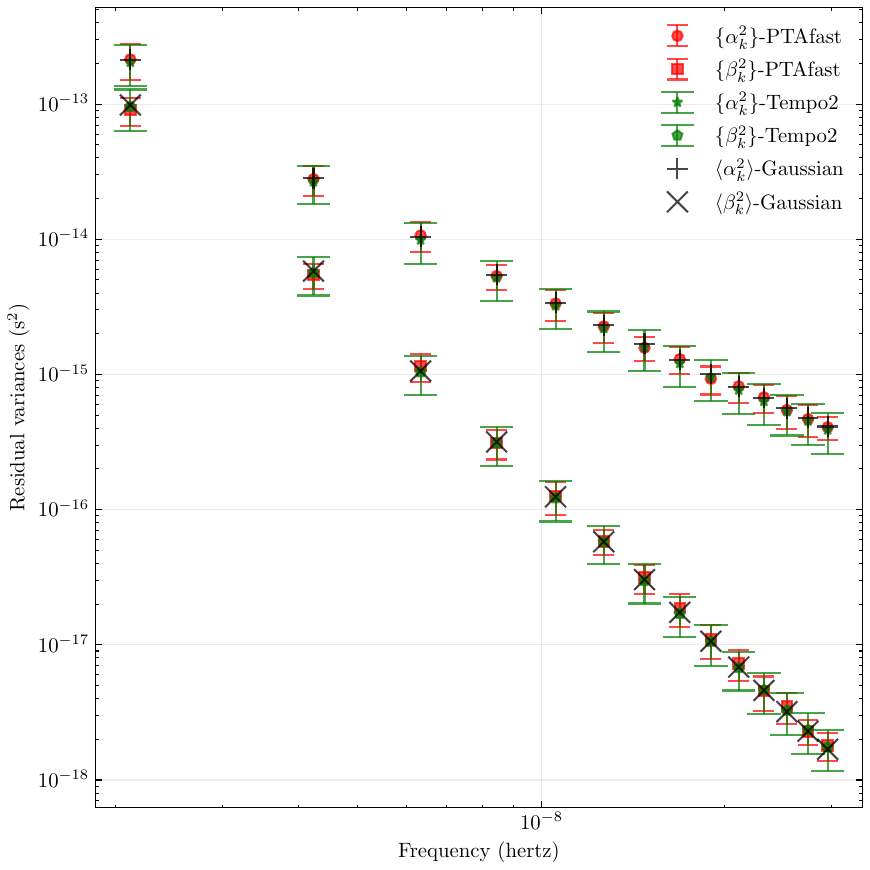}}
	\subfigure[]{
		\includegraphics[width = 0.45 \textwidth]{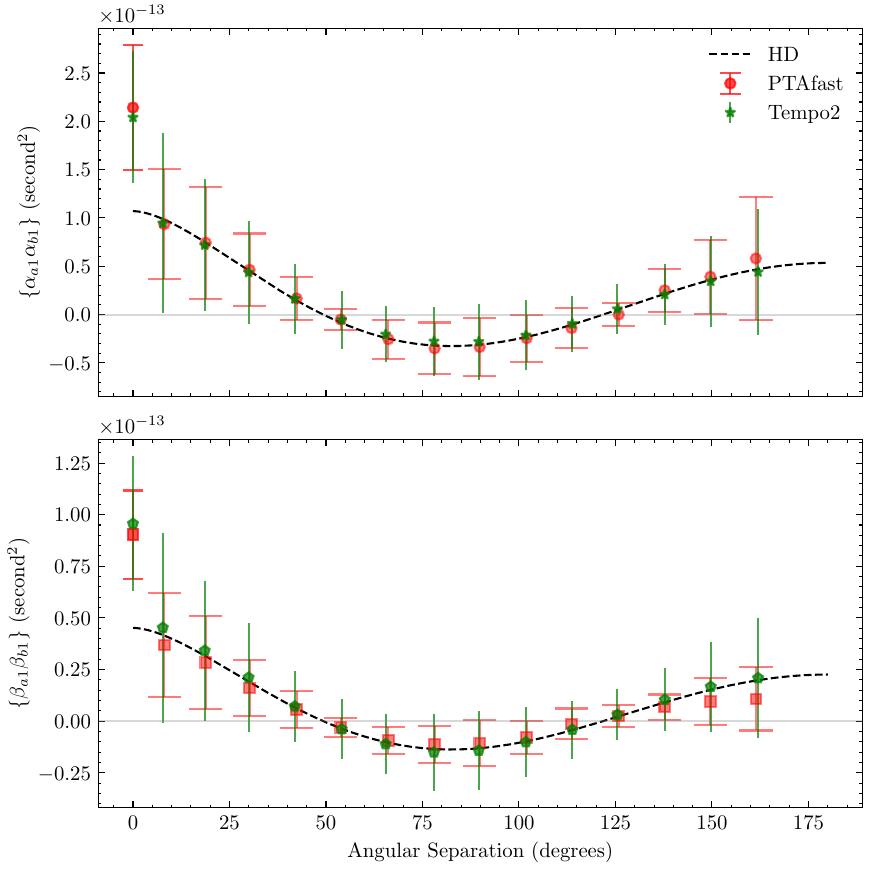}}
    \subfigure[]{
		\includegraphics[width = 0.45 \textwidth]{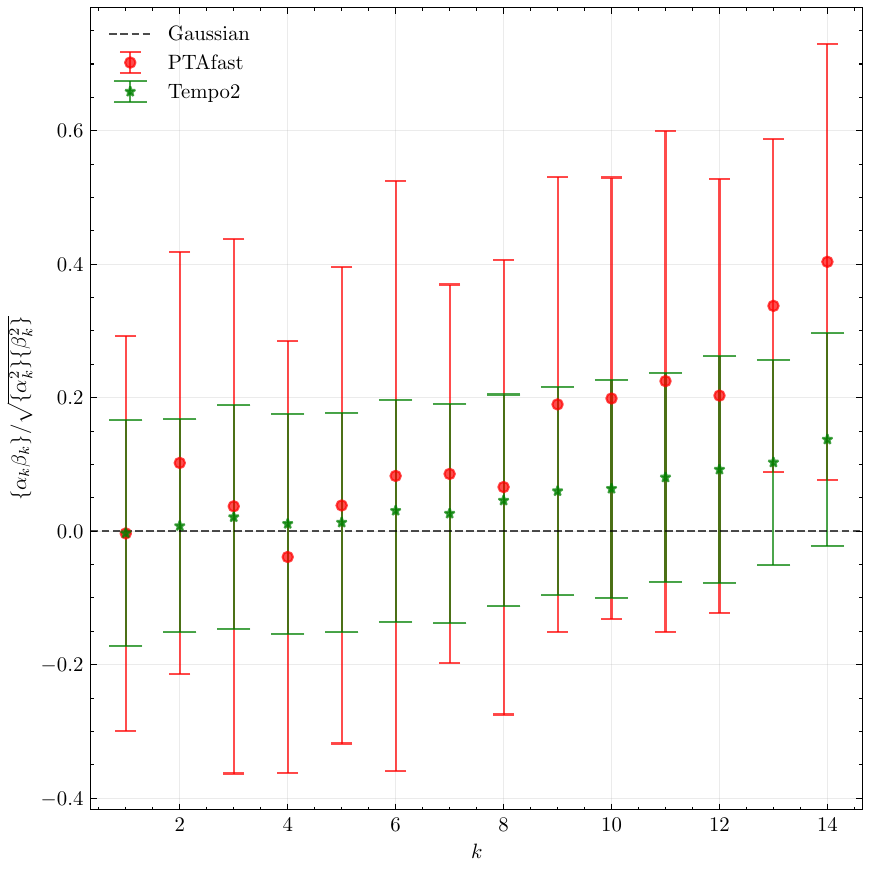}}
    \subfigure[]{
		\includegraphics[width = 0.45 \textwidth]{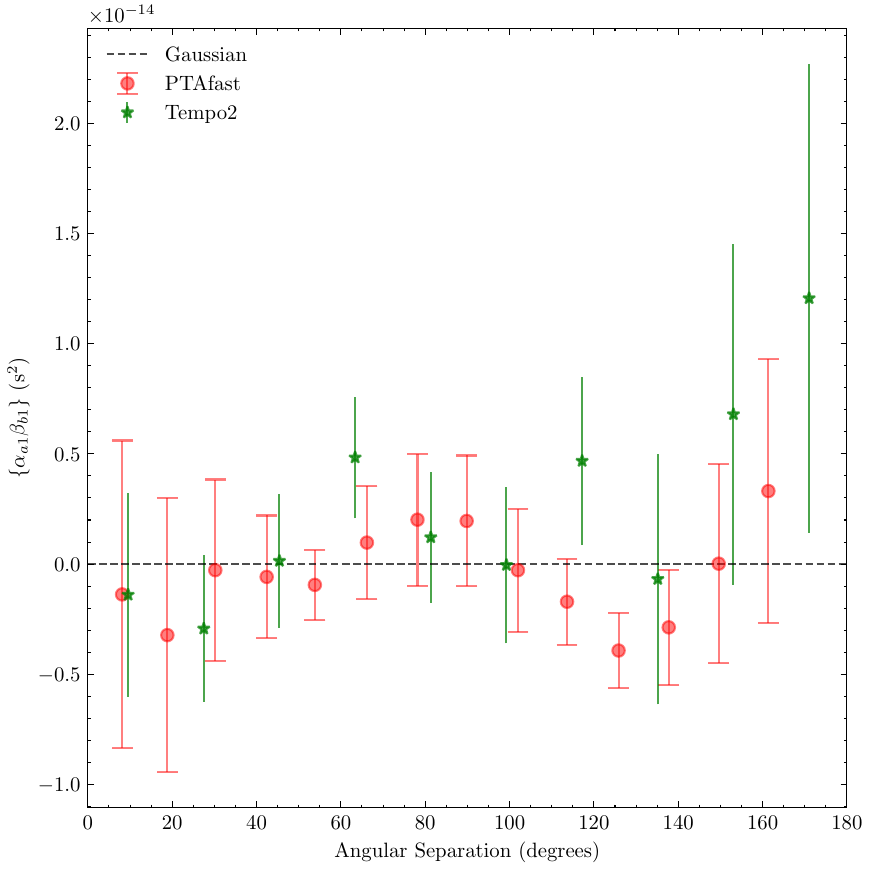}}
    \caption{Fourier components of the timing residual correlation: (a) sample variances, (b) two-point correlation function, (c) cross-Fourier statistic $\{ \alpha_{k} \beta_{k} \}/\sqrt{ \{ \alpha_k^2 \} \{ \beta_k^2 \} }$, and (d) cross-Fourier spatial correlation $\{ \alpha_{a1} \beta_{b1} \}$. Green and red points refer to \texttt{TEMPO2} {(67 pulsars at NG positions \cite{NANOGrav:2023gor})} and \texttt{PTAfast} {(100 pulsars scattered anisotropically)} simulations, respectively, and the error bars in subplots (a-c) represent the standard error across ${\cal O}\left( 10^3 \right)$ realizations. For subplot (d) the error bars shown are the error on the mean. Black points and dashed curves show the Gaussian theoretical expectation values based on Eqs.~(\ref{eq:xy_correlation}-\ref{eq:F_functional}) and Eqs.~(\ref{eq:alpha_filter}-\ref{eq:beta0_filter}) such as the HD curve in (b).}
    \label{fig:residual_spectra_and_correlation}
\end{figure*}

The resulting sample averages and two-point correlation function with ${\cal O}(10^3)$ realizations are shown in green (\texttt{TEMPO2}) and red (\texttt{PTAfast}) points with error bars in Figure~\ref{fig:residual_spectra_and_correlation}. This is shown together with Gaussian theoretical expectation values (black points and curves) obtained using the formalism (\ref{eq:xy_correlation}-\ref{eq:F_functional}) and the transfer functions (\ref{eq:alpha_filter}-\ref{eq:beta_filter}). We refer to $\{ \alpha_k \beta_k \} / \sqrt{ \{ \alpha_k^2 \} \{ \beta_k^2 \} }$ and $\{ \alpha_{ak} \beta_{bk} \}$ as a cross-Fourier statistic and a cross-Fourier spatial correlation, respectively.

The result shows that both \texttt{TEMPO2} and \texttt{PTAfast} simulations are in agreement with the theoretical predictions. The variances of the $\alpha$- and $\beta$-Fourier coefficients diverge in frequency space. This is because the Fourier basis over the observational frequencies $f_k = k/T$, i.e., $\{ \sin(2\pi k t/T), \cos(2\pi k t/T) \}$, does not diagonalize the signal covariance that has a characteristic time scale much larger than the observational window. The transfer functions (\ref{eq:alpha_filter}-\ref{eq:beta_filter}) account for the fact that the signal is only observed in a very short window compared to the signal time scale. The two-point correlation function is consistent with the HD correlation, indicating a GW observation. The cross-Fourier statistic $\sim\{ \alpha_k \beta_k \}$ and the cross-Fourier spatial correlation $\{ \alpha_{a1} \beta_{b1} \}$ are consistent with their Gaussian expectation values at zero.
Both vanish theoretically for all frequency and angular bins since the transfer function between $\alpha$- and $\beta$-Fourier coefficients is zero. (This is a useful reference point for processes that correlate the sine- and cosine-Fourier coeffients across pulsar pairs, such as circular polarization.)

To evaluate the theoretical expectation values, we consider a SMBHB power spectral density \eqref{eq:power_spectral_density_power_law} with a characteristic strain amplitude of $A_{\rm gw}=10^{-15}$ and a spectral index of $\gamma_{\rm gw}=13/3$. When evaluating the integrals, we truncate the lower and upper limits to the input signals's frequencies; so $f_{\rm min}=1$ nHz and $f_{\rm max}=100$ nHz. The integrations are performed in log-space in frequency.

It is worth highlighting that the theoretical Gaussian model shown in Figure~\ref{fig:residual_spectra_and_correlation} (black points and curves) is described by a single parameter, $A_{\rm gw}$; the spectral index is fixed to $\gamma_{\rm gw}=13/3$ for circular SMBHBs. Furthermore, the choice of reference frequency, $f_{\rm ref}=f_{\rm yr}$, is arbitrary; changing the reference frequency, $f_{\rm ref} \rightarrow f_{\rm ref}'$, amounts to shifting the amplitude by $A_{\rm gw} \rightarrow A_{\rm gw}' = A_{\rm gw} \left( f_{\rm ref}'/f_{\rm ref} \right)^{-13/3}$. For a power law power spectrum \eqref{eq:power_spectral_density_power_law}, the appropriate shift will be $A_{\rm gw} \rightarrow A_{\rm gw}' = A_{\rm gw} \left( f_{\rm ref}'/f_{\rm ref} \right)^{-\gamma_{\rm gw}}$. In practice, shifting the reference frequency is going to statistically change the covariance between $A_{\rm gw}$ and $\gamma_{\rm ref}$, especially when errors are huge.
However, if the SGWB signal were interpreted with a diagonal basis utilizing Fourier modes at frequencies $f_k=k/T$, then the choice of reference frequency could turn to be a source of a systematic error on the measured amplitude of the SGWB. See \cite{Bernardo:2024tde, Crisostomi:2025vue} for related systematic errors rooted in approximating the covariance with only diagonal terms and infinitesimally narrow filters.

In PTA analyses, the raw time-of-arrival data undergoes post-fitting; effectively projecting out constant, linear, and quadratic terms in the time series. This reduces the power difference between the ($\alpha$- and $\beta$-) frequency bins and suppresses subnanohertz GW contributions. The resulting observation can be characterized by a transmission function which must be convolved with the power spectrum. Crucially, our results indicate that the distinction of the $\alpha$- and $\beta$-variances persists regardless of post-fitting, reinforcing the necessity of revisiting signal modeling assumptions. While real PTA data is also subject to contamination from white noise, particularly at higher frequency bins, and intrinsic red noise in each pulsar, these factors primarily affect the precision of SGWB measurements.
White noise diminishes the experimental sensitivity at high frequencies, while red noise contributes to the overall measurement uncertainty.

We emphasize that we do not use the variances as direct estimators of the power spectral density, e.g., by constructing the periodogram. Instead, we compute the full covariance function in the frequency- and Fourier-domain, evaluating integrals of the form \eqref{eq:F_functional} to properly account for the dominant frequency correlations \cite{Allen:2024uqs, Crisostomi:2025vue}.

The reference pulsar positions are arbitrary; for example, using Meerkat PTA positions \cite{Miles:2022lkg, Miles:2024rjc} would yield similar results, with differences manifesting only in the ensemble variance. However, the number of pulsars in the array plays a crucial role in determining the statistical variance across realizations—larger arrays provide better sampling and reduce the impact of statistical fluctuations (see Appendix \ref{sec:number_pulsars} for details). 

The choices for the number of frequency and angular bins are also somewhat arbitrary, but are constrained by practical considerations such as the number of pulsars and the cadence of the timing data. In particular, resolving higher frequency bins requires sufficiently dense sampling; for instance, in Figure~\ref{fig:residual_spectra_and_correlation}, the apparent systematic increase in the cross-Fourier statistic at high frequencies for both \texttt{TEMPO2} and \texttt{PTAfast} is attributable to limitations in cadence; ${\cal O}(10^4)$ time-of-arrivals over 15 years were needed to resolve these bins. These considerations highlight the importance of array size and data quality, but do not affect the agreement between simulation and theory, underscoring the robustness of the analysis.

\section{The polarized signal: simulating circular polarization in a PTA}
\label{sec:circular_polarization_in_pta}

We lay down the formalism extending Section \ref{sec:the_signal} to include the signal imprinted by a circularly polarized SGWB on the PTA residuals.

\subsection{Circularly Polarized Basis Tensors and Stokes Parameters}

We consider a Gaussian SGWB with a two-point function of the generalized form,
\be
\label{eq:paa}
\langle h_{A}\left(f,\hat{k}\right) h^*_{A'}\left(f',\hat{k}'\right) \rangle
= \delta\left(f-f'\right) \delta\left(\hat{k}-\hat{k}'\right)P_{AA'}\left(f,\hat{k}\right) \,,
\ee
where the power spectra, $P_{AA'}(f,\hat{k})$, is now anisotropic (direction-dependent); in contrast with Eq.~\eqref{eq:gwb_power_spectrum}. Then, the unpolarized signal can be described in the limit $P_{AA'}\left(f, \hat{k}\right) \rightarrow I(f) \delta_{AA'}$ (Sections \ref{sec:the_signal}-\ref{sec:comparison_with_simulation}). 

Analogous to electromagnetic waves, a suitable way to look into circular polarization is through Stokes parameters. To do so, we consider complex circular polarization basis tensors; defined as
\begin{align}
    \varepsilon_{R} &= \frac{(\varepsilon_{+} + i \varepsilon_{\times})}{\sqrt{2}} \,,
   &\varepsilon_{L} &= \frac{(\varepsilon_{+} - i \varepsilon_{\times})}{\sqrt{2}} \,,
\end{align}
where $\varepsilon_{R}$ ($\varepsilon_{L}$) stands for a right-handed (left-handed) GW with a positive (negative) helicity. The GW amplitudes in Eq.~\eqref{eq:gw_general} in the two different bases are related to each other via
\begin{align}
    h_{R} &= \frac{(h_{+} - i h_{\times})}{\sqrt{2}} \,,
   &h_{L} &= \frac{(h_{+} + i h_{\times})}{\sqrt{2}} \,.
\end{align}
Then, the power spectra or coherency matrix $P_{AA'}\left( f, \hat{k} \right)$ in Eq.~\eqref{eq:paa} can be associated with Stokes parameters, $I$, $Q$, $U$, and $V$ via 
\begin{align}
    I &= \left[ \langle h_R h_R^* \rangle + \langle h_L h_L^* \rangle \right] / 2 \,,\nonumber\\
    Q + iU &=  \langle h_L h_R^*  \rangle \,,\nonumber\\
    Q - iU &=  \langle h_R h_L^*  \rangle \,,\nonumber\\
    V &= \left[ \langle h_R h_R^* \rangle - \langle h_L h_L^* \rangle \right] / 2 \,\label{eq:spIQUV},
\end{align}
which are functions of the frequency $f$ and the propagation direction $\hat{k}$.
The $I\left( f, \hat{k} \right)$ is the intensity, $Q\left( f, \hat{k} \right)$ and $U\left( f, \hat{k} \right)$ represent the linear polarization, and $V\left( f, \hat{k} \right)$ is the circular polarization. In terms of the Stokes parameters, an unpolarized signal is characterized by $I\neq 0$ and $Q=U=V=0$. A circularly polarized signal is characterized by $I\neq 0$ and $V\neq 0$, while $Q=U=0$.

Now, we consider a circularly polarized SGWB. We expand
the Stokes parameters in terms of spherical harmonics as
\begin{align}
    I(f,\hat{k}) &= \sum_{\ell m}I_{\ell m}(f) \; Y_{\ell m}(\hat{k}) \,,\nonumber \\
    V(f,\hat{k}) &= \sum_{\ell m}V_{\ell m}(f) \; Y_{\ell m}(\hat{k}) \,,
\end{align}
where $I(-f,\hat{k})=I(f,\hat{k})$ and $V(-f,\hat{k})=-V(f,\hat{k})$.
Since $I(f,\hat{k})$ and $V(f,\hat{k})$ are real, we have
\begin{align}
I_{\ell m}^*(f)&= (-1)^m I_{\ell -m}(f) \,,\nonumber\\
V_{\ell m}^*(f)&= (-1)^m V_{\ell -m}(f) \,.
\label{IVlmconjugate}
\end{align}

\subsection{Timing Residual Power Spectrum and Correlation}
\label{subsec:spectrum_cp}

Following Sections \ref{subsec:gw_residual}-\ref{subsec:timing_and_orf} with the two-point function \eqref{eq:paa}, the timing residual power spectrum can be shown to be \cite{Liu:2022skj}
\be
\begin{split}
& \langle a_{\ell_1 m_1} a^*_{\ell_2 m_2} \rangle \\
& =
  \intinf \frac{df}{(2\pi f)^2}\,C(f, t_a, t_b) 
   \\
& \phantom{ggg} \sum_{X=\{I,V\}}
   \sum_{\ell m}  X_{\ell m}(f)\, {\tilde\gamma}_{\ell m,\ell_1 m_1 \ell_2 m_2}^{X}(fD_a,fD_b)\,,
\end{split}
\ee
where the coefficients are given by a time integral and Wigner-3j symbols:
\begin{align}
& {\tilde\gamma}_{\ell m,\ell_1 m_1 \ell_2 m_2}^{I,V}(fD_a,fD_b) \nonumber \\
& = \; (-1)^{m_1} \left[1 \pm (-1)^{\ell+\ell_1+\ell_2}\right] J_{\ell_1}(fD_a) J^*_{\ell_2}(fD_b) \nonumber\\
  & \phantom{ggggg} \times \sqrt{\frac{(2\ell +1)(2\ell_1 +1)(2\ell_2 +1)}{4\pi}} \nonumber\\
  & \phantom{gggggg} \times
    \begin{pmatrix}
          \ell &&   \ell_1  &&  \ell_2 \\
            0  &&     -2     &&  2
    \end{pmatrix}
    \begin{pmatrix}
        \ell  && \ell_1  &&  \ell_2 \\
          m &&  -m_1  &&  m_2 
    \end{pmatrix} \,.
\end{align}
The $J_{\ell}(y)$'s are defined as
\be
J_{\ell}(y)= \sqrt{2}\,\pi\, i^{\ell} \sqrt{\frac{(\ell+2)!}{(\ell-2)!}} \int_0^{2\pi y} dx\,e^{ix} \frac{j_{\ell}(x)}{x^2}\,,
\ee
and the time kernel $C(f, t, t')$ is given by Eq.~\eqref{eq:C_filter_def}. Then, the pulsar timing residual correlation can be expressed in the form
\be
\begin{split}
& \langle r(t_a,\hat{e}_a)r(t_b,\hat{e}_b) \rangle \\
& \, =
  \intinf \frac{df}{(2\pi f)^2}\,C(f, t_a, t_b) \\
  & \phantom{ggg}
   \sum_{X=\{I,V\}}
   \sum_{\ell m}  X_{\ell m}(f) \gamma_{\ell m}^{X}(fD_a,fD_b;\hat{e}_a,\hat{e}_b)\,.
\end{split}
\ee
where the ORFs $\gamma_{\ell m}^{I,V}(fD_a,fD_b;\hat{e}_a,\hat{e}_b)$ are defined as
\be
\begin{split}
 & \gamma_{\ell m}^{I,V}(fD_a,fD_b;\hat{e}_a,\hat{e}_b) \\
& = \sum_{\ell_1 m_1 \ell_2 m_2}
{\tilde\gamma}_{\ell m,\ell_1 m_1 \ell_2 m_2}^{I,V}(fD_a,fD_b)\,Y_{\ell_1 m_1}(\hat{e}_a)  Y^*_{\ell_2 m_2}(\hat{e}_b)\,.
\label{IVORF}
\end{split}
\ee

Note that the ORFs are complex in general. As a special case, $\gamma_{00}^I$ is real for $D_a=D_b$ and $\gamma_{00}^V=0$. The latter is expected for a circularly polarized SGWB because of the parity symmetry in a PTA experiment. Under the limit that $fD_a\gg 1$ and $fD_b\gg 1$, $\gamma_{00}^I$ reduces to the HD curve and ${\tilde\gamma}_{\ell m,\ell_1 m_1 \ell_2 m_2}^{I,V}(fD_a,fD_b)$ becomes constant real values. {Since $\gamma^V_{00}=0$, a PTA experiment is insensitive to the circular polarization of an isotropic circularly polarized SGWB. Some degree of anisotropy in the SGWB is needed for circular polarization to be detectable \cite{Kato:2015bye}.}

\subsection{Correlation in the Computational Frame}
\label{subsec:computational}

In the computational frame \cite{Mingarelli:2013dsa, Kato:2015bye, Belgacem:2020nda, Chu:2021krj, Liu:2022skj}, in which pulsar $a$ is placed along the $\hat{\mathbf{z}}$-axis
while pulsar $b$ is in the $\hat{\mathbf{x}}$-$\hat{\mathbf{z}}$ plane, 
the timing-residual correlation becomes
\be
\begin{split}
& \langle r(t_a,\hat{e}_a)r(t_b,\hat{e}_b) \rangle \\
& \, =
  \intinf \frac{df}{(2\pi f)^2}\,C(f, t_a, t_b) \\
  & \phantom{ggg}
   \sum_{X=\{I,V\}}
   \sum_{\ell m}  X_{\ell m}(f) \gamma_{\ell m}^{X}(fD_a,fD_b,\zeta)\,,
\end{split}
\label{residual_com}
\ee
where $\zeta$ is their separation angle ($\cos \zeta = \hat{e}_a \cdot \hat{e}_b$) and
\begin{align}
    & \gamma_{\ell m}^{I,V}(fD_a,fD_b,\zeta) \nonumber \\
  & \, = \sum_{\ell_1 \ell_2} 
    (-1)^m \frac{2\ell_1+1}{4\pi} \left[1 \pm (-1)^{\ell+\ell_1+\ell_2}\right] \phantom{\dfrac{1}{1}} \nonumber\\
    & \phantom{ggg} \times J_{\ell_1}(fD_a) J^*_{\ell_2}(fD_b) \;
  Y_{\ell_2 m}(\zeta,0) \nonumber \\
  & \phantom{gggg} \times \sqrt{(2\ell +1)(2\ell_2 +1)} \phantom{\dfrac{1}{1}} \nonumber \\
  & \phantom{ggggg} \times 
    \begin{pmatrix}
          \ell &&   \ell_1  &&  \ell_2 \\
            0  &&     -2     &&  2
    \end{pmatrix}
    \begin{pmatrix}
        \ell  && \ell_1  &&  \ell_2 \\
          m &&    0       &&  -m
    \end{pmatrix} \,. 
\end{align}
The reflection property~(\ref{reflection}) guarantees that the ORFs in the computational frame have the conjugate relations:
\begin{align}
\gamma_{\ell -m}^{I}&=(-1)^{m} \gamma_{\ell m}^{I}\,,
\\
\gamma_{\ell -m}^{V}&=(-1)^{m+1}\gamma_{\ell m}^{V}\,.
\label{IVconjugate}
\end{align}
Then, using Eq.~\eqref{IVlmconjugate} and Eq.~\eqref{IVconjugate}, we can recast the sum in Eq.~\eqref{residual_com} into
\be
\begin{split}
& \sum_{X=\{I,V\}}\sum_{\ell m}  X_{\ell m}(f) \gamma_{\ell m}^{X}(fD_a,fD_b,\zeta)
\\
& \,\,\,\, =\sum_{\ell m} \bigg[ {\rm Re}[I_{\ell m}(f)] \gamma_{\ell m}^{I}(fD_a,fD_b,\zeta) \\
& \phantom{gggggggg} + i\,{\rm Im}[V_{\ell m}(f)] \gamma_{\ell m}^{V}(fD_a,fD_b,\zeta) \bigg]\,.
\end{split}
\ee
Finally, the timing residual correlation in the computational frame can be expressed as
\bw
\begin{align}
\langle r(t_a,\hat{e}_a)r(t_b,\hat{e}_b) \rangle= &
  \intzeroinf\frac{df}{(2\pi f)^2} \left\{ C(f, t_a, t_b) 
  \sum_{\ell m}  \left[ I_{\ell m}(f) \gamma_{\ell m}^{I}(fD_a,fD_b,\zeta) +  V_{\ell m}(f) \gamma_{\ell m}^{V}(fD_a,fD_b,\zeta) \right] \right.
  \nonumber\\
  & \left. \phantom{ggggggggg} +\, C(f, t_a, t_b)^* 
  \sum_{\ell m}  \left[ I_{\ell m}(f) \gamma_{\ell m}^{I*}(fD_a,fD_b,\zeta) -  V_{\ell m}(f) \gamma_{\ell m}^{V*}(fD_a,fD_b,\zeta) \right]\right\}\,.
\end{align}
To arrive at the last expression, we used $J_{\ell}(-fD)=-J_{\ell}^*(fD)$. In the case $fD_a\gg 1$ and $fD_b\gg 1$, $\gamma_{\ell m}^{I,V}(fD_a, fD_b, \zeta)$ become real functions~\cite{Liu:2022skj}.\footnote{
As a special case, when $t_a=t_b=t$, the pulsar timing residual correlation becomes
\be
\begin{split}
& \langle r(t,\hat{e}_a)r(t,\hat{e}_b) \rangle \\
& =
 2 \intzeroinf {df}\, \frac{\sin^2(\pi f t)}{(\pi f)^2} 
  \sum_{\ell m}  \bigg[ I_{\ell m}(f)\, {\rm Re}\left[\gamma_{\ell m}^{I}(fD_a, fD_b,\zeta)\right] 
  \\
& {\phantom{ggggggggggggggggggggggg}} + i V_{\ell m}(f)\, {\rm Im}\left[\gamma_{\ell m}^{V}(fD_a, fD_b,\zeta)\right] \bigg] \,. 
  \end{split}
\ee
}
In general, with $t_a \neq t_b$, the pulsar timing residual correlation can be expressed as
\be
\begin{split}
\langle r(t_a,\hat{e}_a)r(t_b,\hat{e}_b) \rangle= 
  \intzeroinf\frac{2\,df}{(2\pi f)^2}
  & \bigg[ {\rm Re}[C(f,t_a,t_b)]
  \sum_{\ell m}  I_{\ell m}(f) \gamma_{\ell m}^{I}(fD_a, fD_b,\zeta) \\
  & + {\rm Im}[C(f,t_a,t_b)]
  \sum_{\ell m} i V_{\ell m}(f) \gamma_{\ell m}^{V}(fD_a, fD_b,\zeta) \bigg]\,,
\label{unequal_time_residual}
\end{split}
\ee
\ew
where
\begin{align}
{\rm Re}[C(f,t_a,t_b)]&= 1-\cos(2\pi f t_a) - \cos(2\pi f t_b) \nonumber \\
& \phantom{ggggggggggggggii} + \cos[2\pi f (t_b-t_a)]\,,\nonumber \\
{\rm Im}[C(f,t_a,t_b)]&= \sin(2\pi f t_a) - \sin(2\pi f t_b) \nonumber \\
& \phantom{gggggggggiiii} + \sin[2\pi f (t_b-t_a)]\,.
\label{CStime}
\end{align}
The time kernels ${\rm Re}[C(f,t_a,t_b)]$ and ${\rm Im}[C(f,t_a,t_b)]$ are owed to wave propagation in time and are the generalization of Eq.~\eqref{eq:C_filter_def} to a circularly polarized SGWB.

\begin{figure}[h!]
    \centering
    \subfigure[]{
        \includegraphics[width=0.45\textwidth]{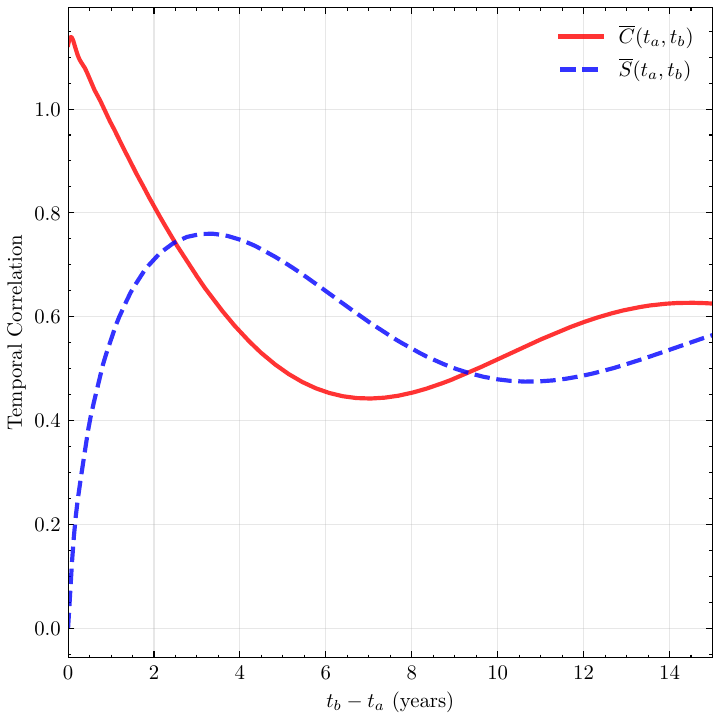}
    }
    \subfigure[]{
        \includegraphics[width=0.45\textwidth]{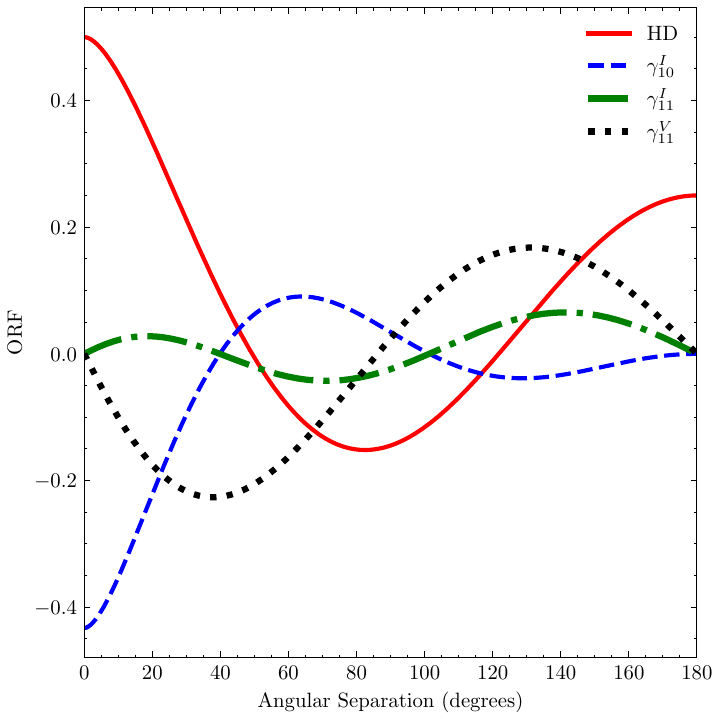}
    }
    \caption{(a) Temporal correlation in the timing residual correlation induced by a circularly polarized SGWB with a flat power spectrum, $I_{lm}(f)\sim$ Constant, $V_{lm}(f)\sim$ Constant, $T = 15$ years; $\overline{C}(t_a, t_b)$ (red solid) and $\overline{S}(t_a, t_b)$ (blue dashed) are temporal correlations associated with intensity and circular polarization, respectively, of the SGWB signal; (b) ORFs, $\gamma^I_{00}(\zeta)=\gamma^{\rm HD}(\zeta)$, $\gamma^I_{10}(\zeta)$, $\gamma^I_{11}(\zeta)$, and $\gamma^V_{11}(\zeta)$, in the computational frame for a circularly polarized SGWB \cite{Kato:2015bye, Sato-Polito:2021efu, Liu:2022skj}.}
    \label{fig:temporal_correlation}
\end{figure}

For visualization, assume that $I_{lm}$'s, $V_{lm}$'s, and their correponding ORFs weakly depend on frequency over the PTA sensitivity range, and that $f D_a, f D_b \gg 1$ is satisfied for all pulsar pairs. 
Then, we have
\be
\begin{split}
\langle r(t_a,\hat{e}_a)r(t_b,\hat{e}_b) \rangle= & \,\,
   {\bar C}(t_a,t_b) \sum_{\ell m}  I_{\ell m} \gamma_{\ell m}^{I}(\zeta)
   \\
   & \, + {\bar S}(t_a,t_b) \sum_{\ell m} i V_{\ell m} \gamma_{\ell m}^{V}(\zeta)\,,
\end{split}
\ee
where the weighted temporal correlations ${\bar C}(t_a,t_b)$ and ${\bar S}(t_a,t_b)$ are defined as $n_\delta \int_{1/T} \left( 2 df/(2\pi f)^2 \right) {\rm Re}[C(f,t_a,t_b)]$ and $n_\delta \int_{1/T} \left( 2 df/(2\pi f)^2 \right) {\rm Im}[C(f,t_a,t_b)]$. The normalization constant $n_\delta= 1/ \int_{1/T} \left( 2 df/(2\pi f)^2 \right)$ is arbitrary and was chosen for illustration purposes.
The integrals were regularized by using an infrared cutoff at $f \sim 1/T$.
Figure \ref{fig:temporal_correlation} shows ${\bar C}(t_a,t_b)$
and ${\bar S}(t_a,t_b)$ as a function of time and the spatial correlations $\gamma^I_{00}(\zeta)=\gamma^{\rm HD}(\zeta)$, $\gamma^I_{10}(\zeta)$, $\gamma^I_{11}(\zeta)$, and $\gamma^V_{11}(\zeta)$.

The pulsar timing residual correlation \eqref{unequal_time_residual} shows that $\bar{C}(t, t')$ and $\bar{S}(t, t')$ are the temporal correlations associated with the intensity and circular polarization of the SGWB signal, respectively. Figure \ref{fig:temporal_correlation} shows that the temporal correlation $\bar{C}(t, t')$ is nonzero for all times $t$ and $t'$, while $\bar{S}(t, t')$ is non-zero only when $t\neq t'$. This means that the contribution by circular polarization vanishes at equal times, i.e., $\bar{S}(t,t)=0$. Figure \ref{fig:temporal_correlation} also shows spatial correlations that would be relevant to a circularly polarized SGWB signal; that is produced by a kinematic dipole (next section). These ORFs admit analytical expressions that were derived {{in the computational frame}} and can be found in \cite{Kato:2015bye}. We utilize these ORFs in the simulations of a circularly polarized SGWB signal in Section \ref{sec:comparison_with_simulation_cp}.

\subsection{Kinematic Dipole}
\label{subsec:kinematic_dipole}

It is instructive to see the circularly polarized signal produced by a kinematic boost, akin to the cosmic dipole.

{
We remind that anisotropy and circular polarization are distinct aspects of a signal, and that in practice it is possible to phenomenologically constrain these components by decomposing the signal in a given basis to tease out known signatures \cite{Mingarelli:2013dsa, Kato:2015bye, Hotinli:2019tpc, Belgacem:2020nda, Sato-Polito:2021efu, Chu:2021krj, Depta:2024ykq, Cruz:2024svc, Cruz:2024esk}). On the other hand, the kinematic dipole is of particular interest to cosmology due to debates about the cosmic dipole \cite{Secrest:2020has, Secrest:2022uvx, Dam:2022wwh, Oayda:2024hnu, vonHausegger:2024jan, Antony:2025tzk, Secrest:2025wyu, Yoo:2025qdq}. As we shall see, the kinematic boost {supports} both anisotropic and circularly polarized components which {can be thought of as} leading order corrections {following the} isotropic {part of the} signal. {We stress out that GW and SGWB signals are naturally polarized to begin with, but that anisotropy is a separate ingredient needed to be experimentally sensitive to polarization.} The advantage of this is that the cosmic dipole's direction and speed are known to a reasonable degree based on cosmological surveys. Thus, in theory, constraining the cosmic dipole in PTAs boils down to searching for well defined correlations modulo the HD curve in the timing residuals \cite{Cruz:2024svc, Cruz:2024esk}.
}

Consider a kinematic boost $\vec{v}=v\hat{v}$; magnitude $v$ and direction $\hat{v}$. When applied, an initially isotropic circularly polarized SGWB will acquire a dipole anisotropy to leading order in $v/c$:
\begin{align}
    I(f,\hat{k}) &= I_{00}(f) \left[1+\beta\, (1-n_I(f))\,  \hat{k}\cdot\hat{v}\right]\,,\nonumber \\
    V(f,\hat{k}) &= V_{00}(f) \left[1+\beta\, (1-n_V(f))\,  \hat{k}\cdot\hat{v}\right] \,,
\end{align}
where $\beta = v/c \ll 1$, and
\be
n_I(f)=\frac{d\ln I_{00}(f)}{d\ln f}\,,\quad n_V(f)=\frac{d\ln V_{00}(f)}{d\ln f}\,.
\ee
Expanding $\hat{k}\cdot\hat{v}$ in spherical harmonics (Appendix~\ref{sec:spinweight}),
\be
\hat{k}\cdot\hat{v}=\frac{4\pi}{3}\sum_{m=-1}^{1} Y_{1m}^*(\hat{v})\, Y_{1m} (\hat{k})\,,
\ee
the dipole components of the Stokes parameters become
\begin{align}
I_{1m}(f)&=\frac{4\pi}{3}\, I_{00}(f)\, \beta\, [1-n_I(f)]\, Y_{1m}^*(\hat{v})\,,\nonumber\\
V_{1m}(f)&=\frac{4\pi}{3}\, V_{00}(f)\, \beta\, [1-n_V(f)]\, Y_{1m}^*(\hat{v})\,.
\label{dipole}
\end{align}

Now, consider a circularly polarized SGWB characterized by these dipole components. Recall that $\gamma_{00}^V=\gamma_{10}^V=0$, $\gamma_{1-1}^V=\gamma_{11}^V$, and $\gamma_{1-1}^I=-\gamma_{11}^I$. Therefore, we have
\begin{equation}
\label{eq:Ilmgammalm}
\begin{split}
& \sum_{\ell m} I_{\ell m}(f) \gamma_{\ell m}^{I}(f, \zeta) \\
& =  I_{00}(f)\bigg[\gamma_{00}^I(f, \zeta) + \beta \left(1-n_I(f)\right)  \\
& \phantom{ggggggggg} \times \bigg( \gamma_{10}^I(f, \zeta) \sqrt{\frac{4\pi}{3}} \cos\theta_v
\\
& \phantom{ggggggggggggg} - \gamma_{11}^I(f, \zeta) \sqrt{\frac{8\pi}{3}}\, \sin\theta_v \cos\phi_v \bigg) \bigg] \,,
\end{split}
\end{equation}
and
\begin{equation}
\label{eq:Vlmgammalm}
\begin{split}
i \sum_{\ell m} V_{\ell m}(f) \gamma_{\ell m}^{V}(f, \zeta) = \, & V_{00}(f)\, \gamma_{11}^V(f, \zeta) \beta \left(1-n_V(f)\right) \\
& \times \sqrt{\frac{8\pi}{3}} \sin\theta_v \sin\phi_v \,.
\end{split}
\end{equation}
We emphasize that in the above expression, the dipole direction $(\theta_v, \phi_v)$ is given in the computational frame \cite{Mingarelli:2013dsa, Kato:2015bye, Belgacem:2020nda, Sato-Polito:2021efu, Chu:2021krj, Liu:2022skj}. This is an important consideration when simulating the circularly polarized signal due to a kinematic dipole and plays a crucial role in constructing the covariance matrix ${\bf \Sigma}$ (see Sections \ref{sec:comparison_with_simulation} and \ref{sec:comparison_with_simulation_cp} or Appendix \ref{sec:simulation}).

\subsection{Transfer Functions}
\label{subsec:filters_cp}

Building on Section \ref{subsec:filters}, we now extend the transfer function formalism to the case of a circularly polarized SGWB, using the timing residual correlation \eqref{unequal_time_residual}. The goal is to express the correlations between Fourier components across frequency bins, as in Section \ref{sec:standard_analysis_of_the_signal}, and to identify the transfer functions that govern the imprint of circular polarization on the PTA data.

The relevant Fourier correlations are given by: 
\begin{align}
& \langle \alpha_{aj} \alpha_{bk} \rangle \nonumber \\
&= \left(2\over T\right)^2 \iint_0^T dt \, dt' 
\langle r(t,\hat{e}_a)r(t',\hat{e}_b) \rangle \sin(\omega_j t) \sin(\omega_k t') \nonumber \\
&=  \intzeroinf \frac{2\,df}{(2\pi f)^2}\, {\cal C}_{j,k}^{\alpha} (f) \sum_{\ell m}  I_{\ell m}(f) \gamma_{\ell m}^{I}(fD_a, fD_b,\hat{e}_a\cdot \hat{e}_b) \,,
\end{align}

\begin{align}
& \langle \beta_{aj} \beta_{bk} \rangle\nonumber \\
& = \left(2\over T\right)^2 \iint_0^T dt \, dt'
\langle r(t,\hat{e}_a)r(t',\hat{e}_b) \rangle \cos(\omega_j t) \cos(\omega_k t') \nonumber \\
&=  \intzeroinf \frac{2\,df}{(2\pi f)^2}\, {\cal C}_{j,k}^{\beta} (f) \sum_{\ell m}  I_{\ell m}(f) \gamma_{\ell m}^{I}(fD_a, fD_b,\hat{e}_a\cdot \hat{e}_b)\,,
\end{align}

\begin{align}
& \langle \alpha_{aj} \beta_{bk} \rangle \nonumber\\
&= \left(2\over T\right)^2 \iint_0^T dt \, dt'
\langle r(t,\hat{e}_a)r(t',\hat{e}_b) \rangle \sin(\omega_j t) \cos(\omega_k t') \nonumber \\
&=  \intzeroinf \frac{2\,df}{(2\pi f)^2}\, {\cal C}_{j,k}^{\alpha\beta} (f) \sum_{\ell m} i V_{\ell m}(f) \gamma_{\ell m}^{V}(fD_a, fD_b,\hat{e}_a\cdot \hat{e}_b)\,,
\end{align}

\begin{align}
& \langle \beta_{aj} \alpha_{bk} \rangle \nonumber \\
&= \left(2\over T\right)^2 \iint_0^T dt \, dt'
\langle r(t,\hat{e}_a)r(t',\hat{e}_b) \rangle \cos(\omega_j t) \sin(\omega_k t') \nonumber \\
&=  \intzeroinf \frac{2\,df}{(2\pi f)^2}\, {\cal C}_{j,k}^{\beta\alpha}(f) \sum_{\ell m} i V_{\ell m}(f) \gamma_{\ell m}^{V}(fD_a, fD_b,\hat{e}_a\cdot \hat{e}_b)\,,
\end{align}
where ${\cal C}^{\alpha}_{j,k}(f)$ and ${\cal C}^{\beta}_{j,k}(f)$ are the transfer functions previously defined in (\ref{eq:alpha_filter}-\ref{eq:beta_filter}), while the new transfer functions corresponding to the circular polarization components, ${\cal C}^{\alpha\beta}_{j,k}(f)$ and ${\cal C}^{\beta\alpha}_{j,k}(f)$, are given by
\begin{align}
& {\cal C}_{j,k}^{\alpha\beta}(f) \nonumber\\
&=\left(2\over T\right)^2 \iint_0^T dt \, dt' \,{\rm Im}[C(f,t_a,t_b)] \sin(\omega_j t) \cos(\omega_k t')\,, \nonumber \\
& {\cal C}_{j,k}^{\beta\alpha} (f) \nonumber\\
&=\left(2\over T\right)^2 \iint_0^T dt \, dt' \,{\rm Im}[C(f,t_a,t_b)] \cos(\omega_j t) \sin(\omega_k t')\,,
\end{align}
with ${\rm Im}[C(f,t_a,t_b)]$ defined in \eqref{CStime}. Importantly, only the terms involving the relative time difference contribute to these integrals, reflecting the physical origin of the circular polarization signal.

Carrying out the time integrations, we arrive at explicit expressions for the circular polarization transfer functions\footnote{
As with the $\alpha$- and $\beta$- transfer functions, the circular polarization transfer functions can be written as
\begin{align*}
{\cal C}_{j,k}^{\alpha\beta} (f) &= - 
\dfrac{4 (-1)^{j+k} f_j f }{(f+f_j)(f+f_k)}{\rm sinc}\left( \pi T \left( f-f_j\right) \right){\rm sinc}\left( \pi T \left( f-f_k\right) \right) \,, \\
{\cal C}_{j,k}^{\beta\alpha} (f) &= \dfrac{4 (-1)^{j+k} f f_k }{(f+f_j)(f+f_k)}{\rm sinc}\left( \pi T \left( f-f_j\right) \right){\rm sinc}\left( \pi T \left( f-f_k\right) \right) \,.
\end{align*}
These are directly related to the imaginary components of the correlation in the exponential Fourier basis (Appendix \ref{sec:a_different_basis}).
}:
\begin{align}
\label{eq:cp_filter}
{\cal C}_{j,k}^{\alpha\beta} (f) &= - \dfrac{4 f_j f }{\pi^2 T^2} \dfrac{\sin^2(\pi f T)}{(f^2 - f_j^2)(f^2 - f_k^2)} \,, \nonumber \\
{\cal C}_{j,k}^{\beta\alpha} (f) &= \dfrac{4 f_k f }{\pi^2 T^2} \dfrac{\sin^2(\pi f T)}{(f^2 - f_j^2)(f^2 - f_k^2)} \,.
\end{align}
An unpolarized SGWB can now be viewed as the limit when the cross terms $\langle \alpha_{aj}\beta_{bk} \rangle$ and $\langle \beta_{aj}\alpha_{bk} \rangle$ vanish. Thus, the transfer functions (\ref{eq:alpha_filter}-\ref{eq:beta_filter}) are associated with the intensity (unpolarized) component of the signal. The new transfer functions \eqref{eq:cp_filter} capture the imprint of circular polarization.

\begin{figure}[h!]
    \centering
    \subfigure[ \ circular polarization transfer functions \eqref{eq:cp_filter} for $j=k$ ]{
        \includegraphics[width = 0.45 \textwidth]{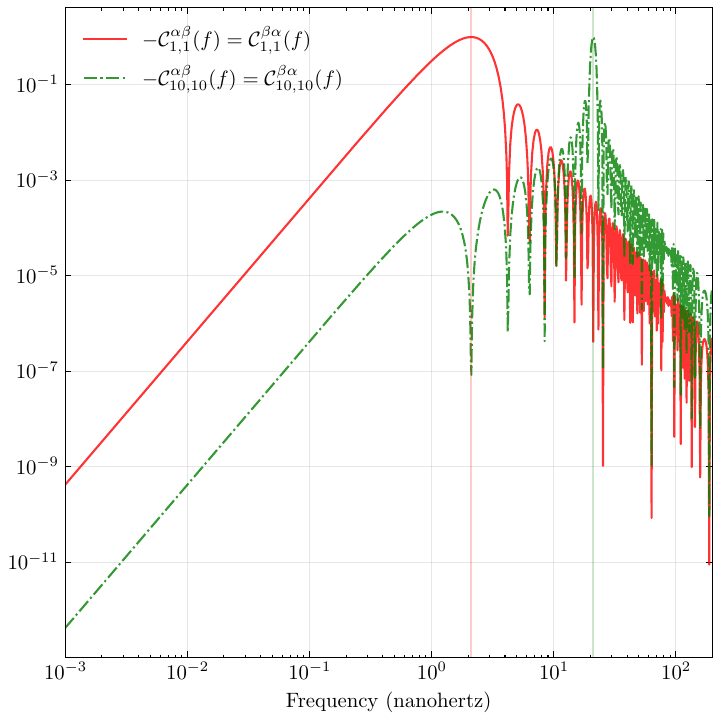}}
    \subfigure[ \ cross frequency transfer functions $j\neq k$ ]{
        \includegraphics[width = 0.45 \textwidth]{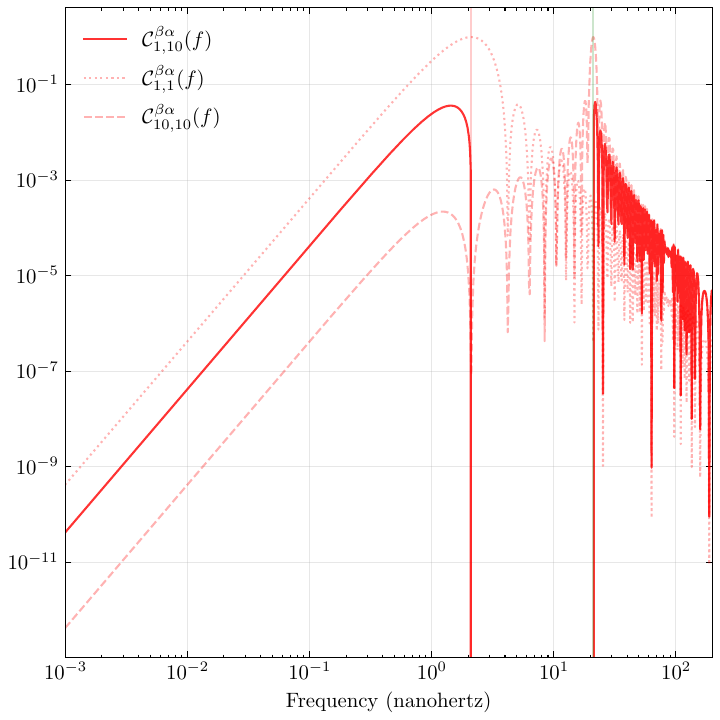}}
    \caption{Representative transfer functions for circular polarization \eqref{eq:cp_filter} in a PTA residuals' Fourier components with $T=15$ years, $f_1 \sim 2.1$ nanohertz, and $f_k \sim k f_1$.}
    \label{fig:filters_cp}
\end{figure}

This formalism reveals how circular polarization introduces distinctive correlations between the sine and cosine Fourier coefficients of the timing residuals, encoded in the nonzero $\langle \alpha_{aj} \beta_{bk} \rangle$ and $\langle \beta_{aj} \alpha_{bk} \rangle$ terms. The structure and frequency dependence of these transfer functions provide a clear pathway for identifying and characterizing anisotropic circular polarization signatures in PTA data, complementing the standard intensity analysis and opening new avenues for probing the polarization content of the SGWB \cite{Mingarelli:2013dsa, Kato:2015bye, Hotinli:2019tpc, Belgacem:2020nda, Sato-Polito:2021efu, Chu:2021krj, Tasinato:2023zcg, Depta:2024ykq, Cruz:2024svc, Cruz:2024esk}.

Figure \ref{fig:filters_cp} illustrates the transfer functions associated with the circular polarization component of the SGWB, as given in Eq.~\eqref{eq:cp_filter}. These transfer functions play an analogous role to those for the intensity component, encoding how a broadband circular polarization signal is distributed across the finite time Fourier bins of the PTA residuals. As with the intensity case, the diagonal elements ($j=k$) exhibit a pronounced mainlobe centered at $f=f_k$, accompanied by sidelobes reminiscent of a far-field radiation pattern. This structure highlights that the dominant contribution to the cross-correlation between sine and cosine Fourier components, $\langle \alpha_{ak} \beta_{bk} \rangle = -\langle \beta_{ak} \alpha_{bk} \rangle$, comes from the adjacent neighborhood (mainlobe) of the frequency $f_k$. The off-diagonal (cross-frequency, $j\neq k$) transfer functions are suppressed, indicating that correlations between different frequency bins due to circular polarization are subdominant, mirroring the behavior found for the intensity transfer functions. This reinforces the general picture developed in the previous sections: the main features of the SGWB signal, whether unpolarized or circularly polarized, are captured by the diagonal transfer functions, while cross-frequency correlations remain small (but nonnegligible, as shown in Ref. \cite{Crisostomi:2025vue}).

Another observation that we wish to spotlight is the the negative sign in the relation $C^{\alpha \beta}_{k,k}(f) = -C^{\beta\alpha}_{k,k}(f)$ between the circular polarization transfer functions. This invokes that $\langle \alpha_{ak} \beta_{bk} \rangle = -\langle \beta_{ak} \alpha_{bk} \rangle$, and so (for $a=b$) imposes that $\langle \alpha_{ak} \beta_{ak} \rangle = \langle \beta_{ak} \alpha_{ak} \rangle = 0$. The auto-correlation terms vanish; or rather, the cross-Fourier components, $\langle \alpha_{aj} \beta_{bk} \rangle$, are only correlated across different pulsars ($a \neq b$). The implies that the signature of circular polarization is distinctly in the cross-correlated power across pulsars or pulsar pairs.

\section{Comparison with simulation}
\label{sec:comparison_with_simulation_cp}

As we did to understand the properties of the unpolarized SGWB signal, we motivate our results for a circularly polarized SGWB by comparing with simulations.

For the simulation, we proceed by extending the framework given in Section \ref{sec:comparison_with_simulation} (or Appendix \ref{sec:simulation}) to include circular polarization components. Specifically, we generalize the covariance matrix $\bf \Sigma$ to
\begin{equation}
\label{eq:covariance_cp}
\Sigma_{ab}^k = \left( \begin{array}{cc}
    \langle \alpha_{ak} \alpha_{bk} \rangle & \langle \alpha_{ak} \beta_{bk} \rangle \\
    \langle \beta_{ak} \alpha_{bk} \rangle & \langle \beta_{ak} \beta_{bk} \rangle
\end{array} \right) \,.
\end{equation}
The Gaussian unpolarized SGWB signal can be recovered by setting $\langle \alpha_{ak} \beta_{bk} \rangle = \langle \beta_{ak} \alpha_{bk} \rangle = 0$; decoupling the sine- and cosine-Fourier coefficients of pulsar pairs. In addition, we consider a standard power-law \eqref{eq:power_spectral_density_power_law} for the intensity and circular polarization power spectra ($n_I = n_V = 2 - \gamma_{\rm gw}$). We define the constant $\Pi=|V_{00}(f)/I_{00}(f)| \leq 1$ as an indicator of the degree of circular polarization of the signal. $\Pi=1$ corresponds to a hundred percent circularly polarized signal. Then, assuming a constant $\Pi$, we write down the components of the covariance matrix as
\begin{eqnarray}
    \langle \alpha_{ak} \alpha_{bk} \rangle
    &=& \Gamma^{I}\left( \hat{e}_a \cdot \hat{e}_b \right) \int_{f_{\min}}^{f_{\max}} df\, S(f)\, {\cal C}^{\alpha}_{k,k}(f) \,, \\
    \langle \beta_{ak} \beta_{bk} \rangle
    &=& \Gamma^{I}\left( \hat{e}_a \cdot \hat{e}_b \right) \int_{f_{\min}}^{f_{\max}} df\, S(f)\, {\cal C}^{\beta}_{k,k}(f) \,, \\
    \langle \alpha_{ak} \beta_{bk} \rangle
    &=& \Gamma^{V}\left( \hat{e}_a \cdot \hat{e}_b \right) \Pi \int_{f_{\min}}^{f_{\max}} df S(f) {\cal C}^{\alpha\beta}_{k,k}(f) \,, \\
    \langle \beta_{ak} \alpha_{bk} \rangle
    &=& \Gamma^{V}\left( \hat{e}_a \cdot \hat{e}_b \right) \Pi \int_{f_{\min}}^{f_{\max}} df S(f) {\cal C}^{\beta\alpha}_{k,k}(f) \,,
\label{eq:covariance_cp_components}
\end{eqnarray}
where the spatial correlation functions are
\bw
\begin{eqnarray}
\label{eq:GammaI}
\Gamma^{I}(\hat{e}_a \cdot \hat{e}_b) &=& \left( 1 + \delta_{ab} \right) \left[{\bf\Gamma}\left(\hat{e}_a \cdot \hat{e}_b\right) + \beta \left(1-n_I\right) \left( {\bf \Gamma}_{10}^I( \hat{e}_a \cdot \hat{e}_b ) \sqrt{\frac{4\pi}{3}} \cos\theta_v
- {\bf \Gamma}_{11}^I( \hat{e}_a \cdot \hat{e}_b ) \sqrt{\frac{8\pi}{3}}\, \sin\theta_v \cos\phi_v \right) \right] \,, \\
\label{eq:GammaV}
\Gamma^V(\hat{e}_a \cdot \hat{e}_b) &=& {\bf \Gamma}_{11}^V(\hat{e}_a \cdot \hat{e}_b) \beta \left(1-n_V\right) \sqrt{\frac{8\pi}{3}} \sin\theta_v \sin\phi_v \,.
\end{eqnarray}
\ew
The ORFs ${\bf\Gamma}_{lm}(\hat{e}_a \cdot \hat{e}_b)$'s are shown in the right panel of Figure~\ref{fig:temporal_correlation}; where the leading term is the HD curve \eqref{eq:hd_curve}. {The bold faced ORFs ${\bf \Gamma}^X_{lm}$ (as shown in Figure 3) are normalized as ${\mathbf \Gamma}=1/2$ in the limit $\hat{e}_a \cdot \hat{e}_b = 1^-$ and $D_a = D_b$.} Their analytical expressions can be found in Ref.~\cite{Kato:2015bye}. Note that the correction terms are controlled by two quantities; the dipole speed ratio $\beta = v/c \sim {\cal O}\left(10^{-3}\right)$ and the degree of circular polarization $\Pi \le 1$. Above all, in this formulation, we emphasize that the terms of the covariance must be built in the computational frame where the ORFs take on the particular shapes drawn in Refs.~\cite{Kato:2015bye, Liu:2022skj, Bernardo:2022rif, Bernardo:2023jhs} (see Figure \ref{fig:temporal_correlation}).

\begin{figure*}[t]
\centering
\subfigure[]{\includegraphics[width=0.9\textwidth]{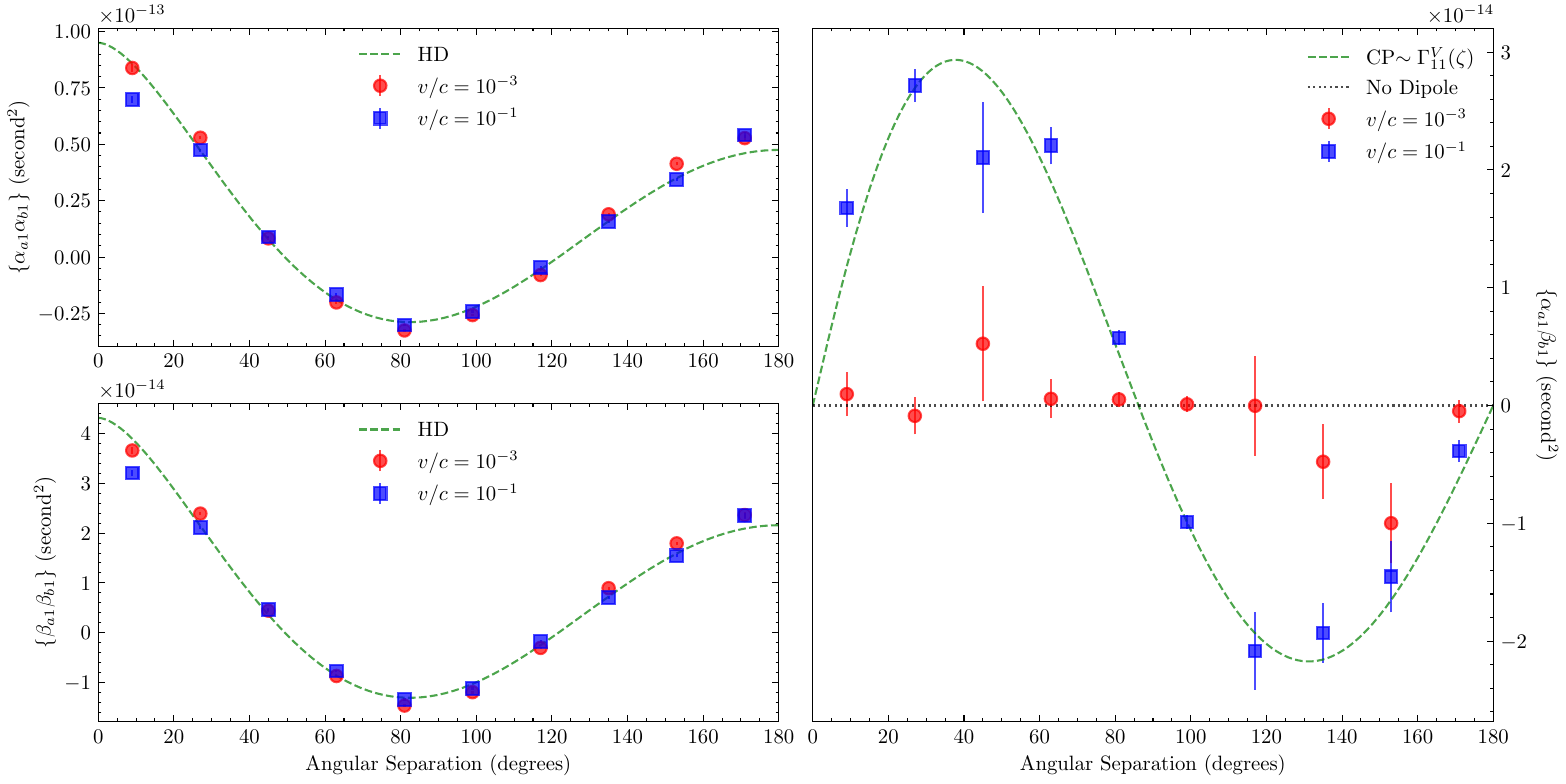}}
\subfigure[]{\includegraphics[width=0.43\textwidth]{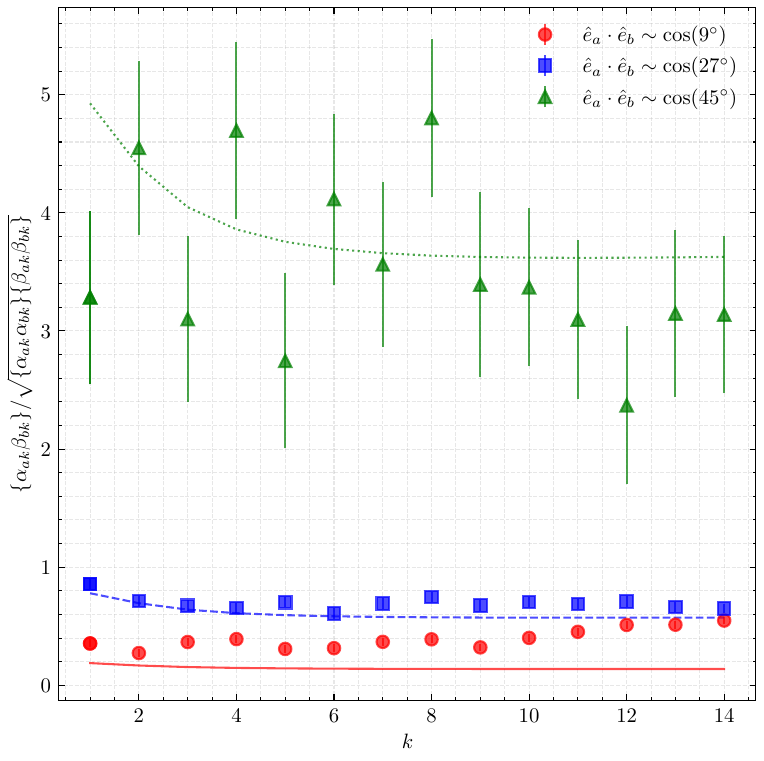}}
\subfigure[]{\includegraphics[width=0.46\textwidth]{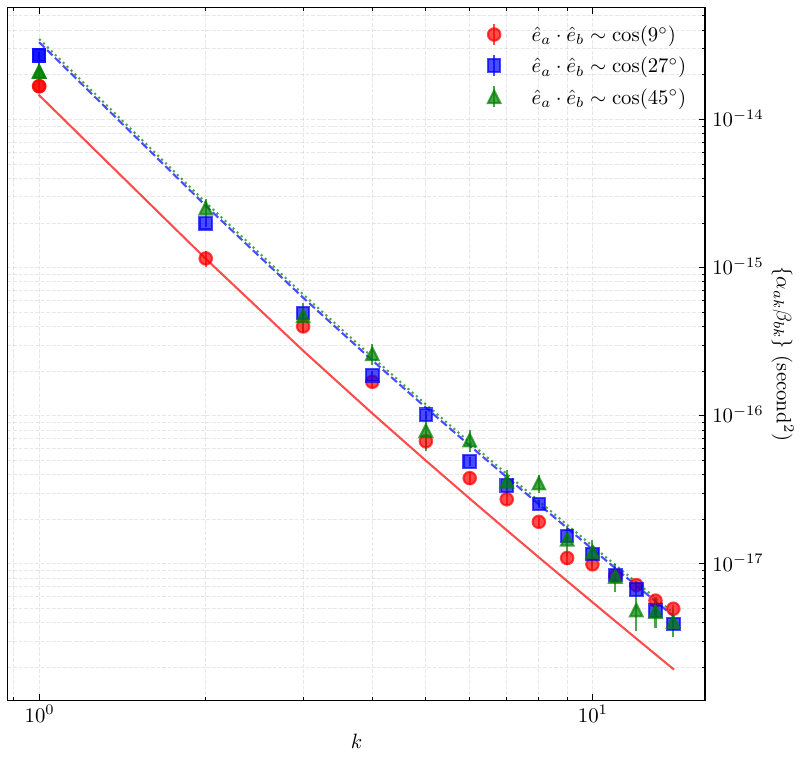}}
\caption{[Top panel] Two-point correlation functions and [bottom panel] the bin power of the Fourier coefficients of pulsar timing residuals for a simulated 15-year PTA data {set} with a circularly polarized SGWB signal. Each one of $10^4$ realizations has 100 pulsars at fixed positions. The error bars shown are the error on the mean. The SGWB characteristic strain and spectral index are $A_{\rm gw}=10^{-15}$ and $\gamma_{\rm gw}=13/3$, respectively. The kinematic dipole (along the $y$ direction) has a speed $\beta=v/c=10^{-3}, 10^{-1}$, and the degree of circular polarization is $\Pi=1$. In the top panel, the green dashed lines represent the Gaussian theoretical expectation values. In the bottom panel, the colored curves (red-solid, blue-dashed, and green-dotted) represent the Gaussian theoretical expectation values of the frequency-bin power of the cross-Fourier statistic $\{ \alpha_{ak} \beta_{bk} \}$ at $9^\circ, 27^\circ, 45^\circ$ angular bins, or $\hat{e}_a \cdot \hat{e}_b \sim \cos(9^\circ), \cos(27^\circ), \cos(45^\circ)$.}
\label{fig:correlation_cp}    
\end{figure*}

Fixing $A_{\rm gw}=10^{-15}$, $\gamma_{\rm gw}=13/3$ (SMBHB), $\beta=v/c=10^{-3}, 10^{-1} $, and a 100~\% circular polarization ($\Pi=1$), we obtain Figure \ref{fig:correlation_cp} using simulated pulsar timing residuals with 100 pulsars at fixed sky positions. The dipole is arbitrarily pointed to the $y$ direction. {The same results are shown in Appendix \ref{sec:stderrs_dipole} or Figure \ref{fig:correlation_cp_stderrs}, but with standard errors to show the variations expected across realizations in the computational frame; as one representative of frame dependent methods.}

The procedure for computing the sample averages is similar to the isotropic case described in Section~\ref{sec:comparison_with_simulation}: we compute the Fourier coefficients and then evaluate the sample statistics as in Eq.~\eqref{eq:binned_data}. However, a key difference for the circular polarization case is the need to work in the computational frame to recover the anisotropic signal.
Specifically, for each realization of the PTA data, we arbitrarily select one pulsar as $a$ and treat the remaining pulsars as $b$'s, thereby forming $n_{\rm psrs}-1$ pairs per realization. This is in contrast to the isotropic case, where all unique pulsar pairs are used, yielding $n_{\rm psrs}(n_{\rm psrs}-1)/2$ pairs and thus greater statistical power.

{The requirement to analyze in the computational frame is limited to the context of this work since the ORFs considered (shown in Figure \ref{fig:temporal_correlation}) have been derived in the computational frame \cite{Kato:2015bye, Mingarelli:2013dsa}. In practice, a frequentist targeted search can be performed to look for non-isotropic signals, by constructing a matched filter response or a detection statistic that is the dot product between the data and an anisotropic distribution of GW power with known polarization content in a fixed reference frame \cite{Ali-Haimoud:2020iyz, NANOGrav:2023tcn}. Other (possibly related) alternatives are to work in celestial coordinates and a bipolar spherical harmonics basis \cite{Hotinli:2019tpc, Liu:2022skj}. Either way, the computational frame is also representative of frame dependent methods, and frame dependent methods do not perform pulsar averaging because this will wash out the sensitivity to anisotropic signals.}

For each $(a,b)$ pair, we rotate pulsar $a$ to the $z$-axis and pulsar $b$ into the $xz$-plane ($\phi=0$), following the standard computational frame convention. The corresponding rotation matrices are then used to determine the dipole direction $(\theta_v, \phi_v)$ in this frame, which is required to evaluate the cross-Fourier spatial correlation (see Eq.~\eqref{eq:GammaV}). To account for the dipole orientation, we compute the normalized statistic $\alpha_{ak} \beta_{bk}/\left( \sin\theta_v \sin\phi_v \right)$ for all $n_{\rm psrs}-1$ pairs in each realization.

An important consequence of this approach is that, because we do not average over all unique pulsar pairs, the ensemble variance of the recovered statistic converges to the larger pulsar variance rather than the cosmic variance~\cite{Roebber:2016jzl, Allen:2022dzg, Allen:2022ksj, Bernardo:2022xzl}. Nevertheless, this method enables us to recover the circularly polarized SGWB signal when the dipole speed is artificially increased, as demonstrated in Figure~\ref{fig:correlation_cp}.

Our simulations demonstrate that the circularly polarized SGWB signal {supported} by a kinematic dipole can be recovered when the dipole speed is artificially increased by one to two orders of magnitude above its physical value ($v/c \sim 10^{-3}$). For realistic dipole speeds, however, the circular polarization signal is suppressed by a factor of $v/c$ relative to the leading isotropic component, rendering its detection extremely challenging in practice. As expected, the HD correlation is robustly recovered in the $\langle \alpha_{ak} \alpha_{bk} \rangle$ and $\langle \beta_{ak} \beta_{bk} \rangle$ Fourier bins (see Figure~\ref{fig:correlation_cp}, top left panel).

In principle, there are also $v$-dependent corrections to the HD correlation (see Eqs.~\eqref{eq:Ilmgammalm} and \eqref{eq:GammaI}) that could, in theory, be used to probe the kinematic dipole. However, these corrections are suppressed by a factor of $v/c \sim 10^{-3}$ relative to the HD curve and are not expected to be detectable due to the dominant cosmic variance~\cite{Roebber:2016jzl, Allen:2022dzg, Allen:2022ksj, Bernardo:2022xzl}. 

When discussing the signal of circular polarization, the focus is on the cross-Fourier components $\langle \alpha_{ak} \beta_{bk} \rangle$ and $\langle \beta_{ak} \alpha_{bk} \rangle$, which become nonzero in the presence of a circularly polarized SGWB. The top right panel of Figure~\ref{fig:correlation_cp} shows that these cross-Fourier components are indeed nonzero and match the Gaussian theoretical expectation, being proportional to ${\bf \Gamma}_{11}^V(\hat{e}_a \cdot \hat{e}_b)$ (see Figure~\ref{fig:temporal_correlation}). The bottom panels of Figure~\ref{fig:correlation_cp} display the frequency-bin power of the cross-Fourier statistic $\{ \alpha_{ak} \beta_{bk} \}$ at selected angular bins, corresponding to $\hat{e}_a \cdot \hat{e}_b \sim \cos(9^\circ), \cos(27^\circ), \cos(45^\circ)$. The bin power due to circular polarization, $\langle \alpha_{ak} \beta_{bk} \rangle$ or $\langle \beta_{ak} \alpha_{bk} \rangle$, falls off with frequency faster than $\langle \alpha_{ak} \alpha_{bk} \rangle$ but slower than $\langle \beta_{ak} \beta_{bk} \rangle$. This behavior reflects the frequency dependence of the circular polarization transfer functions~\eqref{eq:cp_filter}, which lie between the $\alpha$- and $\beta$-transfer functions in their fall off.

Recall that, for a Gaussian unpolarized SGWB, the Fourier coefficients $\alpha_{ak}$ and $\beta_{bk}$ are uncorrelated in a pulsar and in pulsar pairs. In contrast, the presence of a circularly polarized SGWB---such as that {supported} by a kinematic dipole---generates a distinctive, nonzero cross-correlation between the sine and cosine Fourier components of the timing residuals across pulsar pairs. To our knowledge, this cross-Fourier correlation is a unique signature of circular polarization in the SGWB and is not expected to arise from other known astrophysical sources in PTA data.

Our simulations of a Gaussian circularly polarized SGWB signal yield several important insights. First, the reliance on the computational frame restricts us from performing a full-sky average, which significantly limits the statistical power of the analysis. For realistic dipole amplitudes, the circular polarization signal is so weak that it is easily buried in the statistical fluctuations (pulsar variance \cite{Allen:2022dzg, Bernardo:2022xzl}) of any single realization. Even when the dipole speed is artificially amplified, summing over all pulsar pairs in a PTA can obscure the anisotropic signal due to the frame dependence. This highlights a key limitation: the computational frame, while convenient for theoretical calculations, is not optimal for extracting circular polarization signatures in practice. Developing methods to simulate and recover circular polarization without relying on the computational frame such as performing harmonic expansions in the celestial frame (see Eq.~\eqref{IVORF}) or using bipolar spherical harmonics \cite{Liu:2022skj} would be a valuable direction for future work.

While our analysis is statistically limited by the use of the computational frame, our results suggest that the circularly polarized signal associated with the kinematic dipole deserves further investigation. In particular, a robust detection or constraint of this signal could provide an independent probe of the cosmic dipole as PTA sensitivity improves in the coming decade.

{
Lastly, we return to the anisotropic signal corrections due to a kinematic dipole; that appear as extra terms in \eqref{eq:GammaI} following the HD curve. Similar questions have been addressed in \cite{Cruz:2024svc, Cruz:2024esk} where the anisotropic and circularly polarized components were studied separately. At leading order in the dipole's speed with respect to the speed of light, a kinematic dipole sources anisotropic and polarized components in the pulsar timing residual correlation through the SGWB. While our attention was focused on circular polarization due to its unique signature of coupling odd and even Fourier modes through the sample averages of $\alpha_{ak} \beta_{bk}$, we have included as well anisotropic corrections in the simulations. This explains why the sample averages of $\alpha_{ak} \alpha_{bk}$ and $\beta_{ak} \beta_{bk}$ show small deviations relative to the HD curve in Figure \ref{fig:correlation_cp}. However, these leading-order anisotropic terms are suppressed by a factor $v/c$ compared to the HD curve. This is the reason the deviation to the HD curve due to anisotropy remained small even when the dipole's speed was artificially increased to a tenth of the speed of light. That is, the corrections were at most ten percent compared to the dominant isotropic component. Nonetheless, it is possible to dissect the anisotropic corrections due to a kinematic dipole as illustrated in \cite{Cruz:2024svc} by utilizing specific geometric orientations relative to the dipole's direction. Our results in this direction of a purely anisotropic signal are teased out in Appendix \ref{sec:anisotropic_unpolarized}. The level of anisotropic correction ${\cal O}\left(v/c\right)$ and the attention that anisotropic searches receive have motivated us to focus on the circularly polarized correction.
}

\section{Discussion}
\label{sec:discussion}

We wrap up with a Q\&A-like section that goes through very briefly (what we subjectively think are) the most conceptually interesting aspects of the work.

\subsection*{Why are the variances of the sine- and cosine-Fourier coefficients different?}

The short answer is that the Fourier basis $\{ \sin(2\pi k t/T_{\rm obs}), \cos(2\pi k t/T_{\rm obs}) \}$ does not diagonalize the covariance of a signal with a characteristic time scale $T_{\rm s} \gg T_{\rm obs}$. Physically, this is because the observation modes are unavoidably misaligned with the signal's harmonics due to the enourmous time scale difference.

Consider a stationary process $y(t)$ with zero mean, $\langle y(t) \rangle = 0$, and covariance $\langle y(t) y(t') \rangle = {\cal C}(|t - t'|)$. The Wiener-Khinchin theorem relates this to a power spectral density ${\cal S}(f)$ via
\begin{equation}
\label{eq:wienerkhinchinth}
    {\cal C}(\tau) = \int_{0}^\infty df \, {\cal S}(f) \cos\left( 2\pi f \tau \right) \,.
\end{equation}
For a signal with characteristic time scale $T_{\rm s}$ and a set of fundamental frequencies $\{ f_j = j/T_{\rm s} ,\; j \in \mathbb{N} \}$, the time-domain covariance matrix $K_{ij} = C(|t_i - t_j|)$ can be expressed as a Riemann sum over these modes:\footnote{
Equivalently, one may expand ${\cal S}(f)$ as a sum over delta functions at the signal's harmonic frequencies: ${\cal S}(f) = \sum_k s_k \delta(f - f_k)$.
}
\begin{equation}
\begin{split}
    K_{ij} =  \sum_{k=1}^\infty \Delta f_k \, {\cal S}(f_k) \bigg[ & \cos(\omega_k t_i)\cos(\omega_k t_j) \\
    & \,\, + \sin(\omega_k t_i)\sin(\omega_k t_j) \bigg] \,,
\end{split}
\end{equation}
where $\Delta f_k = 1/T_{\rm s}$ and $\omega_k = 2\pi f_k = 2\pi k/T_{\rm s}$. Clearly, the signal covariance is naturally diagonalized in a Fourier basis $\{ \sin(2\pi k t / T_{\rm s}), \cos(2\pi k t/ T_{\rm s}) \}$, with the sine and cosine coefficients both having variances proportional to ${\cal S}(f_k)$.

However, in practice, observations are limited to a finite time window $T_{\rm obs} \ll T_{\rm s}$, and the observational frequency basis $\{ f_k = k / T_{\rm obs} \}$ no longer aligns with the signal's intrinsic modes. Consequently, the covariance matrix will not be diagonal in the observational basis $\{ \sin(2\pi k t/T_{\rm obs}), \cos(2\pi k t/T_{\rm obs}) \}$, and the sine and cosine Fourier coefficients will generally have unequal variances. This is especially relevant for PTA searches, where the gravitational-wave (GW) time scale is $\mathcal{O}(10^9)$ years, vastly longer than the typical observational baseline of $\mathcal{O}(10)$ years \cite{Allen:2024uqs}.

{We emphasize that the signal is not periodic. Above, the time scale $T_{\rm s}$ parametrizes the harmonic space frequencies of a signal, such that each adjacent harmonics are separated by $T_{\rm s}^{-1}$, i.e., $\Delta f_{k}= f_{k}-f_{k-1} = T_{\rm s}^{-1}$. In general, a signal's harmonic frequencies don't have to lie on an equally spaced lattice, the arguments can be generalized, and our simplified usage only highlights physical concepts at play.}

A key implication is that approximating the covariance as diagonal in the observational Fourier basis \cite{Taylor:2021yjx} neglects the frequency correlations, leading to systematic biases in parameter estimation \cite{Bernardo:2024tde, Crisostomi:2025vue}. The GW signal characteristic time scale can be estimated using Peters' time scale \cite{Peters:1964qza, Zwick:2019yjl}; for an equal mass binary (each with mass $m$ and a separation $r$) this gives $T_{\rm gw} \sim 5 c^5 r^4 / (256 G^3 m^3) \sim 5.8 \times 10^8 \, {\rm yr} \left( r/{1\,{\rm pc}} \right)^4 \left( 10^{10} M_\odot / m \right)^3$.

\subsection*{Are the so-called transfer functions discussed throughout special for PTA?}

No. The transfer functions defined in Eqs.~(\ref{eq:alpha_filter}--\ref{eq:beta_filter}) are not exclusive to PTAs but arise generically when a stationary, temporally correlated process is observed over a finite time window. This situation arises in time series analyses involving correlated processes such as intrinsic red noise in pulsars.

These transfer functions encapsulate how the signal is projected onto a Fourier basis with a finite time window. As the duration of observation increases and approaches the signal’s intrinsic time scale, the transfer functions asymptotically approach delta functions centered at the signal’s natural frequencies, thereby recovering the ideal decomposition.

This can also be realized by writing Eqs. (\ref{eq:alpha_filter}-\ref{eq:beta_filter}) in terms of the finite time approximation to the delta function, $\delta_T(x)=T \, {\rm sinc}(\pi xT)$. As $T\rightarrow\infty$, the $\alpha$-transfer function approaches the $\beta$-transfer function. Then, $\langle \alpha_{aj} \alpha_{bk} \rangle / \langle \beta_{aj} \beta_{bk} \rangle \rightarrow 1$ and hence the Fourier mode decomposition of the covariance function $\langle r(t)r(t') \rangle$ in Eq. \eqref{eq:timing_residual_correlation_fourier_decomposition} reduces to a sum of $\cos(w_k(t-t'))$.

\subsection*{What are the prospects for detecting circular polarization in the SGWB?}

Our results indicate that detecting circular polarization in the SGWB will be challenging. Let us begin with the encouraging aspects: circular polarization induces a unique cross-correlation between the sine and cosine Fourier coefficients of different pulsars (see Figure~\ref{fig:correlation_cp}). This coupling produces a distinctive spatial correlation pattern and associated power spectrum that, in principle, could be extracted from data. Furthermore, such a signal might contain information about the anisotropic distribution of sources \cite{Sato-Polito:2021efu}.

On the downside, the signal {supported} by a kinematic dipole with velocity $v/c \sim 10^{-3}$ is expected to lie well below the cosmic variance threshold set by the HD correlation \cite{Roebber:2016jzl, Allen:2022dzg, Allen:2022ksj}, and subsequently also below instrumental noise. 

Despite these challenges, the circularly polarized component of the SGWB can still be constrained phenomenologically by measuring the cross-correlated power between the sine and cosine Fourier coefficients across pulsar pairs. This motivates a likelihood analysis that does not depend on the computational frame, enabling full sky averaging with PTA data to resolve circular polarization. Developing such methods remains an open area for future research. We also refer the reader to the important works \cite{Mingarelli:2013dsa, Kato:2015bye, Hotinli:2019tpc, Belgacem:2020nda, Sato-Polito:2021efu, Chu:2021krj, Depta:2024ykq, Cruz:2024svc, Cruz:2024esk} for leading the way to this direction.

{Anisotropy, polarization and the kinematic dipole are independent concepts, and in this work we used the kinematic dipole (motivated cosmologically through CMB and large scale structure observations) to tease out leading order corrections following an isotropic PTA signal. We refer the reader to the works cited in the preceding paragraphs for applications beside a kinematic dipole.}

\subsection*{Can a PTA be used to probe the cosmic dipole?}

In theory, yes. If the SGWB originates from cosmological sources, a leading order dipolar anisotropy arising from our motion relative to the cosmic rest frame should imprint a characteristic correction of the order of $v/c \sim 10^{-3}$. This kinematic dipole is of considerable cosmological interest \cite{Secrest:2020has, Dalang:2021ruy, Secrest:2022uvx, Dam:2022wwh, Oayda:2024hnu, vonHausegger:2024jan, Antony:2025tzk, Secrest:2025wyu, CosmoVerseNetwork:2025alb, Yoo:2025qdq,  Wagenveld:2025ewl}, especially in light of ongoing efforts to test the consistency of the cosmological principle using independent large scale observables such as the CMB and quasar catalogs {(see e.g., \cite{Secrest:2025wyu} and section 2.3.6 of \cite{CosmoVerseNetwork:2025alb})}.

PTAs offer a novel opportunity to detect this dipolar signature via its characteristic spectral and spatial correlations in the timing residuals. However, this is contingent upon the SGWB being predominantly cosmological. If instead the SGWB is dominated by astrophysical sources, then the intrinsic anisotropies are likely to be much larger, potentially obscuring the kinematic dipole \cite{Tasinato:2023zcg, Cruz:2024svc, Cruz:2024esk}.

Detecting the cosmic dipole thus remains an ambitious goal for PTA science.

\section{Conclusions}
\label{sec:conclusions}

In this work, we have revisited the formulation of the stochastic gravitational wave background (SGWB) signal in pulsar timing arrays (PTAs), providing a unified frequency- and Fourier-domain framework that connects theoretical predictions with simulation results. By comparing Gaussian theoretical expectation values with both standard point-source simulations and our own covariance-matrix-based approach, we have demonstrated the robustness of this framework for both isotropic unpolarized (Sections \ref{sec:the_signal}-\ref{sec:comparison_with_simulation}) and circularly polarized (Sections \ref{sec:circular_polarization_in_pta}-\ref{sec:comparison_with_simulation_cp}) SGWB signals.

A central result of our analysis is the identification and explicit construction of transfer functions in the Fourier basis that factor in frequency correlations that occur due to the observing a temporally correlated signal in a finite time window. We have shown that the correlations of the Fourier coefficients of pulsar timing residuals can be accurately described by convolving these transfer functions with the relevant power spectra or power spectral densities. This approach, detailed in Eqs.~(\ref{eq:xy_correlation}--\ref{eq:F_functional}) and Eqs.~(\ref{eq:alpha_filter}--\ref{eq:beta0_filter}), provides a simple yet powerful prescription for modeling the SGWB signal in the frequency- and Fourier-domain, complementary to time-domain methods that have been mainly utilized in PTA analysis. Our results further indicate that static and cross-frequency effects, while present, are subdominant. Cross-frequency effects were nonetheless given the spotlight in Ref. \cite{Crisostomi:2025vue}, showing their importance in mitigating biased parameter estimates.

{
We emphasize that the diagonals considered in this work fully account for inter-frequency correlations through the accommodation of transfer functions; and are distinct to the diagonals of the diagonal covariance approximation often adopted for PTA analysis. The diagonal covariance approximation was a milestone in PTA analysis for a practical reason: computing the off diagonal terms take significantly more time compared to the diagonals. It is the same practicality that has driven us to focus on the diagonals, and we could take the agreement between the point source-based {\texttt{TEMPO2}} simulations and our results as a validation. However, this is not a guarantee that the off-diagonals can be left out of practical applications and it is important to pay attention to their contribution to avoid systematic biases. This is a work that we have in mind in the future.
}

Importantly, the transfer functions do not rely on any new assumptions about the SGWB, but instead emerge directly from the temporal structure of any correlated signal; on equal footing with the Hellings-Downs curve and the power spectrum in characterizing the SGWB in PTA data.

Looking ahead, several promising directions emerge from this work. First, incorporating transfer functions into post-fit analyses would help mitigate systematic biases in SGWB parameter estimation in the frequency- and Fourier-domain. Second, {exploring} simulation strategies for circularly polarized signals that do not rely on the computational frame {and other frame dependent strategies} would enable more robust statistical analyses and facilitate full sky averaging. {Frame dependent frequentist targeted searches are often used in practice which utilizes estimators of the spherical harmonic components of the GW power with known spatial distribution and polarization.} Third, the transfer function formalism motivates a renewed examination of subnanohertz GW contributions in PTA data, which may reveal new physical insights as PTA sensitivity improves \cite{DeRocco:2022irl, DeRocco:2023qae}. Finally, extending this framework to include intrinsic pulsar red noise, white measurement noise, and higher-order statistics~\cite{Bernardo:2024uiq, Xue:2024qtx} will be important for future PTA analyses and for testing the Gaussianity of the SGWB signal.

Overall, our results highlight the importance of temporal correlations and transfer functions in the accurate modeling and interpretation of SGWB signals in PTAs, and provide a foundation for future developments in both theoretical and simulation-based PTA analyses.

\acknowledgements
The authors thank Stephen Appleby for relevant discussion on correlated variables, Adrian Villanueva on GW time scales, Guillem Dom\'{e}nech for pointing out typos in a preliminary draft, and Xiao Xue for important comments on detecting circular polarization in a PTA. The authors are grateful to Bruce Allen, Rutger van Haasteren, and Boris Goncharov for various important discussions that have influenced the development of this work. RCB was supported by an appointment to the JRG Program at the APCTP through the Science and Technology Promotion Fund and Lottery Fund of the Korean Government, and was also supported by the Korean Local Governments in Gyeongsangbuk-do Province and Pohang City. This work was supported in part by the National Science and Technology Council of Taiwan, Republic of China, under Grant No. NSTC 113-2112-M-001-033. RCB is grateful to the Institute of Physics, Academia Sinica for the hospitality that enabled the completion of this work. The authors are grateful to the participants of the external program APCTP-GW2025 [or APCTP2025-E05] held at Academia Sinica, Taipei, Taiwan for fruitful discussions.

\appendix

\section{A choice of basis functions}
\label{sec:a_different_basis}

The pulsar timing residual, $r(t, \hat{e})$, or any time series, in a domain $t \in (t_i, t_i + T)$ for some initial time $t_i$ can be interpreted as Eq. \eqref{eq:residual_fourier_series}. This can be done because the sines and cosines over the domain $t \in (0, T)$ form a complete basis;
\begin{eqnarray}
\int_0^T \dfrac{2dt}{T} \sin(\omega_j t)\sin(\omega_k t) &=& \delta_{jk} \,,\\
\int_0^T \dfrac{2dt}{T} \cos(\omega_j t)\cos(\omega_k t) &=& \delta_{jk} \,,\\
\int_0^T \dfrac{2dt}{T} \sin(\omega_j t)\cos(\omega_k t) &=& 0 \,,
\end{eqnarray}
where $\omega_j = 2\pi j/T${, and $j,k$ are positive integers ($j,k \in {\mathbb{N}}$)}. This choice of basis functions is however arbitrary, and another common way a time data can be interpreted is by
\begin{equation}
\label{eq:residual_fourier_series_exp}
    r(t, \hat{e})=\sum_{k=-\infty}^\infty \psi_k(\hat{e}) e^{i\omega_k t} \,,
\end{equation}
utilizing an exponential Fourier basis functions, e.g., in PTA science this has been used in Refs. \cite{Allen:2024uqs, Crisostomi:2025vue}. In this case, the orthonormality of the basis functions can be expressed as
\begin{equation}
\int_{-T/2}^{T/2} \dfrac{dt}{T} e^{i (\omega_j - \omega_k) t} = \delta_{jk} \,.
\end{equation}
Then, the Fourier coefficients $\psi_k(\hat{e})$ can be obtained as
\begin{equation}
    \psi_k(\hat{e})=\int_{-T/2}^{T/2} \dfrac{dt}{T} e^{-i \omega_k t} r(t, \hat{e}) \,.
\end{equation}
Take note that the domain of the data has been shifted to $t\in(-T/2, T/2)$. This is always possible in practice. The two descriptions are equivalent, as will be shown.

In the exponential Fourier basis, the correlation between the Fourier coefficients of the pulsar timing residuals' (following analogous steps done in the main text, Section \ref{sec:the_signal}) can be shown to be
\begin{equation}
\begin{split}
    & \langle \psi_j(\hat{e}_a) \psi_k^*(\hat{e}_b) \rangle \\
    & \,\,\,\, = \int_{f_{\rm min}}^{f_{\rm max}} \frac{2\,df}{(2\pi f)^2}\, {\cal C}_{j,k}(f) I(f) \gamma(fD_a, fD_b,\hat{e}_a,\hat{e}_b) \,,
\end{split}
\end{equation}
where the corresponding transfer function is given by \cite{Allen:2024uqs, Crisostomi:2025vue}
\begin{equation}
\label{eq:sincsinc}
    C_{j,k}(f) = {\rm sinc}\left(\pi T(f-f_j)\right) {\rm sinc}\left(\pi T(f-f_k)\right) \,.
\end{equation}
An important observation is that in this basis there is only one transfer function, $C_{j,k}(f)$, due to the phase symmetry of the exponential sum, unlike with sines and cosines.

Naturally, the Fourier components of the timing residual correlation, or the covariance matrix ${\bf \Sigma}$ (Eqs. \eqref{eq:covariance_matrix_unpolarized} or \eqref{eq:covariance_cp_components}), can be expressed in terms of $\langle \psi_j(\hat{e}_a) \psi_k^*(\hat{e}_b) \rangle$. It can be shown that
\begin{eqnarray}
    \alpha_{0}(\hat{e})/2 &=& \psi_0(\hat{e}) \,, \nonumber \\
    \alpha_k(\hat{e}) &=& i \left( \psi_k(\hat{e}) - \psi_k^*(\hat{e}) \right) (-1)^k \,, \nonumber \\
    \beta_k(\hat{e}) &=& \left( \psi_k(\hat{e}) + \psi_k^*(\hat{e}) \right) (-1)^k \,.
\end{eqnarray}
The factor $(-1)^k$ is due to shifting the time coordinate $t \rightarrow t' = t - T/2$ to match the origins of Eqs. \eqref{eq:residual_fourier_series} and \eqref{eq:residual_fourier_series_exp}. Following the notation in the main text, we obtain
{
\begin{eqnarray}
    \dfrac{\langle \alpha_{aj} \alpha_{bk} \rangle}{2(-1)^{j+k}} &=&  {\rm Re}\left[ \langle \psi_{aj} \psi_{bk}^* \rangle \right] - {\rm Re}[ \langle \psi_{aj} \psi_{b(-k)}^* \rangle ] \,, \nonumber \\
    \dfrac{\langle \beta_{aj} \beta_{bk} \rangle}{2(-1)^{j+k}} &=& {\rm Re}\left[ \langle \psi_{aj} \psi_{bk}^* \rangle \right] + {\rm Re}[ \langle \psi_{aj} \psi_{b(-k)}^* \rangle ] \,, \nonumber \\
    \dfrac{\langle \alpha_{aj} \beta_{bk} \rangle}{2(-1)^{j+k}} &=-&  {\rm Im}\left[ \langle \psi_{aj} \psi_{bk}^* \rangle \right] - {\rm Im}[ \langle \psi_{aj} \psi_{b(-k)}^* \rangle ]  \,, \nonumber \\
    \dfrac{\langle \beta_{aj} \alpha_{bk} \rangle}{2(-1)^{j+k}} &=&  {\rm Im}\left[ \langle \psi_{aj} \psi_{bk}^* \rangle \right] - {\rm Im}[ \langle \psi_{aj} \psi_{b(-k)}^* \rangle ]  \,. \nonumber \\
\end{eqnarray}
}
Then, it is straightforward to confirm that the transfer functions of Eq. \eqref{eq:residual_fourier_series} (Sections \ref{subsec:filters} and \ref{subsec:filters_cp}) can be derived from the symmetric transfer function of Eq. \eqref{eq:residual_fourier_series_exp} [Eq. \eqref{eq:sincsinc}] as
\begin{widetext}
\begin{align}
{\cal C}_{j,k}^\alpha (f) &=& 2(-1)^{j+k} \left( \dfrac{C_{j,k}(f)+C_{-j,-k}(f)}{2} - \dfrac{C_{j,-k}(f)+C_{-j,k}(f)}{2} \right) &=& \dfrac{4 (-1)^{j+k} f_j f_k }{(f+f_j)(f+f_k)}C_{j,k}(f) \,, \nonumber \\
{\cal C}_{j,k}^\beta (f) &=& 2(-1)^{j+k} \left( \dfrac{C_{j,k}(f)+C_{-j,-k}(f)}{2} + \dfrac{C_{j,-k}(f)+C_{-j,k}(f)}{2} \right) &=& \dfrac{4 (-1)^{j+k} f^2 }{(f+f_j)(f+f_k)} C_{j,k}(f) \,, \nonumber \\
{\cal C}_{j,k}^{\alpha\beta} (f) &= -& 2(-1)^{j+k} \left( \dfrac{C_{j,k}(f)-C_{-j,-k}(f)}{2} + \dfrac{C_{j,-k}(f)-C_{-j,k}(f)}{2} \right) &=& - 
\dfrac{4 (-1)^{j+k} f_j f }{(f+f_j)(f+f_k)}C_{j,k}(f) \,, \nonumber \\
{\cal C}_{j,k}^{\beta\alpha} (f) &=& 2(-1)^{j+k} \left( \dfrac{C_{j,k}(f)-C_{-j,-k}(f)}{2} - \dfrac{C_{j,-k}(f)-C_{-j,k}(f)}{2} \right) &=& \dfrac{4 (-1)^{j+k} f f_k }{(f+f_j)(f+f_k)}C_{j,k}(f) \,.
\end{align}
\end{widetext}
The above relations establish the equivalence and give a dictionary to translate between Eq. \eqref{eq:residual_fourier_series} or Eq. \eqref{eq:residual_fourier_series_exp} when interpreting a signal in the Fourier domain.

As a consequence, the variance of the sine- and cosine-Fourier coefficients of the time correlated process is shown to be different, due to $C^\alpha_{j,k}(f) \neq C^\beta_{j,k}(f)$ (see \eqref{eq:alpha_beta_relation} for the exact relation).
To realize this, consider a very narrow signal in frequency space. In the extreme infinitesimally narrow limit, we have $I(f) \sim \delta (f - \overline{f})$. Then, the variances of the sine- and cosine-Fourier coefficients will be related to each other by ${\rm Var} \left[ \beta_k^2 \right] / {\rm Var} \left[ \alpha_k^2 \right] \sim {\overline f}^2 / f_k^2$.

We now specialize back to PTA science; where the intensity of the signal can be associated with $\langle \alpha_{aj} \alpha_{bk} \rangle$ and $\langle \beta_{aj} \beta_{bk} \rangle$, and the circular polarization to $\langle \alpha_{aj} \beta_{bk} \rangle$ and $\langle \beta_{aj} \alpha_{bk} \rangle$. The above expressions make it transparent that the intensity of a signal can be associated with the real part of $\langle \psi_{aj} \psi_{bk}^* \rangle$ and the circular polarization to the imaginary part of $\langle \psi_{aj} \psi_{bk}^* \rangle$. In other words, a circularly polarized SGWB signal in a PTA can be identified either by $\langle \alpha_{aj} \beta_{bk} \rangle \neq 0$ in the sine-cosine basis \eqref{eq:residual_fourier_series} or ${\rm Im}\left[ \langle \psi_{aj} \psi_{bk}^* \rangle \right] \neq 0$ in the exponential basis \eqref{eq:residual_fourier_series_exp}.

\bw
\section{Spin-Weighted Spherical Harmonics}
\label{sec:spinweight}
The explicit form of the spin-weighted spherical harmonics that we use is
\begin{align}
{}_{s}Y_{\ell m}(\theta,\phi) = 
    (-1)^{m}
    e^{im\phi}
    \sqrt{\frac{(2\ell+1)}{(4\pi)}\frac{(\ell+m)!(\ell-m)!}{(\ell+s)!(\ell-s)!}}
    \sin^{2\ell}\!\left(\frac{\theta}{2}\right)
    \sum_{r}
    \binom{\ell-s}{r} \binom{\ell+s}{r+s-m}
    (-1)^{\ell-r-s}
    \cot^{2r+s-m}\!\left(\frac{\theta}{2}\right) \,.
\end{align}
\ew
When $s=0$, it reduces to the ordinary spherical harmonics,
\be
Y_{\ell m}(\hat{n})=\sqrt{\frac{(2\ell+1)}{(4\pi)}\frac{(\ell-m)!}{(\ell+m)!}}P^m_\ell(\cos \theta) e^{i m \phi} \,.
\ee

Spin-weighted spherical harmonics satisfy the orthogonal relation,
\be
    \int_{S^2} d{\hat{n}}\; {}_{s}Y^*_{\ell m}(\hat{n}){}_{s}Y_{\ell' m'}(\hat{n})
    = \delta_{\ell \ell'} \delta_{m m'} \,,
\ee
and the completeness relation,
\begin{align}
    \sum_{\ell m} {}_{s}Y^*_{\ell m}(\hat{n}){}_{s}Y_{\ell m}(\hat{n}')
    &= \delta(\hat{n}-\hat{n}') \nonumber \\
    &= \delta(\phi-\phi')\delta(\cos\theta-\cos\theta')\,.
\end{align}
Its complex conjugate is
\be
{}_{s}Y^*_{\ell m}(\hat{n}) =  (-1)^{s+m} {}_{-s}Y_{\ell -m}(\hat{n}) \,,
\label{Yconjugate}
\ee
and its parity is given by
\be
{}_{s}Y_{\ell m}(-\hat{n}) \equiv {}_{s}Y_{\ell m}(\pi-\theta,\phi+\pi)=(-1)^{\ell} {}_{-s}Y_{\ell m}(\hat{n}) \,.
\label{Yparity}
\ee
Also, we have the spherical wave expansion:
\be
\label{eq:swexpansion}
e^{i\vec{k}\cdot\vec{r}} = 4\pi \suml \summ i^\ell j_\ell(kr) Y_{\ell m}^*(\hat{k}) Y_{\ell m}(\hat{r}) \,,
\ee
where $j_\ell(x)$ is the spherical Bessel function. 

We can calculate the integral of a product of three spin-weighted spherical harmonics using the formula:
\bw
\be
    \int_{S^2} d\hat{e} \;
    {}_{s_1} Y_{\ell_1 m_1}(\hat{e})\; {}_{s_2}\!Y_{\ell_2 m_2}(\hat{e})\; {}_{s_3}\!Y_{\ell_3 m_3}(\hat{e})=
    \sqrt{\frac{(2\ell_1+1)(2\ell_2+1)(2\ell_3+1)}{4\pi}}
    \begin{pmatrix}
          \ell_1 &&   \ell_2  &&  \ell_3 \\
        -s_1 && -s_2  &&  -s_3 
    \end{pmatrix}
    \begin{pmatrix}
        \ell_1  && \ell_2  &&  \ell_3 \\
       m_1 && m_2  &&  m_3 
    \end{pmatrix} \,,
\label{eq:threeJ}
\ee
\ew
which involves two Wigner-3j symbols representing the coupling coefficients between different spherical harmonics~\cite{doi:10.1142/0270}. The Wigner-3j symbol is zero unless it satisfies: $\ell_1$, $\ell_2$, and $\ell_3$ have to meet the triangular condition, i.e.~$\ell_1 + \ell_2 \ge \ell_3 \ge |\ell_1- \ell_2|$, while $m_1+m_2+m_3=0$; when $m_1=m_2=m_3=0$, $\ell_1+\ell_2+\ell_3$ is even.  The Wigner-3j symbols have the reflection property and the summation relation:
\bw
\begin{align}
 &
 \begin{pmatrix}
        \ell_1 && \ell_2  &&  \ell_3 \\
       s_1  && s_2  && s_3
  \end{pmatrix} 
  = (-1)^{\ell_1 + \ell_2 +  \ell_3}
  \begin{pmatrix}
        \ell_1 && \ell_2  &&  \ell_3 \\
       -s_1  && -s_2  &&  -s_3
  \end{pmatrix}
  = (-1)^{\ell_1 + \ell_2 +  \ell_3}
  \begin{pmatrix}
        \ell_1 && \ell_3  &&  \ell_2 \\
       s_1  && s_3  && s_2
  \end{pmatrix}\,, 
\label{reflection} \\
& \nonumber \\
&
(2\ell +1)\sum_{m_1 m_2}
\begin{pmatrix}
        \ell  && \ell_1  &&  \ell_2 \\
          m &&  -m_1  &&  m_2 
\end{pmatrix}
\begin{pmatrix}
        \ell'  && \ell_1  &&  \ell_2 \\
          m' &&  -m_1  &&  m_2 
 \end{pmatrix}
=  \delta_{\ell \ell'} \delta_{m m'}\,.
\label{sum3j}
\end{align}
\ew

\section{Pulsar timing residuals due to a Gaussian SGWB signal}
\label{sec:simulation}

The pulsar timing residual simulation of \texttt{PTAfast} is a fork of the open-source \texttt{fakepta}~\cite{Babak:2024yhu}. The original \texttt{fakepta} was designed to simulate post-fit pulsar timing residuals due to various astrophysical signals, including the SGWB. The pulsar, time-of-arrival generation, and noise modules of \texttt{fakepta} are carried over to \texttt{PTAfast}. This includes but not limited to the \texttt{Pulsar} class (consisting of a simulated pulsar's position, distance, and noise properties) and functions for distributing pulsars across the sky (isotropic or anisotropic), generating time-of-arrival data, and simulating intrinsic pulsar white and colored noises. 

\texttt{PTAfast} repurposes this as a tool for simulating pre-fit pulsar timing residuals with a precise and efficient SGWB signal modelling in a finite time window. Modifications of \texttt{PTAfast} are focused on the generation of pulsar timing residuals in a finite observation window due to a Gaussian SGWB signal.


We define the SGWB signal by its continuous support in the frequency domain, i.e., the frequency range of the SGWB signal, $f \in (f_{\rm min}, f_{\rm max})$, and its power spectral density. For simplicity, we assume a power-law power spectral density \eqref{eq:power_spectral_density_power_law}. Then, the amplitude of the SGWB signal is given by $A_{\rm gw}$, and the spectral index is given by $\gamma_{\rm gw}$. 

Once a set of pulsar objects are created, for each frequency bin, a timing data is generated as Eq.~\eqref{eq:residual_fourier_series} where the vector of Fourier coefficients, {\boldmath$\alpha$} and {\boldmath$\beta$}, are obtained as follows. First, we draw a vector {\boldmath$\xi$}$\sim {\cal N}\left( 0, 1 \right)$ with length $2 n_{\rm psrs}$ where $n_{\rm psrs}$ is the length of both {\boldmath$\alpha$} and {\boldmath$\beta$}. We spatially correlate the pulsars according to Eqs.~(\ref{eq:xy_correlation}-\ref{eq:F_functional}) and Eqs.~(\ref{eq:alpha_filter}-\ref{eq:beta0_filter}) by constructing the covariance matrix ${\bf\Sigma}$, generally,
\begin{equation}
    \Sigma^{{j}k}_{ab} \equiv \left( \begin{array}{cc}
    \langle \alpha_{a{j}} \alpha_{bk} \rangle & \langle \alpha_{a{j}} \beta_{bk} \rangle \\[0.2cm]
    \langle \beta_{a{j}} \alpha_{bk} \rangle & \langle \beta_{a{j}} \beta_{bk} \rangle
    \end{array} \right)  \,,
\end{equation}
for all pulsar pairs $(a,b)$ and frequency bins $j,k$. For a fixed frequency bin $k$, the covariance matrix ${\bf \Sigma}$ is a $2n_{\rm psrs}\times 2n_{\rm psrs}$ matrix. The Fourier components of the correlation such as $\langle \alpha_{a{j}} \alpha_{bk} \rangle$ and $\langle \beta_{a{j}} \beta_{bk} \rangle$ are computed by convolving the transfer functions (\ref{eq:alpha_filter}-\ref{eq:beta0_filter}) with the power spectral density across the support of the signal.
Finally, the input Fourier coefficients {\boldmath$\alpha$} and {\boldmath$\beta$} are obtained as {\boldmath$\alpha$}, {\boldmath$\beta$}$\sim {\bf L}${\boldmath$\xi$} where ${\bf L}$ is the lower triangular matrix of the Cholesky decomposition of the covariance matrix ${\bf \Sigma}$, i.e., ${\bf \Sigma} \equiv {\bf L} {\bf L}^{\rm T}$. See also Appendix C of \cite{Sato-Polito:2021efu}.

Simulated pulsar timing residuals following the above prescription are shown in Figure~\ref{fig:residuals_gauss}. In this case, we consider a power law power spectrum \eqref{eq:power_spectral_density_power_law} and the Hellings-Downs curve $\Gamma^{\rm HD}\left({\hat{e}_a \cdot \hat{e}_b}\right) = \left( 1 + \delta_{ab} \right) {\bf \Gamma}\left(f \right)$ for the spatial correlation. This corresponds to an unpolarized SGWB signal; with a continuous support in $f \in (1, 100)$ nHz. Then, the nontrivial elements of the covariance matrix are given by
\begin{equation}
    \langle \alpha_{a{j}} \alpha_{bk} \rangle
    = \Gamma^{\rm HD}\left( \hat{e}_a \cdot \hat{e}_b \right) \int_{f_{\rm min}}^{f_{\rm max}} df S(f) {\cal C}^{\alpha}_{{j},k}(f)  
\end{equation}
and
\begin{equation}
    \langle \beta_{a{j}} \beta_{bk} \rangle
    = \Gamma^{\rm HD}\left( \hat{e}_a \cdot \hat{e}_b \right) \int_{f_{\rm min}}^{f_{\rm max}} df S(f) {\cal C}^{\beta}_{{j},k}(f)\,, 
\end{equation}
where $f_{\rm min}=1$ nHz and $f_{\rm max}=100$ nHz. Note that we have pulled out the HD curve, $\Gamma^{\rm HD}\left({\hat{e}_a \cdot \hat{e}_b}\right)$, out of the integral. Also, $\langle \alpha_{a{j}} \beta_{bk} \rangle = \langle \beta_{a{j}} \alpha_{bk} \rangle = 0$ for an unpolarized SGWB signal.

For comparison, pulsar timing data simulations with and without transfer functions as well as with and without white noise are presented in Figure~\ref{fig:residuals_gauss}. Three representative pulsars are shown. For white noise, we draw the usual pulsar noise parameters EFAC, EQUAD, and ECORR \cite{Babak:2024yhu} from a uniform distribution, i.e., $\text{EFAC} \sim U(0.8, 1.1)$, $\text{EQUAD} \sim U(-8, -6)$, and $\text{ECORR} \sim U(-8, -6)$. The white noise is added to the simulated residuals after the SGWB signal is generated. Generally, a white noise component is uncorrelated in time and frequency, and across pulsars. This is the case with EFAC and EQUAD. As a special case, ECORR is associated to pulsar jitter noise which is correlated across frequency subbands.

Figure \ref{fig:residuals_gauss} demonstrates that the SGWB signal is clearly visible in the timing residuals, even when white noise is present. However, the inclusion of transfer functions is crucial for accurately capturing the high-frequency features of the SGWB in the residuals, as illustrated by the differences between the top and bottom panels of Figure~\ref{fig:residuals_gauss}. At higher levels of white noise, these high-frequency components can be obscured. Moreover, transfer functions are essential for recovering the correct amplitude of the SGWB signal: without them, the simulated residuals have amplitudes roughly half those obtained when transfer functions are included. Notably, the implementation with transfer functions yields SGWB signal amplitudes and statistics that are in agreement with those produced by direct superposition of gravitational waves from individual sources \cite{Hobbs:2009yn, Bernardo:2024tde}.

\section{Number of pulsars}
\label{sec:number_pulsars}

\begin{figure*}[t]
\centering
\subfigure[]{
    \includegraphics[width=0.45\textwidth]{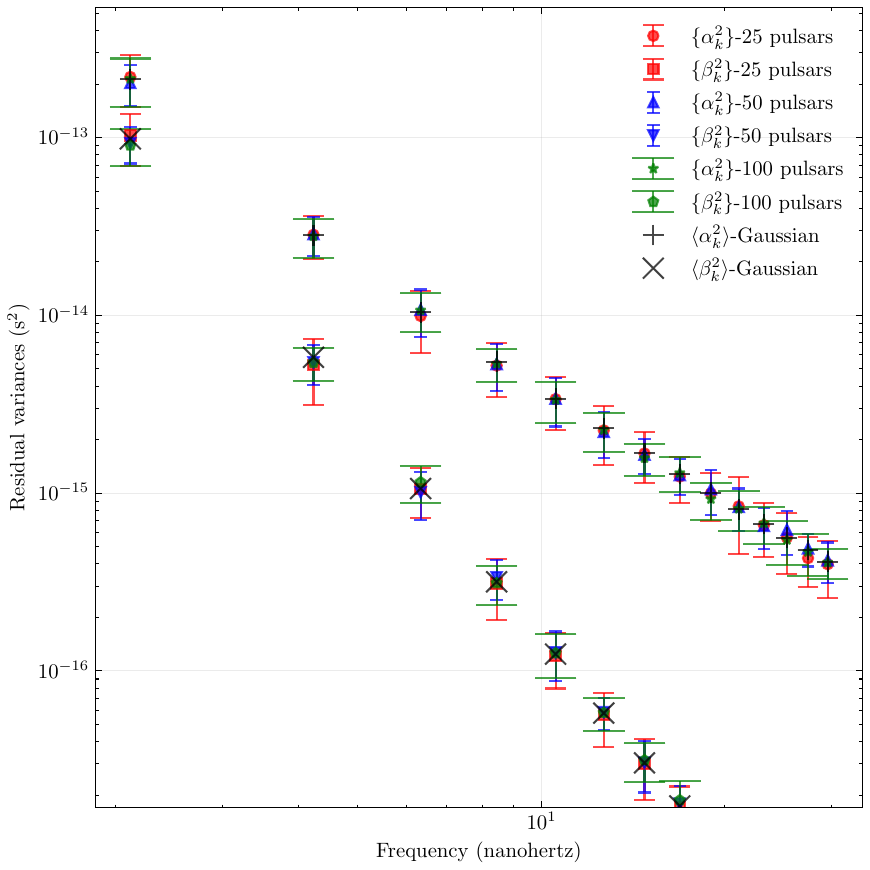}
}
\subfigure[]{
    \includegraphics[width=0.45\textwidth]{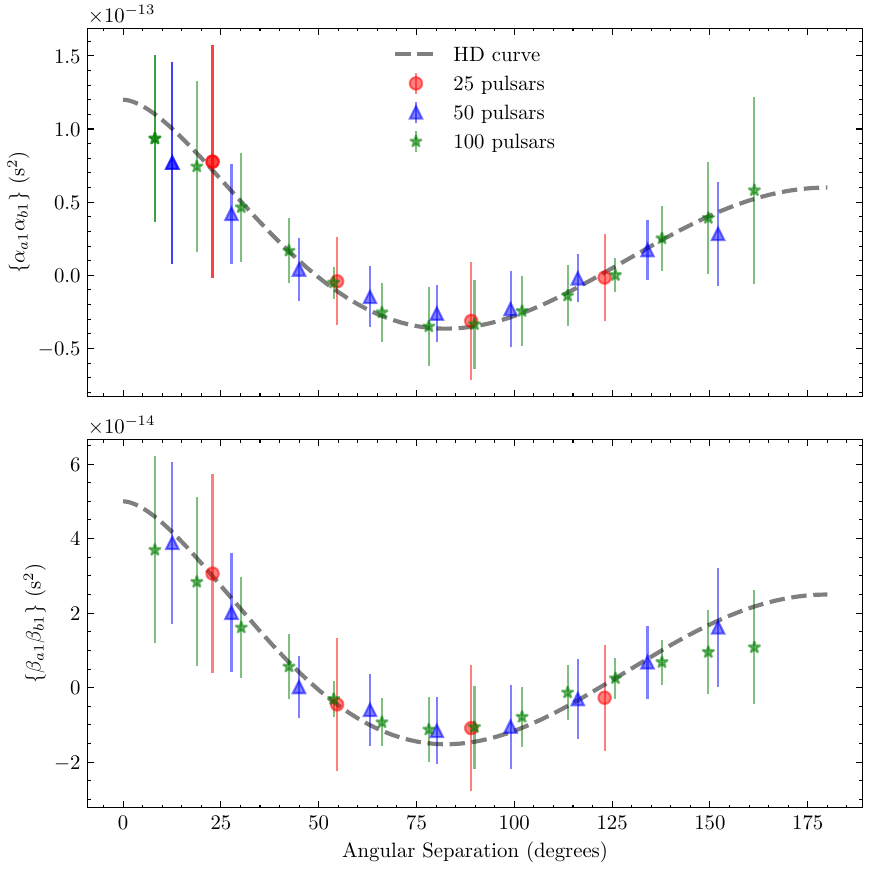}
}
\subfigure[]{
    \includegraphics[width=0.45\textwidth]{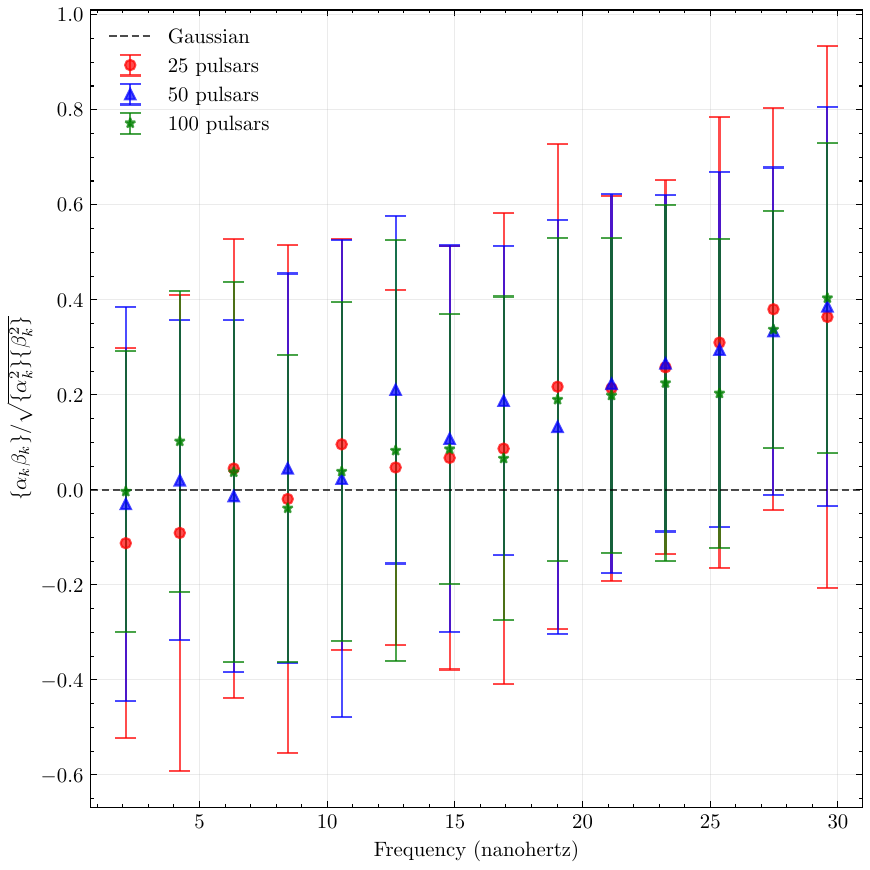}
}
\subfigure[]{
    \includegraphics[width=0.45\textwidth]{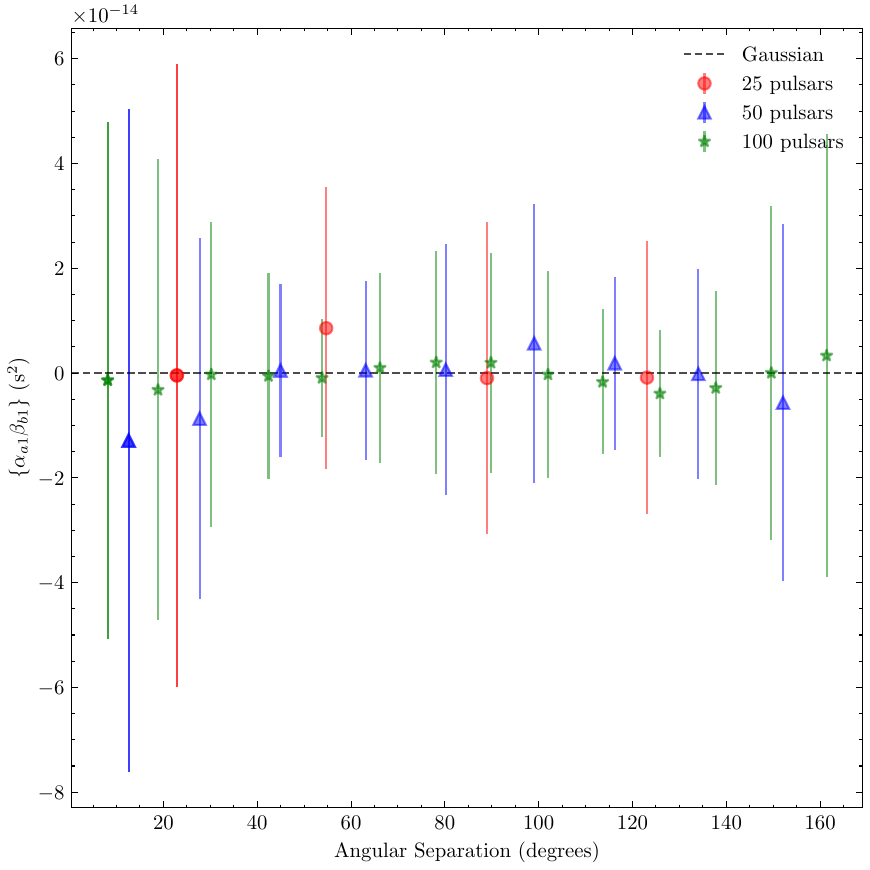}
}
\caption{Fourier components of timing residual correlation due to a Gaussian SGWB signal for different number of pulsars in 15-year \texttt{PTAfast} simulations: (a) sample variances, (b) two-point correlation function, (c) cross-Fourier statistic $\{ \alpha_{k} \beta_{k} \}/\sqrt{ \{ \alpha_k^2 \} \{ \beta_k^2 \} }$, and (d) cross-Fourier spatial correlation $\{ \alpha_{a1} \beta_{b1} \}$. Red, blue, and green points refer to 25, 50, and 100 pulsars, respectively, and the error bars represent the standard error across $50$ realizations. Black points and dashed curves show the Gaussian theoretical expectation values based on Eqs.~(\ref{eq:xy_correlation}-\ref{eq:F_functional}) and Eqs.~(\ref{eq:alpha_filter}-\ref{eq:beta0_filter}).}
\label{fig:residual_variances_and_correlation_npsrs}
\end{figure*}

We present results complementary to Section \ref{sec:comparison_with_simulation} for varying number of pulsars in Figure~\ref{fig:residual_variances_and_correlation_npsrs}.

The number of pulsar pairs for a $n_{\rm psrs}$ array is given by $n_{\rm psrs}(n_{\rm psrs}-1)/2$. The number of pulsar pairs is important for the significance of the SGWB signal in a realization. In Figure~\ref{fig:residual_variances_and_correlation_npsrs}, we show the Fourier components of timing residual correlation due to a Gaussian SGWB signal for different number of pulsars in 15-year \texttt{PTAfast} simulations. The results are shown for 25, 50, and 100 pulsars, respectively. The error bars represent the standard error across $50$ realizations. The black points and dashed curves show the Gaussian theoretical expectation values based on Eqs.~(\ref{eq:xy_correlation}-\ref{eq:F_functional}) and Eqs.~(\ref{eq:alpha_filter}-\ref{eq:beta0_filter}). See Appendix \ref{sec:simulation} for details on the simulation procedure.

All cases are consistent with the theoretical expectation values of the variances and the Hellings-Downs curve. However, the size of the ensemble variances (visible as error bars in Figure~\ref{fig:residual_variances_and_correlation_npsrs}) depends on the number of pulsars in the array. With more pulsars, there are more independent pulsar pairs contributing to each angular bin, which improves the statistical sampling and reduces the variance across different realizations. As a result, the error bars shrink as the number of pulsars increases. In the limit of a very large array ($n_{\rm psrs} \to \infty$), the ensemble variance approaches the cosmic variance \cite{Roebber:2016jzl, Allen:2022dzg, Allen:2022ksj, Bernardo:2022xzl}. For a finite array, the SGWB signal in a given realization can deviate from the theoretical mean by {down} to the cosmic variance, but increasing the number of pulsars systematically reduces this variation, as seen in Figure~\ref{fig:residual_variances_and_correlation_npsrs}.

\section{{Quick recipe for computing the elements of the covariance matrix}}
\label{sec:summary_covariance}

{
We put together formulae for computing the elements of the covariance matrix in the frequency- and Fourier-domain (Section \ref{sec:standard_analysis_of_the_signal}).
}

{
For an {\it unpolarized} SGWB, the relevant Fourier components of the correlation are given by
\begin{equation}
\label{eq:aa_correlation_summary}
    \langle \alpha_{aj} \alpha_{bk} \rangle = {\cal F}\left[ \sum_{\ell m}  I_{\ell m}(f) \gamma_{\ell m}^{I}(fD_a, fD_b, \hat{e}_a, \hat{e}_b), \mathcal{C}_{j,k}^{\alpha}(f) \right] \,,
\end{equation}
and
\begin{equation}
\label{eq:bb_correlation_summary}
    \langle \beta_{aj} \beta_{bk} \rangle = {\cal F}\left[ \sum_{\ell m}  I_{\ell m}(f) \gamma_{\ell m}^{I}(fD_a, fD_b,\hat{e}_a, \hat{e}_b), \mathcal{C}_{j,k}^{\beta}(f) \right] \,,
\end{equation}
where the functional ${\cal F}$ is defined as
\begin{equation}
\label{eq:F_functional_summary}
    {\cal F}[G(f, y), {\cal C}(f)] = \int_{f_{\rm min}}^{f_{\rm max}} \frac{2\,df}{(2\pi f)^2} G(f, y) {\cal C}(f) \,,
\end{equation}
and ${\cal C}_{j,k}^{\alpha/\beta}(f)$ are given by
\begin{align}
{\cal C}_{j,k}^\alpha (f) &= \dfrac{4 f_j f_k }{\pi^2 T^2} \dfrac{\sin^2(\pi f T)}{(f^2 - f_j^2)(f^2 - f_k^2)} \,, \\
{\cal C}_{j,k}^\beta (f) &= \dfrac{4 f^2 }{\pi^2 T^2} \dfrac{\sin^2(\pi f T)}{(f^2 - f_j^2)(f^2 - f_k^2)} \,.
\end{align}
For a {\it circularly polarized} SGWB, the relevant Fourier components of the correlation pertaining to circular polarization are given by
\begin{equation}
\label{eq:cp_correlation_summary}
    \langle \alpha_{aj} \beta_{bk} \rangle = {\cal F}\left[ \sum_{\ell m} iV_{\ell m}(f) \gamma_{\ell m}^{V}(fD_a, fD_b,\hat{e}_a, \hat{e}_b), \mathcal{C}_{j,k}^{\alpha\beta}(f) \right] \,,
\end{equation}
where ${\cal C}_{j,k}^{\alpha\beta}(f)$ are given by
\begin{align}
{\cal C}_{j,k}^{\alpha\beta} (f) &= - \dfrac{4 f_j f }{\pi^2 T^2} \dfrac{\sin^2(\pi f T)}{(f^2 - f_j^2)(f^2 - f_k^2)} \,, \nonumber \\
{\cal C}_{j,k}^{\beta\alpha} (f) &= \dfrac{4 f_k f }{\pi^2 T^2} \dfrac{\sin^2(\pi f T)}{(f^2 - f_j^2)(f^2 - f_k^2)} \,.
\end{align}
The $I_{lm}(f)$'s and $V_{lm}(f)$'s are spherical harmonic components of the Stokes parameters of the SGWB power spectra. The overlap reduction functions (ORFs) $\gamma^{X}_{lm}(f, x)$'s can be calculated as described in Refs.~\cite{Chu:2021krj, Liu:2022skj} or in the computational frame \cite{Kato:2015bye,Sato-Polito:2021efu} using \texttt{PTAfast} \cite{2022ascl.soft11001B, Bernardo:2022rif, Bernardo:2022xzl, Bernardo:2023jhs}.
}

\begin{figure*}[t]
\centering
\subfigure[]{\includegraphics[width=0.9\textwidth]{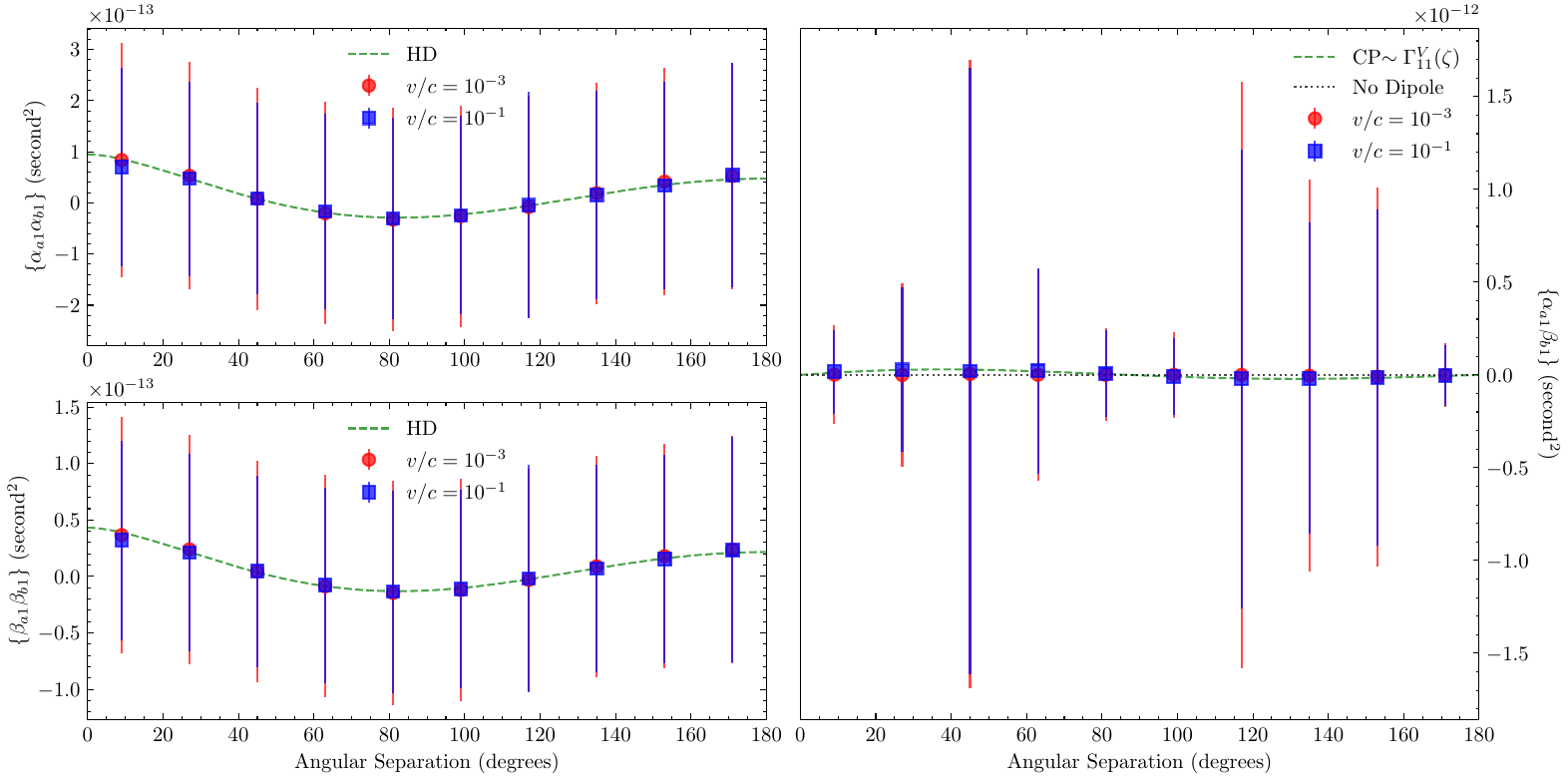}}
\subfigure[]{\includegraphics[width=0.43\textwidth]{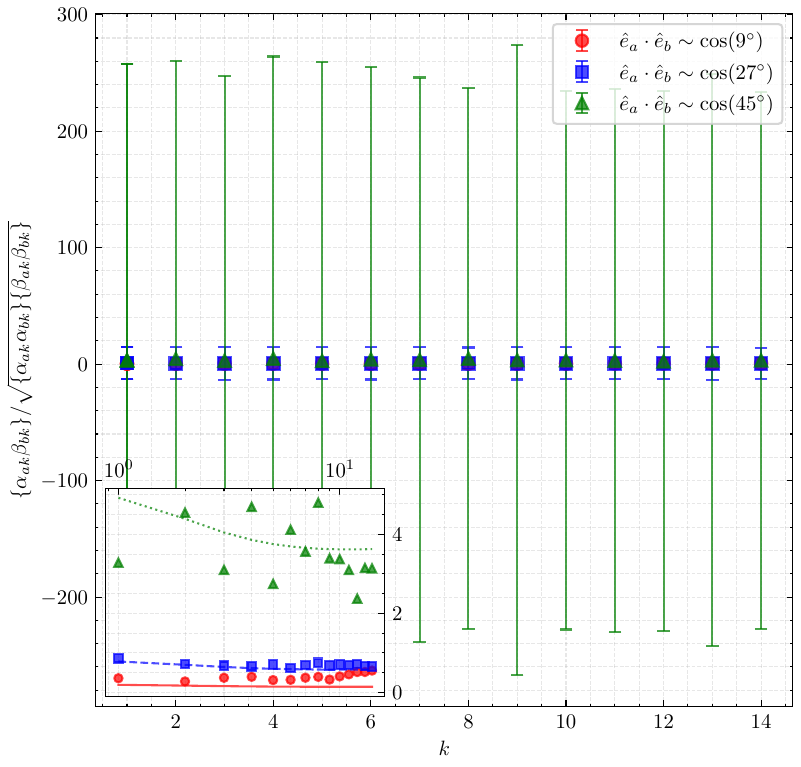}}
\subfigure[]{\includegraphics[width=0.43\textwidth]{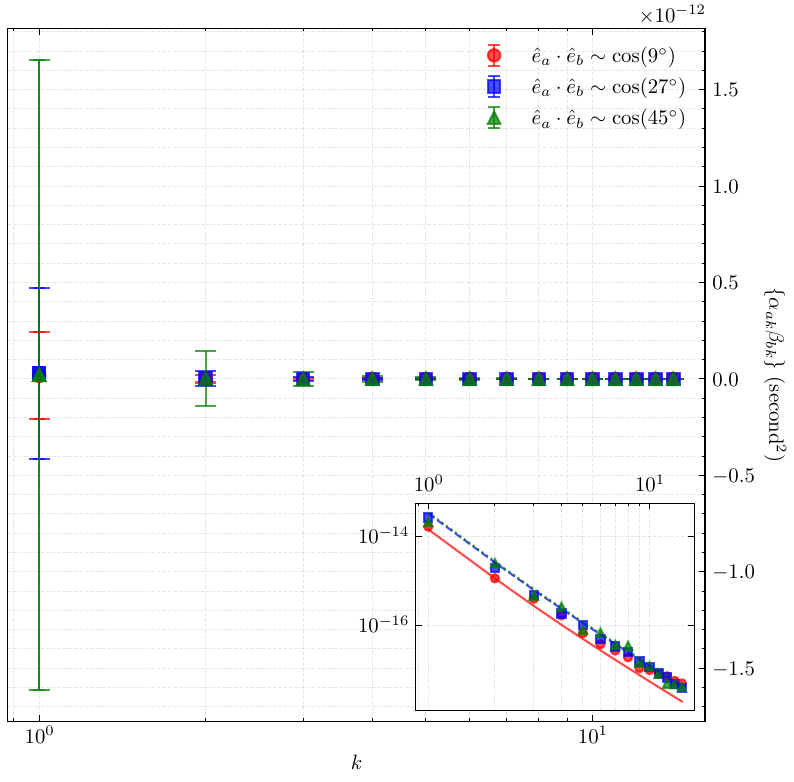}}
\caption{{[Top panel] Two-point correlation functions and [bottom panel] the bin power of the Fourier coefficients of pulsar timing residuals for a simulated 15-year PTA data {set} with a circularly polarized SGWB signal. The parameters are exactly the same as Figure \ref{fig:correlation_cp}, except the error bars shown are the standard errors (variations across realizations); the inset plots at the bottom panel show the means. In the top panel, the green dashed lines are the theoretical expectation values. In the insets of the bottom panel, the colored curves (red-solid, blue-dashed, and green-dotted) represent the theoretical expectation values of the frequency-bin power of the cross-Fourier statistic $\{ \alpha_{ak} \beta_{bk} \}$ at $9^\circ, 27^\circ, 45^\circ$ angular bins.}}
\label{fig:correlation_cp_stderrs}    
\end{figure*}

\section{{Standard errors of the kinematic dipole correlations}}
\label{sec:stderrs_dipole}

{
The results of simulations of a circularly polarized PTA signal in Section \ref{sec:comparison_with_simulation_cp} or Figure \ref{fig:correlation_cp} with standard errors are shown in Figure \ref{fig:correlation_cp_stderrs}. The standard errors SE are related to the standard error on the mean SEM via SEM$=$SE$/\sqrt{n_{\rm reals}}$ where $n_{\rm reals}$ is the number of independent realizations of an experiment or a simulation. The SEM is useful for highlighting the existence of a signal, since it can be reduced indefinitely by increasing the number of independent realizations. On the other hand, standard errors are representative of the variation across realizations for multiple repeated simulations of an experiment.
}

\section{{The anisotropic unpolarized signal}}
\label{sec:anisotropic_unpolarized}

\begin{figure*}[t]
\centering
\subfigure[]{\includegraphics[width=0.45\textwidth]{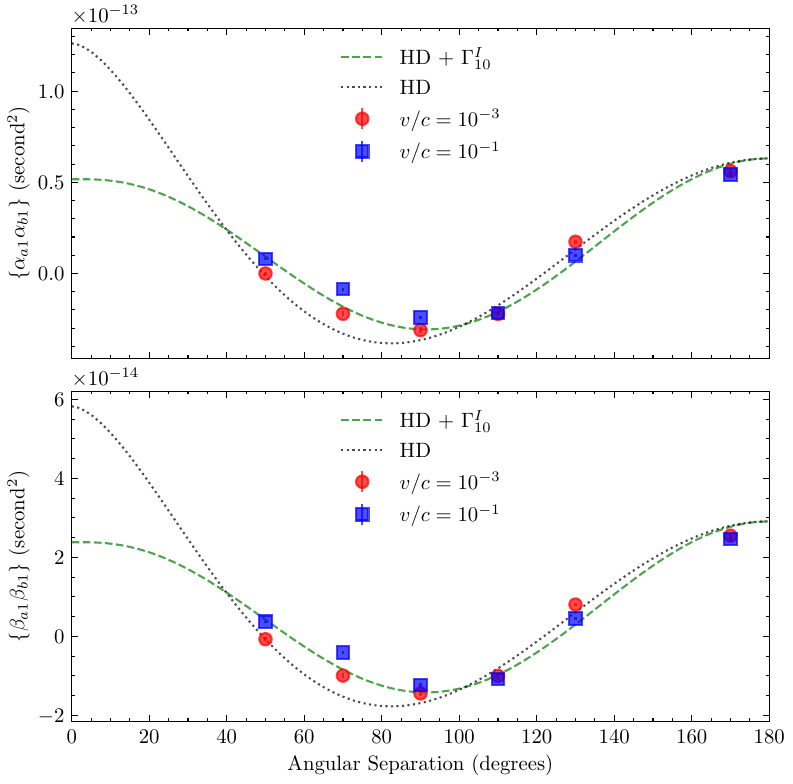}}
\subfigure[]{\includegraphics[width=0.45\textwidth]{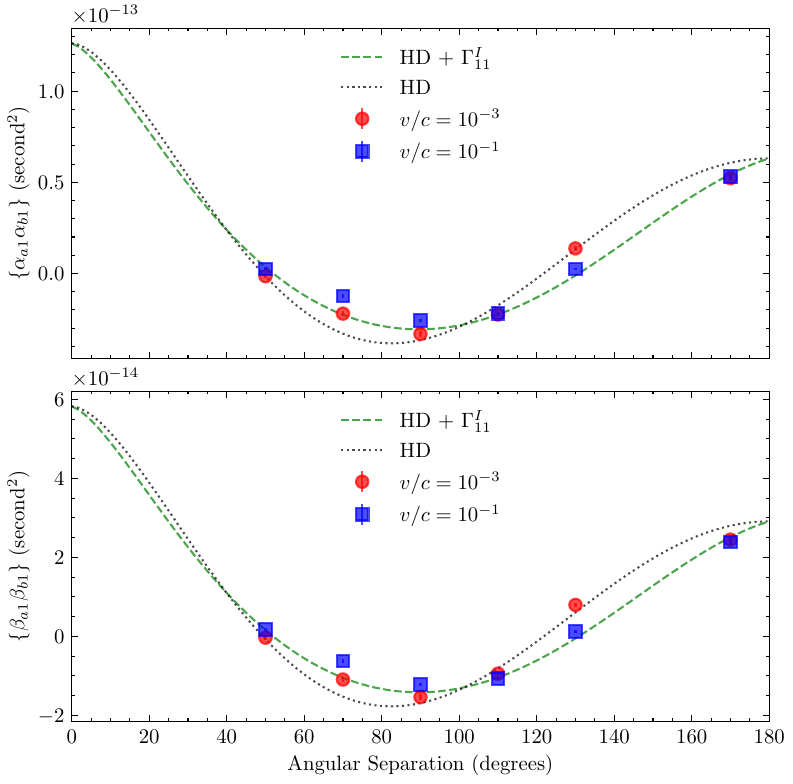}}
\caption{{Correlation functions in the purely anisotropic case with (a) $\Gamma_{10}^I$ and (b) $\Gamma_{11}^I$ in simulated 15-year PTA data. HD+$\Gamma_{10}^I$ and HD+$\Gamma_{11}^I$ in the legends are shorthands for the full terms in \eqref{eq:GammaI} or \eqref{eq:Ilmgammalm}; with kinematic dipole speeds $\beta=v/c=10^{-3}$ (round points) and $10^{-1}$ (square points). Each of $20000$ realizations has 10 pulsars at fixed positions; in (a) these are arranged so that $\cos(\theta_v)=1$, $\sin(\theta_v)=0$, in (b) $\cos(\theta_v)=0$, $\sin(\theta_v)=\cos(\phi_v)=1$. The error bars are the error on the mean. The SGWB strain and spectral index are $A_{\rm gw}=10^{-15}$ and $\gamma_{\rm gw}=13/3$, respectively. The green dashed lines represent the Gaussian theoretical expectation values and the dotted line is the HD curve.}}
\label{fig:correlation_anisotropy}    
\end{figure*}

{
It might be instructive to look at the anisotropic unpolarized case, in addition to the anisotropic and circularly polarized one presented in the main text. These were generated as a special case of the methodology discussed in the main text, simply by setting $V_{00}(f) = 0$ or $\Pi = 0$ (zero circular polarization); the relevant corrections to the correlation appear in \eqref{eq:GammaI} or \eqref{eq:Ilmgammalm} whereas the terms that are contained in \eqref{eq:GammaV} or \eqref{eq:Vlmgammalm} reduce to zero with $\Pi=0$. In this case, sines and cosines will be uncorrelated, and so we may focus on the sample averages $\{ {\alpha_{ak} \alpha_{bk}} \}$ and $\{ {\beta_{ak} \beta_{bk}} \}$ in each realization. The results in the first frequency bin with a SGWB characteristic strain $A_{\rm gw}=10^{-15}$, spectral index $\gamma_{\rm gw}=13/3$, kinematic dipole speeds $v/c=10^{-3}$ and $10^{-1}$ are shown in Figure \ref{fig:correlation_anisotropy}. For this set of simulations, in each realization, a mock 15-year PTA data with 10 pulsars at fixed positions are generated. A total of 20000 realizations were considered.  
}

{
The reason for considering 10 pulsars and 20000 realizations is the following. Two anisotropic corrections (${\cal O}(v/c)$ terms in \eqref{eq:GammaI} or \eqref{eq:Ilmgammalm}) with distinct computational frame ORFs (${\bf \Gamma}_{10}^I(\hat{e}_a\cdot\hat{e}_b)$ and ${\bf \Gamma}_{11}^I(\hat{e}_a\cdot\hat{e}_b)$ trail the HD curve in the sine-sine and cosine-cosine correlations). Thus, to look into each one, rather than both at the same time, we utilized pulsars that are in a very specific configuration relative to the direction of the kinematic dipole. In particular, to obtain the HD correlation plus the leading anisotropic ${\bf \Gamma}_{10}^I(\hat{e}_a\cdot\hat{e}_b)$ correction, we considered a pulsar-dipole configuration with $\cos(\theta_v)=1$, $\sin(\theta_v)=0$ (pulsars on the plane of the dipole and pulsar $a$). For the ${\bf \Gamma}_{11}^I(\hat{e}_a\cdot\hat{e}_b)$ correction, we considered $\cos(\theta_v)=0$, $\sin(\theta_v)=\cos(\phi_v)=1$ (dipole and pulsar $a$ point to the same direction whereas the rest of the pulsars lie on a plane). However, these specific configurations correspond to pulsars that lie on a plane instead of scattered anisotropically. In a fixed frame, points that are uniformly distributed on the two-sphere follow a distribution function $(\sin \theta) / 2$, and that the fraction that would be on a plane (e.g., positive $xz$ plane) will be $\sin( \Delta \theta / 2)$ where $\Delta \theta$ is a resolution. With an angular resolution of a degree, this fraction is approximately $\sin(\pi/360) \sim 0.009$. Should there be a 1000 pulsars that are distributed uniformly, then around 10 can be expected to lie on any fixed plane given a one-degree resolution. Then we utilize only 10 pulsars based on this rough estimate to simulate the anisotropic corrections with correlation functions ${\bf \Gamma}_{10}^I(\hat{e}_a\cdot\hat{e}_b)$ and ${\bf \Gamma}_{11}^I(\hat{e}_a\cdot\hat{e}_b)$. To confront the limited sample statistics with a small number of pulsars, we have increased the number of realizations to 20000, which is twice compared to that used throughout the main text. 
}

{
Figure \ref{fig:correlation_anisotropy} shows that both anisotropic correlations ${\bf \Gamma}^{I}_{10}(\hat{e}_a\cdot\hat{e}_b)$ and ${\bf \Gamma}^{I}_{11}(\hat{e}_a\cdot\hat{e}_b)$ are recovered consistently in simulations when the dipole's speed is artificially increased to $v/c=10^{-1}$ (a tenth of the speed of light). When the dipole speed is $v/c=10^{-3}$, the results reduce to the HD correlation. It must be mentioned that because there are only ten pulsars that there are only nine pulsar pairs in a frame dependent search, and fewer angular bins. Arguably, the results with anisotropy are expected to be challenging to distinguish to the HD curve even when the dipole's speed has already been increased unrealistically; again, this is because the terms are suppressed by $v/c$ relative to the HD curve in the sine-sine and cosine-cosine correlations. This is one reason we preferred to focus on circular polarization in the main text, because the dipole signal of circular polarization alone correlates the sine and cosine coefficients.
}


%

\end{document}